\documentclass[smallextended]{svjour3}       

\usepackage{amsthm}
\theoremstyle{definition}

\usepackage{subcaption}

\def\I {\mathbbm i}
\newcommand{\mb}[1]{{#1}}
\DeclareMathAlphabet{\mathbb}{U}{msb}{m}{n}
\DeclareMathAlphabet{\mathcalprime}{OMS}{zplm}{m}{n}

\usepackage{bbm}
\def~{\hphantom{0}}

\RequirePackage{fix-cm}
\usepackage{mathrsfs}
\usepackage{ogonek}
\newtheorem{condition}{Condition}
\newtheorem{thm}{Theorem}
\newtheorem{prop}{Proposition}
\newtheorem{prpty}{Property}

\smartqed  
\usepackage{graphicx}
\usepackage{amsmath}
\usepackage{mathptmx}      

\usepackage{enumerate}
\usepackage[misc]{ifsym}
\newcommand{\CovC}{\mathscr{C}}
\usepackage{xcolor}
\definecolor{darkblue}{rgb}{0.0,0.0,0.6}
 \definecolor{darkbrown}{rgb}{0.6, 0.25, 0.15}
\usepackage{hyperref}
\hypersetup{colorlinks,breaklinks,linkcolor=darkblue,urlcolor=darkblue,
            anchorcolor=darkblue,citecolor=darkblue}
\journalname{Mathematical Geosciences}

\usepackage[authoryear]{natbib}

\makeatletter
\providecommand{\@LN}[2]{}
\makeatother
\usepackage{multirow}
\usepackage{url}
\usepackage{amssymb}

\begin{document}

\title{Multivariate Confluent Hypergeometric Covariance Functions with Simultaneous Flexibility over Smoothness and Tail Decay}

\titlerunning{Multivariate Confluent Hypergeometric}

\author{Drew Yarger    \and
        Anindya Bhadra}
\institute{Drew Yarger (\Letter) ORCID: 0000-0002-0420-7346 \at
              Department of Statistics, Purdue University,\\
              150 N. University St., West Lafayette, Indiana 47907-2066, U.S.A. \\ 
              \email{anyarger@purdue.edu} 
           \and
           Anindya Bhadra (\Letter) ORCID: 0000-0003-0636-5273 \at
              Department of Statistics, Purdue University,\\
              150 N. University St., West Lafayette, Indiana 47907-2066, U.S.A. \\ 
              \email{bhadra@purdue.edu}
}
\date{Received: date / Accepted: date}

\maketitle

\begin{abstract}
Spatially-indexed multivariate data appear frequently in geostatistics and related fields including oceanography and environmental science.
To take full advantage of this data structure, cross-covariance functions are constructed to describe the dependence between any two component variables at different spatial locations. 
Modeling of multivariate spatial random fields requires these constructed cross-covariance functions to be valid, which often presents challenges that lead to complicated restrictions on the parameter space. 
The purpose of this work is to present techniques using multivariate mixtures for establishing validity that are simultaneously simplified and comprehensive. 
In particular, cross-covariances are constructed for the recently-introduced {confluent hypergeometric (CH)} class of covariance functions, which has slow (polynomial) decay in the tails of the covariance that better handles large gaps between observations in comparison with other covariance models. 
In addition, the spectral density of the confluent hypergeometric covariance is established and used to construct new valid cross-covariance models. 
The approach leads to valid multivariate cross-covariance models that inherit the desired marginal properties of the confluent hypergeometric model and outperform the multivariate Mat\'ern model in out-of-sample prediction under slowly-decaying correlation of the underlying multivariate random field.  
The model captures heavy tail decay and dependence between variables in an oceanography dataset of temperature, salinity and oxygen, as measured by autonomous floats in the Southern Ocean.

\end{abstract}

\section{Introduction}

Spatially-indexed multivariate data appear frequently in 
geosciences, for example, in the fields of mineral resource modeling \citep{maleki2017joint, avalos_spatial_2023, emery2022new} and groundwater hydrology \cite[][and references therein]{dowd_many_2024}.
The book by \cite{wackernagel_multivariate_2010} overviews geostatistical methodology for such data, 
which are also common in climate and the environmental sciences \citep{mitchell_l_krock_modeling_2023, gneiting2010matern}.
However, multivariate model construction poses some unique challenges: it is not enough to construct a valid univariate covariance function, which must be positive semidefinite, but the entire matrix of marginal and cross-covariances must be valid. 
Moreover, one ideally should retain flexibility in the marginal covariances (for example, the covariances' origin and/or tail behaviors) and introduce flexibility in the cross-covariances.
The second goal often conflicts with the first; to construct a valid {joint} model, flexibility of the individual covariance functions may be lost. 
Concerned with this aspect, much of the work in the literature \citep[see, for example,][]{gneiting2010matern, apanasovich2012valid} has focused on simple parametric covariance functions, such as Mat\'ern, and considered how to build a valid cross-covariance using this component as the marginal model. 
The challenge with this approach is that establishing validity for the {joint} model often leads to complicated and artificial restrictions which must be worked out in a dimension-specific manner. 
This happens because one considers two processes at a time, and the desired validity conditions are combined post hoc. 
Thus, the construction of a valid covariance function for two correlated series, for example, temperature and oxygen over a spatial region, must be re-done when one adds a third, for example, temperature, oxygen, {and salinity}, over the same region.

To address this issue, the current work establishes validity conditions primarily deploying multivariate mixtures. 
Although the mixture approach has attracted some recent attention \citep[e.g.,][]{emery2022new, emery_schoenberg_2023}, it remains relatively obscure compared to the two-at-a-time approaches mentioned above. 
However, the chief benefit here is that (a) conditions of validity can be established in an identical manner regardless of the dimension and (b) it is still possible to allow considerable flexibility in the marginal and cross-covariance terms.
There are three major aspects of flexibility established in this work.
First, the marginal covariances are no more constrained when modeling them jointly compared to treating them independently.
In addition, this work establishes flexibility in the parameters governing cross-covariances to allow processes to be more or less dependent at different spectral frequencies. 
Finally, the work introduces asymmetric cross-covariances such that the dependence between two processes varies based on the direction of lag between the two processes.

The main concrete running illustration concerns the recently proposed univariate {confluent hypergeometric (CH)} covariance class \citep{ma2022beyond} that rectifies one significant limitation of the Mat\'ern class \citep{matern2013spatial}. 
The main reason for the popularity of Mat\'ern over other covariances is the precise control over the degree of mean square differentiability of the associated random process \citep[see, e.g., Chapter 2 of][]{stein1999interpolation}. 
However, the Mat\'ern class possesses exponentially decaying tails, which could be too fast a rate of decay for some applications. 
For example, if there are large gaps in space between sampled data, this class is challenged in modeling the dependence across these gaps. 
The importance of modeling slow decay in covariance functions has been recognized in geostatistical time series
\citep{montillet2015modeling,wkeglarczyk2009studies} and spatial data \citep{kleiber2015nonstationary,alegria2024hybrid}.
Other covariances such as the generalized Cauchy class admit polynomial tails at the expense of allowing no control over the degree of mean square differentiability. 
The CH class allows the best of both worlds: it contains two parameters that control the degree of mean squared differentiability and the polynomial tail decay rate, respectively, and independently of each other. 
With control over the tail decay, the CH class covers both long-range and short-range dependence \citep[see definitions in][]{de_oliveira_simple_2023}.
The CH class has demonstrated considerable success in one dimension over the Mat\'ern class in simulations and in the analysis of atmospheric CO$_2$ data \citep{ma2022beyond}. 
A valid multivariate generalization, however, has so far remained elusive. 
The current article provides a solution.

By constructing a multivariate CH model as a continuous mixture of a multivariate Mat\'ern model, this work constructs a valid model that provides full flexibility in the origin behavior, tail behavior, and scale of each of the marginal covariances, as well as flexibility in the cross-covariances. 
In addition, the spectral density of the CH covariance is established at all frequencies, whereas \citet{ma2022beyond} only established its tail behavior. 
This gives additional tools to establish cross-covariance models. 
For example, \cite{yarger2023multivariate} proposed using the spectral density to establish asymmetric multivariate Mat\'ern models, and the approach is used to construct new asymmetric CH cross-covariances here. 
Throughout, this work limits its attention to stationary models.

\begin{figure}[!t]
    \centering
    \begin{subfigure}[t]{0.48\textwidth}
    \includegraphics[width = .88\textwidth]{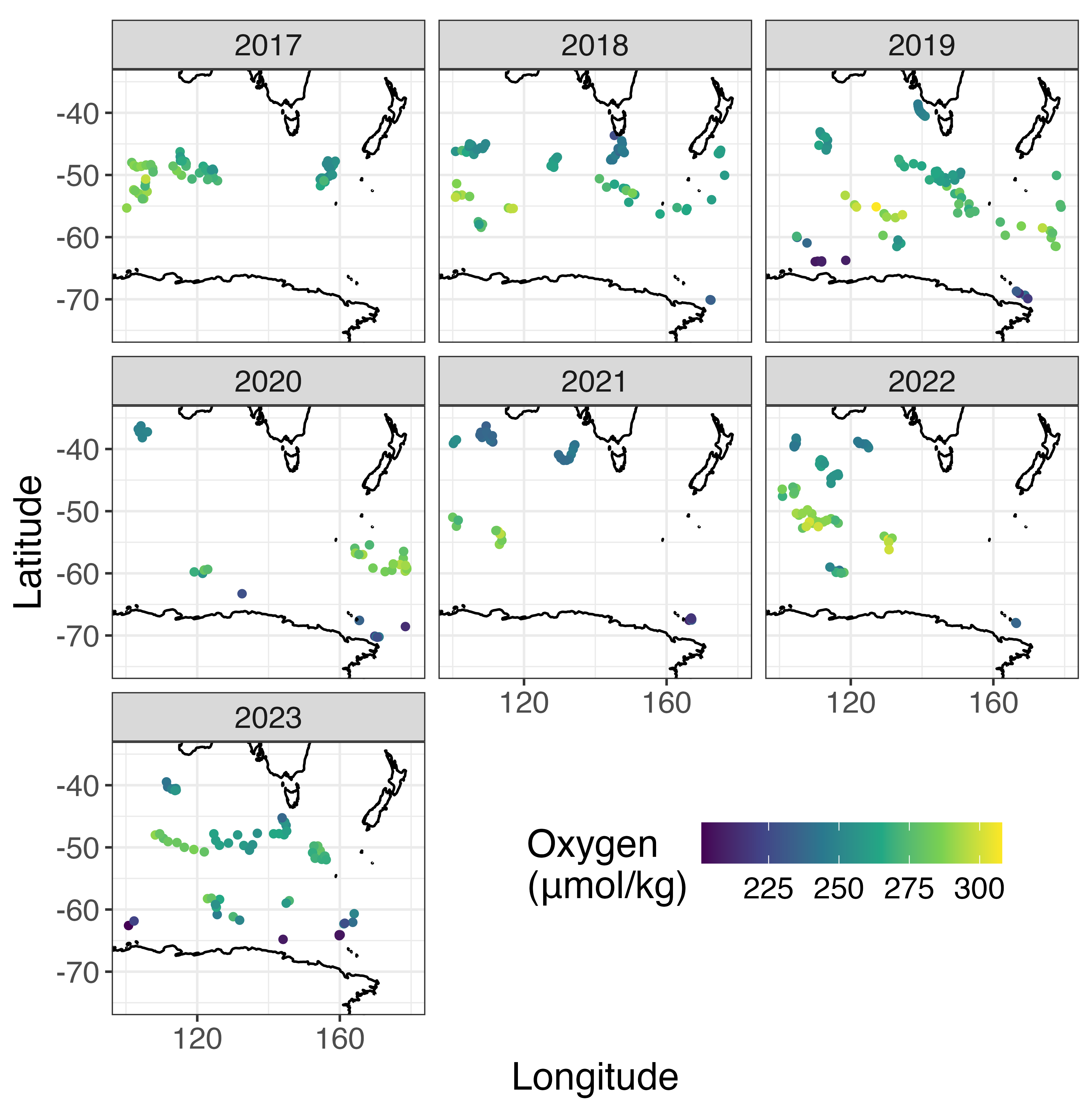}
    \caption{}
    \end{subfigure}
    \begin{subfigure}[t]{0.48\textwidth}
    \includegraphics[width = .86\textwidth]{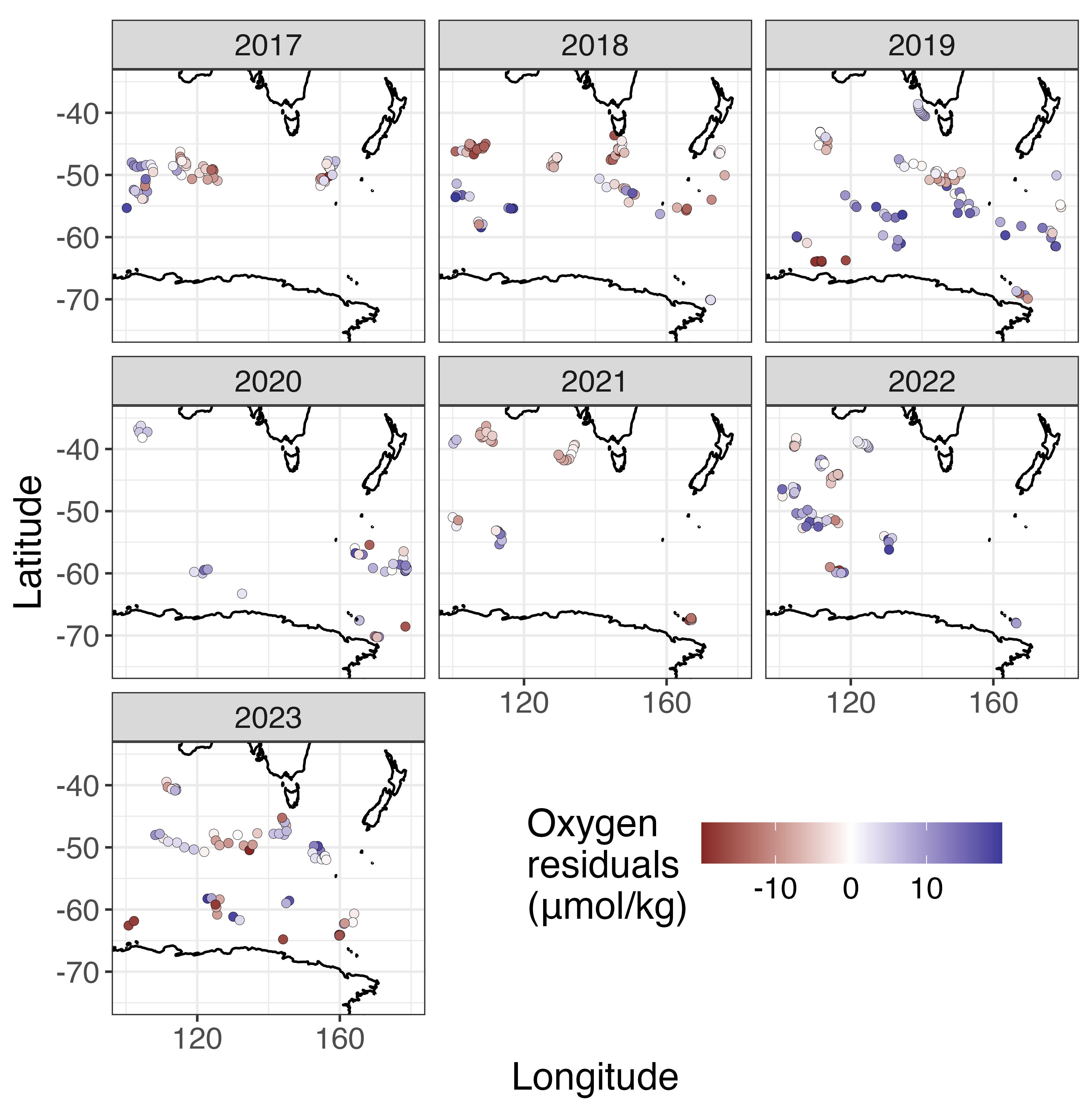}
    \caption{}
    \end{subfigure}

        \begin{subfigure}[t]{0.48\textwidth}
    \includegraphics[width = .86\textwidth]{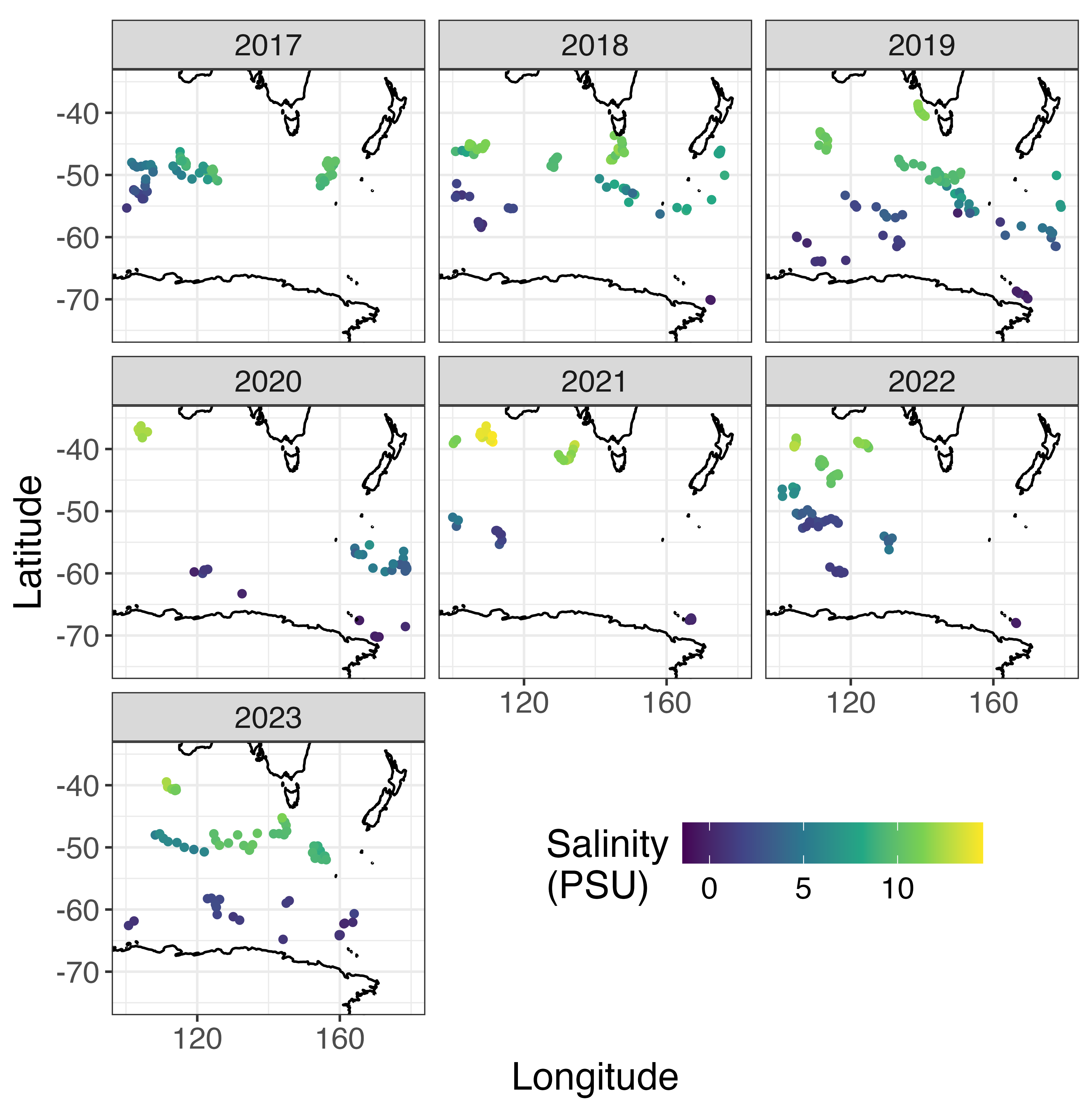}
    \caption{}
    \end{subfigure}
    \begin{subfigure}[t]{0.48\textwidth}
    \includegraphics[width = .86\textwidth]{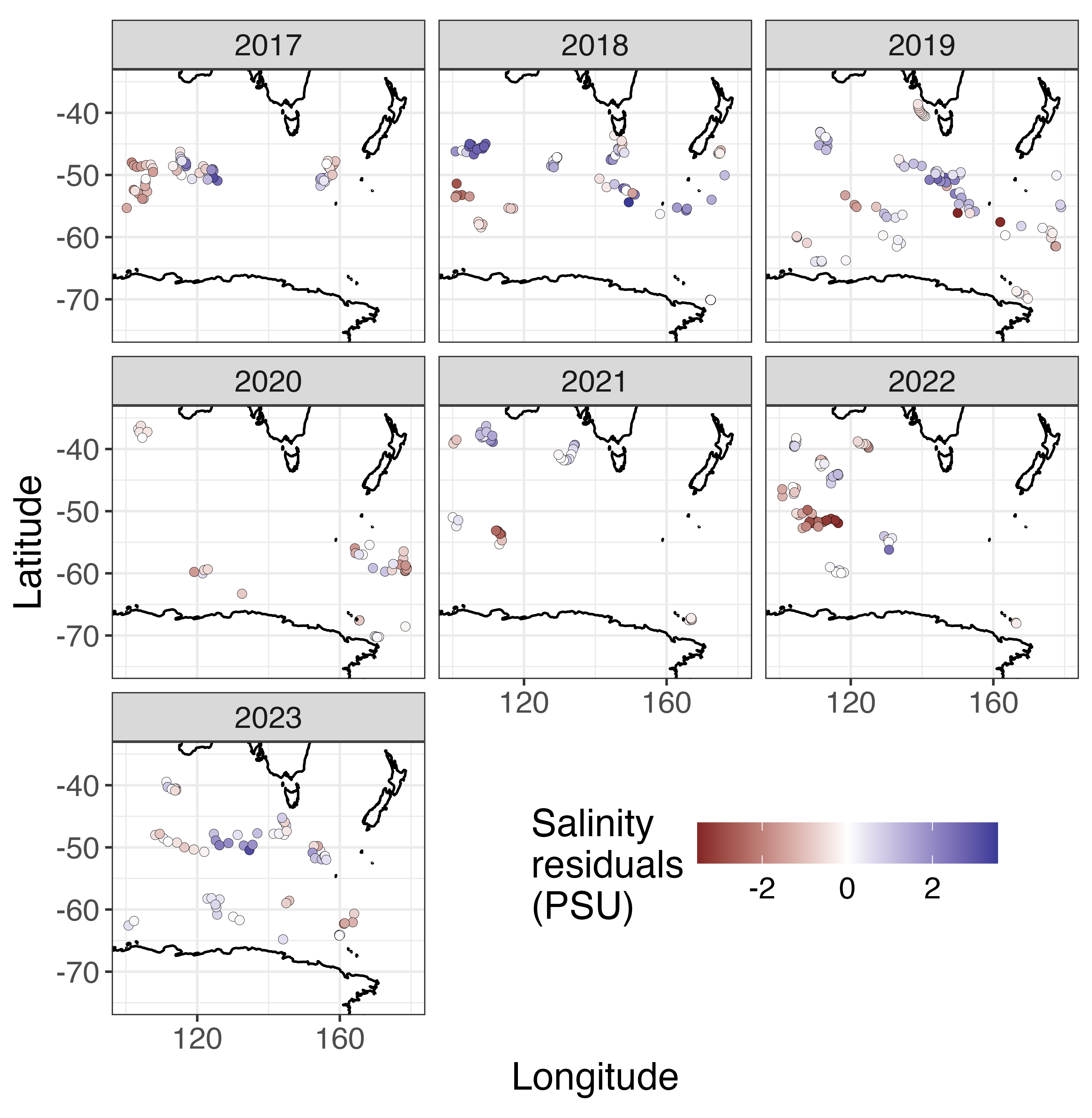}
    \caption{}
    \end{subfigure}

        \begin{subfigure}[t]{0.48\textwidth}
    \includegraphics[width = .86\textwidth]{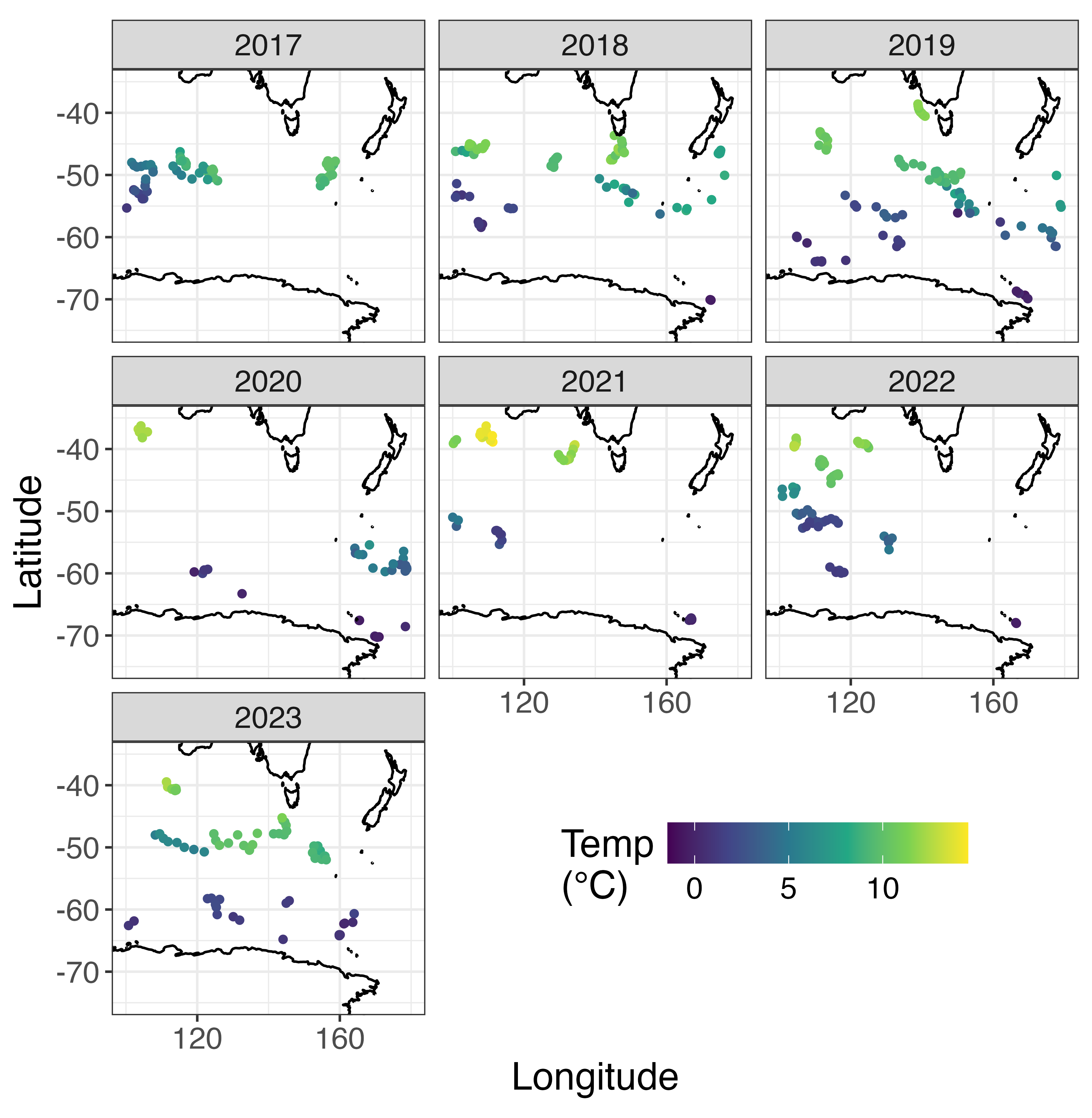}
    \caption{}
    \end{subfigure}
    \begin{subfigure}[t]{0.48\textwidth}
    \includegraphics[width = .86\textwidth]{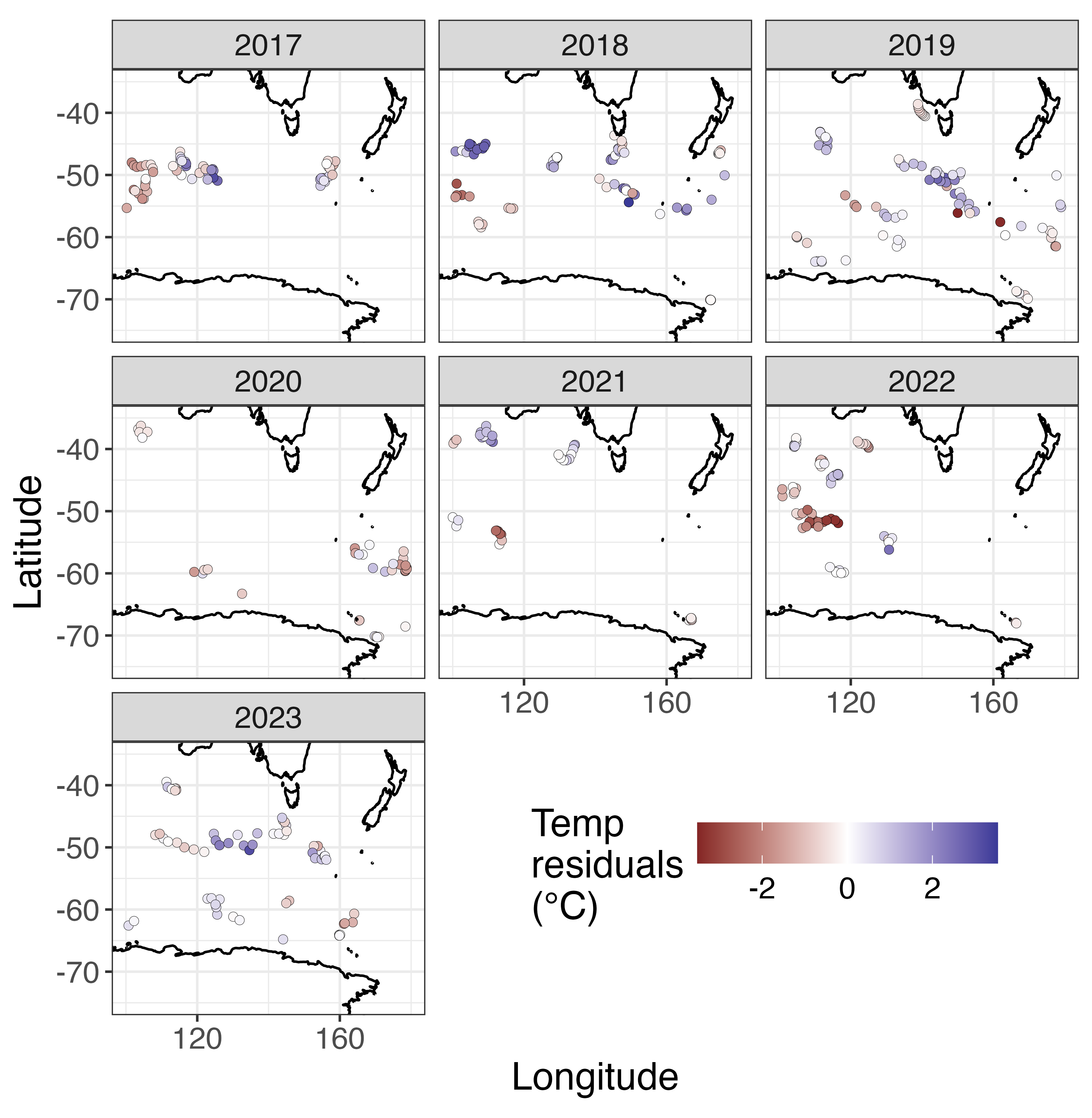}
    \caption{}
    \end{subfigure}
    
    \caption{(Left) Oxygen (a), salinity (c), and temperature (e) measurements taken at 150 meters deep in the ocean during February, March, and April, which were used in the data application (Right) Residuals based on a local polynomial fit}
    \label{fig:argo_intro}
\end{figure}

The multivariate CH and Mat\'ern models are compared in simulation studies and in analysis of an oceanography data set. 
The simulation studies demonstrate the multivariate CH covariance's flexibility compared to the multivariate Mat\'ern in its tail behavior and the generalized Cauchy in its origin behavior.
The data analysis focuses on oceanographic temperature, salinity, and oxygen data collected by devices called floats. 
The data used is plotted in Fig. \ref{fig:argo_intro}. 
Since there are a limited number of such floats, gaps between observations can be hundreds of kilometers. In such a setting, the multivariate CH covariance proves advantageous. 
The proposed technique leverages available temperature and salinity data, which are more abundant, to improve predictions of oxygen, while accounting for polynomial tail decay behavior of the covariances and cross-covariances and ensuring validity.


Some notation used in this work is described here. 
Let $\odot$ denote the Hadamard, or entry-wise product, of two matrices of the same dimension. 
In some sections, as specified later, this notation is omitted and all matrix operations should be taken entry-wise, unless specified otherwise. 
Let $f(z) \overset{z\to \infty}{\asymp} g(z)$ to mean $\lim_{z \to \infty} {f(z)}/{g(z)} = c$ for some $c\in (0,\infty)$, while the notation $f(z) \overset{z\to \infty}{\sim} g(z)$ is reserved for the case $c=1$. 
Let $\lVert x \rVert = ({\sum_{i=1}^d x_i^2})^{1/2}$ be the Euclidean norm in $d$ dimensions.
For a scalar $a\in \mathbb{R}$, let $\lfloor a \rfloor$ denote the floor function, the largest integer less than or equal to $a$. 
Let $\Gamma(a)$ denote the Gamma function and  $B(a,b) = \Gamma(a) \Gamma(b)/\Gamma(a+b) $ denote the Beta function. 
Let $\I$ be the imaginary unit, $\Re(z)$ and $\Im(z)$ denote the real and imaginary parts of $z \in \mathbb{C}$, $\mathbb{I}(\cdot)$ be the indicator function, and $\textrm{sign}(z) = \mathbb{I}(z > 0) - \mathbb{I}(z < 0)$ be the sign function.

The rest of the paper is organized as follows. 
In Sect. \ref{sec:review-ch}, background material on the univariate CH and Mat\'ern classes for random fields is reviewed. 
Here, the new result on the spectral density of the CH class in Sect. \ref{sec:ch-spectral} is also given. 
The main topic of this paper, valid multivariate generalizations of the CH class, is discussed in Sect. \ref{sec:mult-ch}. Extensive simulations supporting the theory are discussed in Sect. \ref{sec:sims}, followed by the analysis of multivariate oceanography data in the Southern Ocean in Sect. \ref{sec:real}. The paper concludes with some future directions in Sect. \ref{sec:disc}.

\section{Univariate CH and Mat\'ern Classes and New Results on CH Spectral Density}\label{sec:review-ch}
\subsection{Preliminaries on Univariate CH and Mat\'ern Classes}\label{sec:ch-mat-intro}

The Mat\'ern covariance model is a celebrated and commonly-used class of covariance functions in spatial statistics; see \cite{porcu2023mat}.
For a process $Y(s)$ for $s\in \mathbb{R}^d$, a commonly-used assumption in geostatistics and spatial statistics is second-order stationarity, which states that $E\{Y(s)\} = \mu$ and $E\{(Y(s) - \mu)(Y(t) - \mu)\} = C(t-s; \theta)$ for all $s, t \in \mathbb{R}^d$ and $C(\cdot; \theta)$ is a covariance function with parameters $\theta$. 
Without a loss of generality, take $\mu = 0$.
For a vector $h \in \mathbb{R}^d$ and parameters $\nu > 0$, $\phi>0$, and $\sigma > 0$, the Mat\'ern model is \begin{align*}
    C(h; \theta) = \mathcalprime{M}(h; \nu, \phi, \sigma) &= \sigma \frac{2^{1-\nu}}{\Gamma(\nu)}\left(\frac{\lVert h\rVert}{\phi}\right)^{\nu} \mathcalprime{K}_{\nu}\left(\frac{\lVert h\rVert}{\phi}\right),
\end{align*}where $\mathcalprime{K}_\nu(\cdot)$ is the modified Bessel function of the second kind \cite[see, e.g.,][]{stein1999interpolation}.
This is a slightly different parameterization of the Mat\'ern model compared to \cite{ma2022beyond}. 
Therefore, later results and construction for the confluent hypergeometric covariance is slightly different than in \cite{ma2022beyond}. 
The spectral density of the Mat\'ern model is \begin{align}
    f_{\mathcalprime{M}}(x; \nu, \phi, \sigma) &= \sigma \frac{\Gamma(\nu + \frac{d}{2})}{\pi^{\frac{d}{2}}\Gamma(\nu)\phi^{2\nu}} \left(\phi^{-2} + \lVert x\rVert^2\right)^{-\nu - \frac{d}{2}} \label{eq:spec_matern1},
\end{align} so that $
    \mathcalprime{M}(h; \nu, \phi, \sigma)= \int_{\mathbb{R}^d} e^{\I h^\top x}f_{\mathcalprime{M}}(x; \nu, \phi, \sigma) dx $.
In the Mat\'ern model, the parameter $\sigma$ controls the marginal variance of the resulting process: $\mathcalprime{M}(0; \nu, \phi, \sigma) = \sigma$; although another common notation takes the variance parameter as $\sigma^2$, this work uses $\sigma$ to match more analogously to the multivariate case.
The parameter $\nu$ is called the smoothness parameter due to the property that $Y(s)$ is $\lfloor \nu \rfloor$-times mean-square differentiable. 
Finally, the parameter $\phi$ is a range parameter controlling how fast the covariance decays. 
In particular, since $\mathcalprime{K}_\nu(z) \overset{\lvert z\rvert\to \infty}{\sim}
\{\pi/(2\lvert z\rvert)\}^{1/2} \textrm{exp}(-\lvert z\rvert)$ \citep{NIST:DLMF}, the asymptotic expansion is  \begin{align*}
    \mathcalprime{M}(h; \nu, \phi, \sigma) \overset{\lVert h \rVert\to\infty}{\sim} \sigma (2\pi)^{\frac{1}{2}}2^{ - \nu}\ \left(\frac{\lVert h\rVert}{\phi}\right)^{\nu - \frac{1}{2}}\textrm{exp}\left(-\frac{\lVert h\rVert}{\phi}\right).
\end{align*}
For large $\lVert h \rVert$, the covariance then primarily decays exponentially. 

However, in many settings, covariances with slower decay may be desired. 
With this in mind, \cite{ma2022beyond} introduced a confluent hypergeometric covariance class, defined as \begin{align*}
    \mathcalprime{CH}(h; \nu, \alpha, \beta, \sigma) &=\frac{\beta^{2\alpha}}{2^\alpha \Gamma(\alpha)} \int_0^\infty   \mathcalprime{M}(h; \nu, \phi, \sigma) \phi^{-2\alpha-2} \textrm{exp}\left(- \frac{\beta^2}{2\phi^2}\right)d\phi^2 \\
    &= \sigma \frac{\Gamma(\nu+ \alpha)}{\Gamma(\nu)} \mathcalprime{U}\left(\alpha, 1 - \nu, \frac{\lVert h\rVert^2}{2\beta^2}\right),
\end{align*}where $\mathcalprime{U}(a, b, z)$ is the confluent hypergeometric function of the second kind \citep[see, for example, Chapter 13 of][]{NIST:DLMF}. 
The CH covariance is obtained as a mixture of the Mat\'ern covariance over the parameter $\phi^2$ with respect to an inverse gamma distribution with parameters $\alpha$ and $\beta^2/2$. 
When the marginal covariance of $Y(s)$ is CH, $\nu$ continues to control the smoothness of the process, while the parameter $\alpha$ controls the tail decay of the covariance or the long range dependence of the covariance. 
In particular, \begin{align*}
    \mathcalprime{CH}(h; \nu, \alpha, \beta, \sigma) \overset{\lVert h \rVert \to \infty}{\sim} \sigma \frac{\Gamma(\nu+ \alpha)2^\alpha}{\Gamma(\nu)} 
 \left(\frac{\lVert h\rVert}{\beta}\right)^{-2\alpha },
\end{align*}using 13.2.6 of \cite{NIST:DLMF}, which establishes that $\mathcalprime{U}(a,b, z) \overset{z \to \infty}{\sim} z^{-a}$ for $z \in \mathbb{R}$. 
This decay matches the first-order term in \cite{ma2022beyond}, up to the differences in parameterizations of the CH covariance.
Higher-order terms may be computed using 13.5.2 of \cite{abramowitz1968handbook}, which can then be made to more closely match the result derived in Theorem 2 of \cite{ma2022beyond}.
The range parameter $\beta$ in the CH covariance plays a role similar to $\phi$ in the Mat\'ern covariance. 

\subsection{Spectral Density of the Univariate Isotropic CH Class}\label{sec:ch-spectral}
\begin{figure}
    \centering
        \begin{subfigure}[t]{0.32\textwidth}
\includegraphics[width = .98\textwidth]{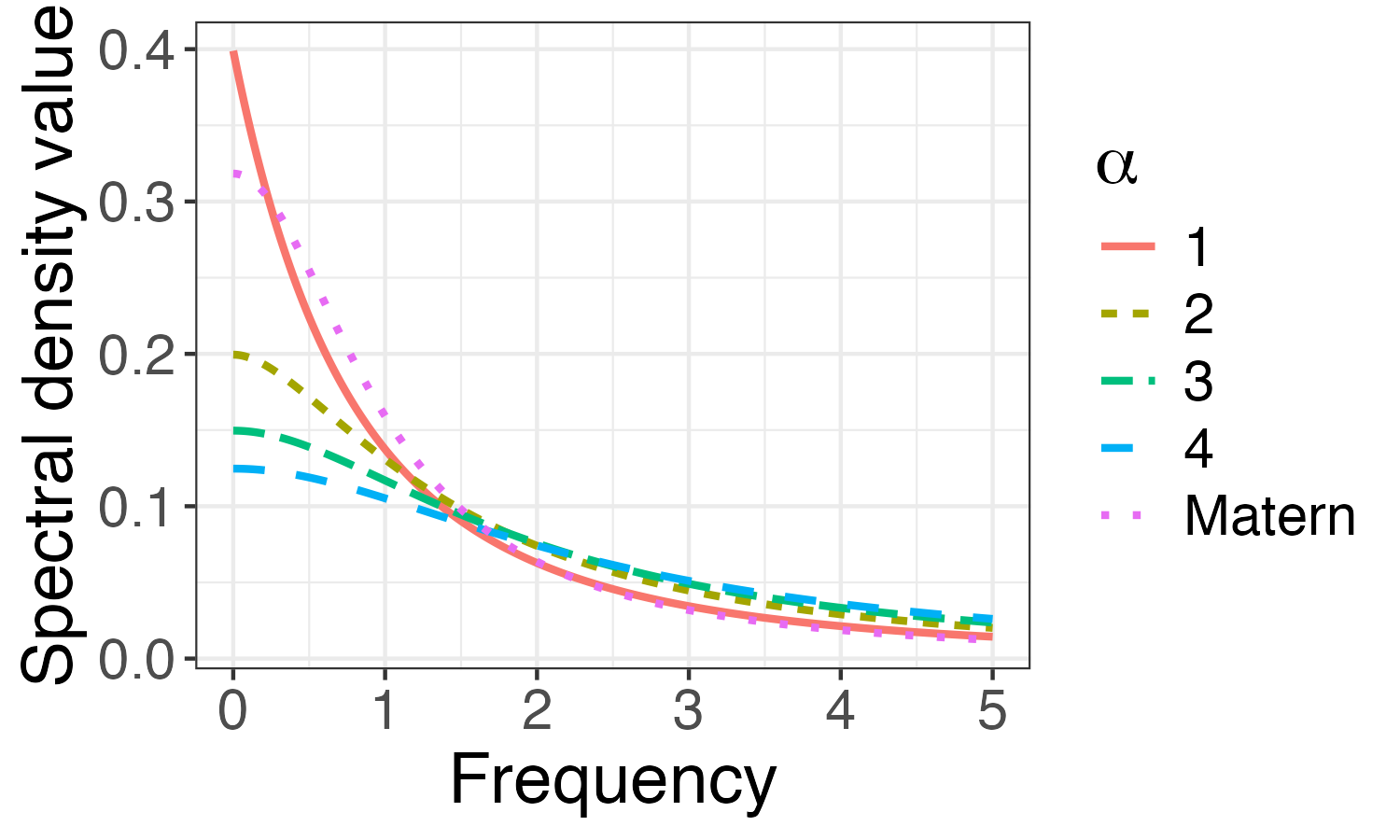}
    \caption{}
    \end{subfigure}
            \begin{subfigure}[t]{0.32\textwidth}
\includegraphics[width = .98\textwidth]{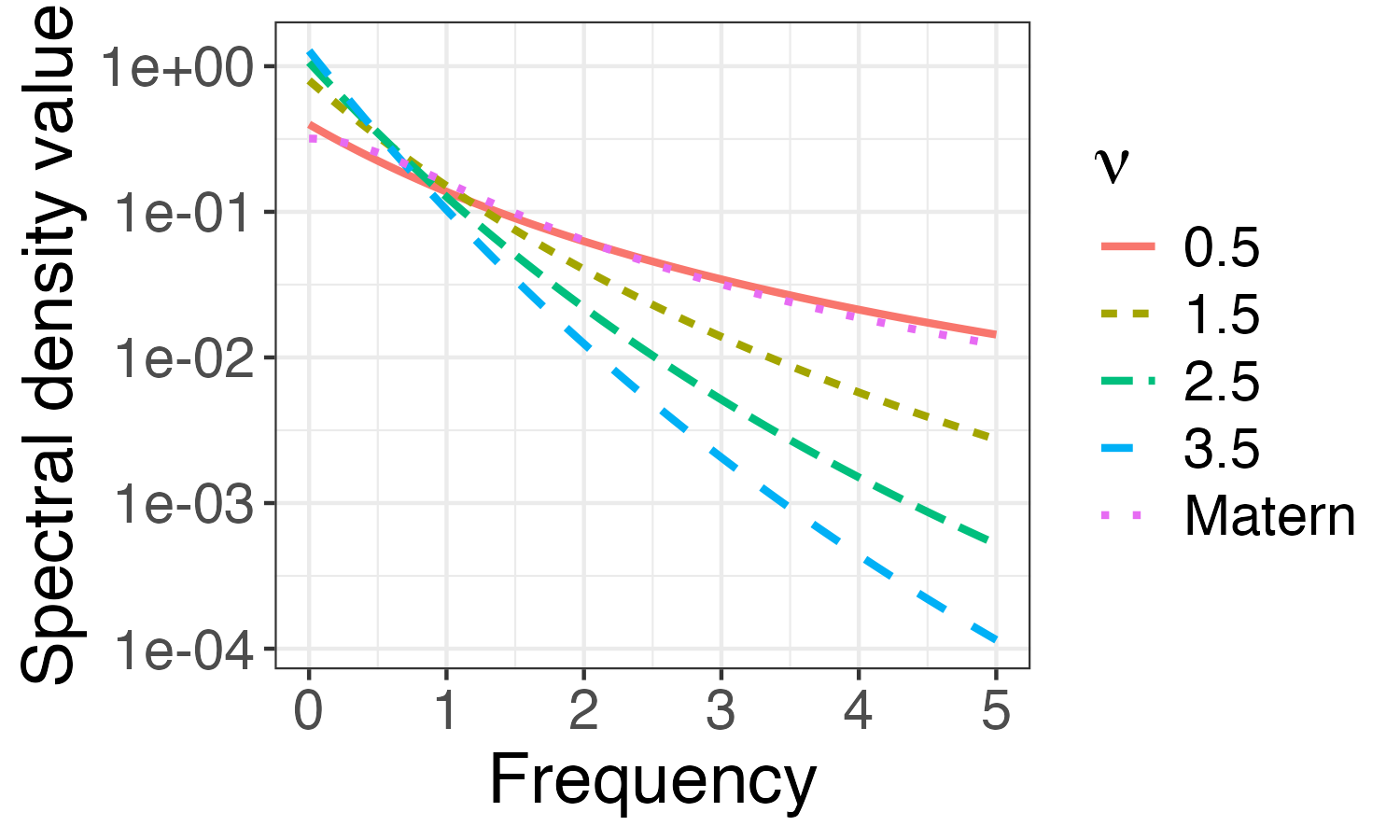}
    \caption{}
    \end{subfigure}
            \begin{subfigure}[t]{0.32\textwidth}
\includegraphics[width = .98\textwidth]{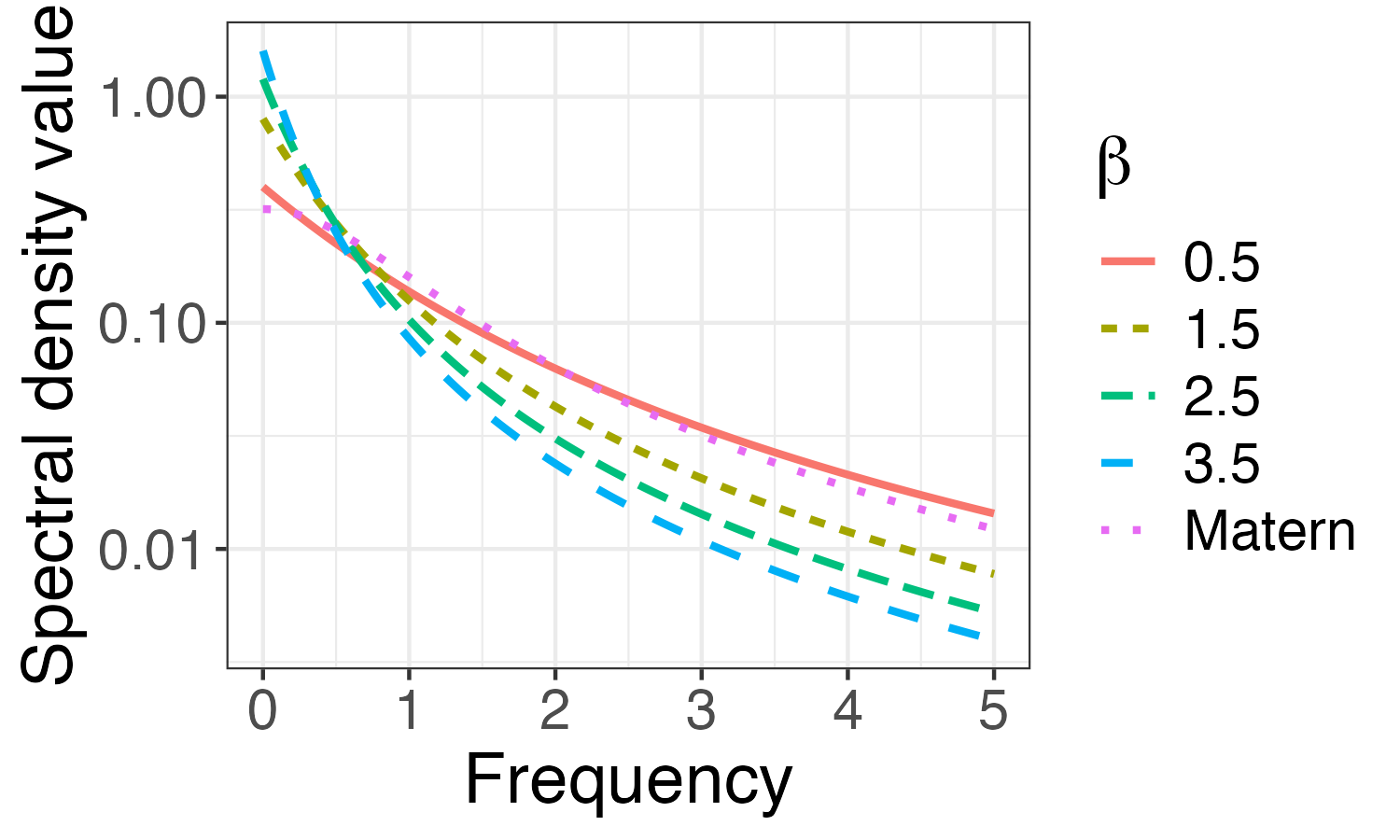}
    \caption{}
    \end{subfigure}
    \caption{Spectral density of the CH class. Unless otherwise specified, let $d=1$, $\sigma = 1$, $\alpha = 1$, $\nu = 0.5$, and $\beta = 1$, and compare with the Mat\'ern spectral density with $\sigma = 1$, $\nu = 0.5$, and $\phi = 1$ (a) Varying $\alpha$; (b) varying $\nu$ on a log scale; (c) varying $\beta$ on a log scale  }
    \label{fig:spec_dens}
\end{figure}

While the tail decay of the spectral density of the CH covariance was presented in \cite{ma2022beyond}, the spectral density of the CH covariance when $\alpha > d/2$ is presented here. 
\begin{prop}[Spectral density of confluent hypergeometric covariance]\label{prop:spectral_density}
Suppose that $\alpha > d/2$. The spectral density of the univariate confluent hypergeometric covariance is \begin{align*}
f_{\mathcalprime{CH}}(x; \nu, \alpha, \beta, \sigma)&=\sigma \frac{\Gamma(\nu + \frac{d}{2})\beta^{d}}{(2\pi)^{\frac{d}{2}}B(\alpha, \nu) } \mathcalprime{U}\left(\nu + \frac{d}{2} , 1 - \alpha + \frac{d}{2}, \frac{\beta^2\lVert x\rVert^2}{2}\right).
\end{align*}
\end{prop}
A concise proof is provided in the Supplement. 
As noted in the Proof of Proposition 1 in \cite{ma2022beyond}, the spectral density does not exist when $0 < \alpha \leq d/2$, as the covariance function is not absolutely integrable over $\mathbb{R}^{d}$ for these values of $\alpha$. 
This lack of integrability of the covariance function is sometimes used to define long-range dependence.
The CH class thus exhibits long-range dependence when $0 < \alpha \leq d/2$ and short-range dependence when $\alpha > d/2$ \citep{de_oliveira_simple_2023}.
Some results in this work only apply to the short-range dependence case, while some apply to both settings. 
Due to the asymptotic expansion of $\mathcalprime{U}(a, c, z) \overset{z \to \infty}{\sim} z^{-a}$ \citep[13.2.6,][]{NIST:DLMF}, the expression matches the tail of the spectral density found in \cite{ma2022beyond}.
The CH covariance and its spectral density have opposite forms in the arguments of $\mathcalprime{U}$. 
While $\nu$ controls the origin behavior of the covariance and the tail behavior of the spectral density, $\alpha$ controls the covariance's tail and the spectral density's origin. 
Examples of spectral densities are plotted in Fig. \ref{fig:spec_dens}. 
The spectral density proves valuable for the construction, simulation, and estimation of multivariate CH models in the subsequent sections. 

\section{Valid CH Cross-Covariances via Multivariate Mixtures and Their Properties}\label{sec:mult-ch}

\subsection{Preliminaries on Cross-Covariance Construction via Multivariate Mixtures}

Here, a few approaches for constructing valid multivariate Mat\'ern models are discussed, which will later be used to develop multivariate CH models. 
In the multivariate case, $\mb{Y}(s)$ is vector-valued: $\mb{Y}(s) = \{Y_1(s), \dots, Y_p(s)\}^\top$. 
Similarly, instead of a scalar-valued covariance function $C(h;\theta)$, one uses a matrix-valued covariance function $E\{\mb{Y}(s+h)\mb{Y}(s)^\top\} = \CovC(h; \mb{\theta}) = \{C_{jk}(h; \mb{\theta})\}_{j,k=1}^p$, where $C_{jj}(h; \mb{\theta})$ are referred to as the marginal covariance functions, and $C_{jk}(h; \mb{\theta})$ for $j\neq k$ are the cross-covariance functions that describe the covariance between $Y_j(s+h)$ and $Y_k(s)$.

A primary concern in the construction of multivariate Mat\'ern models has been ensuring the validity of the model \citep[see, for example,][]{gneiting2010matern,yarger2023multivariate}. 
If the spectral densities for each component are available, that is,\begin{align*}
    \CovC(h; \mb{\theta})&= \int_{\mathbb{R}^d} e^{\I h^\top x } \mb{f}(x; \mb{\theta}) dx,
\end{align*}where $\mb{f}(x;\mb{\theta} ) = \{f_{jk}(x; \mb{\theta})\}_{j,k=1}^p$, $\CovC(h; \mb{\theta})$ may be shown to be valid if $\mb{f}(x; \mb{\theta})$ is Hermitian and positive semidefinite for all $x \in \mathbb{R}^d$, a multivariate version of Bochner's theorem. 
The spectral density $\mb{f}(x;\mb{\theta})$ is now matrix-valued. 

\cite{gneiting2010matern} proposed multivariate Mat\'ern models with Mat\'ern cross-covariances \begin{align}
    \CovC(h; \mb{\theta}) &= \{\mathcalprime{M}(h; \nu_{jk}, \phi_{jk}, \sigma_{jk})\}_{j,k=1}^p ;\label{eq:matern_like_mm}
\end{align}for parameters $\{\sigma_{jk}, \nu_{jk}, \phi_{jk}\}_{j, k=1}^p$ and $\sigma_{jk} = \sigma_{kj}$, $\nu_{jk} = \nu_{kj}$, and $\phi_{jk} = \phi_{kj}$ for all $j$ and $k$. 
\cite{gneiting2010matern}, and later \cite{apanasovich2012valid}, used the spectral density to establish a valid model, though the corresponding parameter restrictions are quite technical.
For the bivariate model ($p=2$), \cite{gneiting2010matern} directly analyzed $|f_{12}(x; {\theta})|/\sqrt{f_{11}(x; {\theta}) f_{22}(x;{\theta})}$ to show validity. 
\cite{gneiting2010matern} also established conditions for a valid model for general $p$ using $f(x; \theta)$, with the simplification $\phi_{jk} = \phi^*$ for all $j$ and $k$ as well as $\nu_{jk} = (\nu_{jj} + \nu_{kk})/2$. 
\cite{apanasovich2012valid} improved upon this by using the Schur product theorem and existing results on positive semidefinite and conditionally negative semidefinite matrices. 

Recently, \cite{yarger2023multivariate} proposed to instead construct multivariate Mat\'ern models beginning with the spectral density, referred to as spectrally-generated cross-covariance functions. 
In particular, they proposed taking $$
    f_{jk}(x; \mb{\theta})= \sigma_{jk}f_{\mathcalprime{M}}(x; \nu_{j}, \phi_{j}, 1)^{\frac{1}{2}}f_{\mathcalprime{M}}(x; \nu_{k}, \phi_{k}, 1)^{\frac{1}{2}}.$$ 
\cite{yarger2023multivariate} also introduced cross-covariance functions that have asymmetric form. 
This form considers $f(x) = P(x) \mb{\sigma} P(x)^\top $ under usual matrix multiplication, where $P(x)$ is diagonal with entries $f_{\mathcalprime{M}}(x; \nu_{j}, \phi_{j}, 1)^{\frac{1}{2}}$ and $\mb{\sigma}=\{\sigma_{jk}\}$. 
The construction makes validity conditions of the model immediate and simple: one needs the matrix $\mb{\sigma}$ to be positive semidefinite. 
However, closed-form expressions of the covariance may not be attainable and can be computed through fast Fourier transforms. 

The validity of multivariate Mat\'ern models with form Eq. \eqref{eq:matern_like_mm} has also been shown using a mixture representation of the Mat\'ern covariance \citep{emery2022new}. 
In particular, consider \begin{align*}
    \CovC(h; \mb{\theta}) &= \int_0^\infty \CovC_*(h; \mb{\theta}, u)\odot \mb{p}(u \mid \mb{\theta}) du ,
\end{align*}where $\CovC_*(h; \mb{\theta}, u)$ is a multivariate covariance function that has parameter $u$, and $\mb{p}(u\mid \mb{\theta})$ is a matrix with nonnegative entries, each consisting of a mixing density for $u$. 
Then, if $\CovC_*(h; \mb{\theta}, u)$ is a valid multivariate covariance for all $u \in (0, \infty)$, and the matrix of $\mb{p}(u\mid \mb{\theta})$ is positive semidefinite for all $u \in (0, \infty)$, the covariance $\CovC(h; \mb{\theta})$ is also valid \citep{emery2022new}. 
This mixture approach, combined with the matrix tools used in \cite{apanasovich2012valid}, established more general and flexible conditions for the validity of the multivariate Mat\'ern model.
Similarly, valid multivariate CH covariances will be constructed with a mixture of valid multivariate Mat\'ern covariances. 

Next, a class of models is constructed in the same way as Eq. \eqref{eq:matern_like_mm} by using cross-covariances that are proportional to CH covariance functions. 

\subsection{A Parsimonious Multivariate Confluent Hypergeometric Model}\label{sec:valid_ch}

Consider a multivariate covariance with entries $\mathcalprime{CH}(h; \nu_{jk}, \alpha_{jk}, \beta_{jk}, \sigma_{jk})$.
Throughout, assume $\alpha_{jk} >0$, $\beta_{jk} > 0$, $\nu_{jk} > 0$, and $\sigma_{jj}>0$ for each $j$ and $k$. 
Let $\mb{\sigma} = \{\sigma_{jk}\}$, $\mb{\nu} = \{\nu_{jk}\}$, $\mb{\alpha} = \{\alpha_{jk}\}$, and $\mb{\beta} = \{\beta_{jk}\}$ be $p\times p$ symmetric, real matrices represented by the parameters. 
With this form, one must have $\alpha_{jk} = \alpha_{kj}$ (and likewise for $\beta_{jk}$, $\nu_{jk}$, and $\sigma_{jk}$) due to the requirement of $C_{jk}(h;\theta) = C_{kj}(-h;\theta)$ for all $h \in \mathbb{R}^d$ for real processes \citep[see Eq.~(4.3) and ensuing discussion in][]{yaglom_correlation_1987}.
In the following, treat $\nu$, $\alpha$, $\beta$, and $\sigma$ as matrices, using, for example, $\nu_{jk}$ or $\nu^*$ as scalars. 
In Sects. \ref{sec:valid_ch}, \ref{sec:other_conditions}, and \ref{sec:equiv}, take all matrix operations elementwise.
For example,  $\Gamma(\mb{\nu}) = \{\Gamma(\nu_{jk})\}$ and $
        \mathcalprime{CH}(h; \mb{\nu}, \mb{\alpha}, \mb{\beta}, \mb{\sigma}) = \{\mathcalprime{CH}(h; \nu_{jk}, \alpha_{jk}, \beta_{jk}, \sigma_{jk})\}_{j,k=1}^p$. 
Before discussing validity conditions, a few properties of the multivariate covariance are discussed. 

\begin{prpty}[Tail decay]\label{prop:one}
    The tail decay of multivariate covariance is, \begin{align*}
        \mathcalprime{CH}(h; \mb{\nu}, \mb{\alpha}, \mb{\beta}, \mb{\sigma}) \overset{\lVert h\rVert\to \infty}{\sim} \mb{\sigma}\frac{2^{\mb{\alpha}}\Gamma(\mb{\nu}+ \mb{\alpha})}{\Gamma(\mb{\nu})} \left(\frac{\lVert h \rVert}{\mb{\beta}}\right)^{-2\mb{\alpha}}.
    \end{align*}
\end{prpty}
\begin{prpty}[Smoothness]\label{prop:two}
The $j$th process is $\lfloor \nu_{jj} \rfloor$ times continuously differentiable. 
\end{prpty}
\begin{prpty}[Spectral density]\label{prop:three}
    The multivariate spectral density of the model, when the entries of $\alpha$ satisfy $\alpha_{jk} > d/2$ for all $j$ and $k$, is $$ \mb{\sigma}\frac{\Gamma(\mb{\nu} + \frac{d}{2}) \mb{\beta}^{d}}{(2\pi)^{\frac{d}{2}}B(\mb{\alpha}, \mb{\nu})} \mathcalprime{U}\left(\mb{\nu} + \frac{d}{2}, 1 - \mb{\alpha} + \frac{d}{2}, \mb{\beta}^2 \frac{\lVert x \rVert^2}{2}\right).$$
\end{prpty}
\begin{prpty}[Tail behavior of spectral density]\label{prop:four}
 Suppose the entries of $\alpha$ satisfy $\alpha_{jk} > d/2$ for all $j$ and $k$. As $\lVert x \rVert \to \infty$, the spectral density decays as \begin{align}
        \mb{\sigma}\frac{\Gamma(\mb{\nu} + \frac{d}{2})2^{\mb{\nu}}}{\mb{\beta}^{2\mb{\nu}}B(\mb{\alpha}, \mb{\nu})\pi^{\frac{d}{2}}} \lVert x \rVert^{-2\mb{\nu} - d}. \label{eq:tail_spec_density}
    \end{align}
\end{prpty}

\begin{prpty}[Symmetry]\label{prop:five}
The model is symmetric: \begin{align*}
         \mathcalprime{CH}(h; \mb{\nu}, \mb{\alpha}, \mb{\beta}, \mb{\sigma}) &=  \mathcalprime{CH}(-h; \mb{\nu}, \mb{\alpha}, \mb{\beta}, \mb{\sigma})
    \end{align*}for all $h \in \mathbb{R}^d$.
\end{prpty}

Properties \ref{prop:one}, \ref{prop:three}, \ref{prop:four}, and \ref{prop:five} should be interpreted entry-wise.
Property \ref{prop:one} follows from the asymptotic expansion of $\mathcalprime{U}(a, b, z)$ discussed above and matches the results in \cite{ma2022beyond}; Property \ref{prop:two} follows from \cite{ma2022beyond}; Property \ref{prop:three} follows directly from Proposition \ref{prop:spectral_density}; Property \ref{prop:four} follows from 13.2.6 of \cite{NIST:DLMF}; Property \ref{prop:five} follows from the fact that the covariance only depends on $h$ through $\lVert h\rVert$. 
Property \ref{prop:five} is one limitation of the cross-covariance models introduced in this Section through Section \ref{sec:equiv}.
Thus, the model may not be appropriate for processes with cross-dependence that is lagged in $h$.
Asymmetric cross-covariance models are introduced in Section \ref{sec:spectral}.
For easier notation, set $\nu_{j} = \nu_{jj}$, $\alpha_j = \alpha_{jj}$, and $\beta_{j} = \beta_{jj}$.

\begin{thm}\label{thm:pars_like}
    Suppose that $\nu_{jk} = (\nu_{j} + \nu_{k})/2$, $\alpha_{jk}=(\alpha_{j} + \alpha_{k})/2$, and $\beta_{jk}^2 = (\beta_{j}^2 + \beta_{k}^2)/2$ for all $j \neq k$. 
    Then the multivariate model defined by $\mathcalprime{CH}(h; \nu_{jk}, \alpha_{jk}, \beta_{jk}, \sigma_{jk})$ is valid if the matrix \begin{align}
        \mb{\sigma} \mb{\beta}^{2\mb{\alpha}} \frac{\Gamma(\mb{\nu} + \frac{d}{2})}{\Gamma(\mb{\nu})\Gamma(\mb{\alpha})}=\left\{\sigma_{jk} \left(\frac{\beta_{j}^2 + \beta_{k}^2}{2}\right)^{\frac{\alpha_{j} + \alpha_{k}}{2}}
        \frac{\Gamma(\frac{\nu_{j} + \nu_{k}}{2} +\frac{d}{2})}{\Gamma(\frac{\nu_{j} + \nu_{k}}{2})\Gamma(\frac{\alpha_{j} + \alpha_{k}}{2})} \right\}_{j,k=1}^p \label{eq:pars_condition}
    \end{align}is positive semidefinite. 
\end{thm}

A proof is in the Supplementary Material.
This choice gives simple validity conditions and eliminates the need to estimate additional parameters $\nu_{jk}$, $\alpha_{jk}$, and $\beta_{jk}$ for $j\neq k$, while allowing the different covariances to have different $\nu_j$, $\alpha_j$, and $\beta_{j}$ values.
In some ways, it is similar to the parsimonious multivariate Mat\'ern model of \cite{gneiting2010matern}, but this model also allows the tail behavior and the scale parameter to vary for each process. 
These conditions suggest that processes with different smoothnesses, tail decay, or scale parameters cannot be perfectly correlated. 
The conditions for validity are similar to previous work on the multivariate Mat\'ern in \cite{gneiting2010matern}, \cite{apanasovich2012valid}, and \cite{emery2022new}. 
For example, the term $\Gamma(\mb{\nu} + d/2)/\Gamma(\mb{\nu})$ results immediately from mixing the valid parsimonious multivariate Mat\'ern model.

\begin{figure}[!ht]
    \centering
    \includegraphics[width = .97\textwidth]{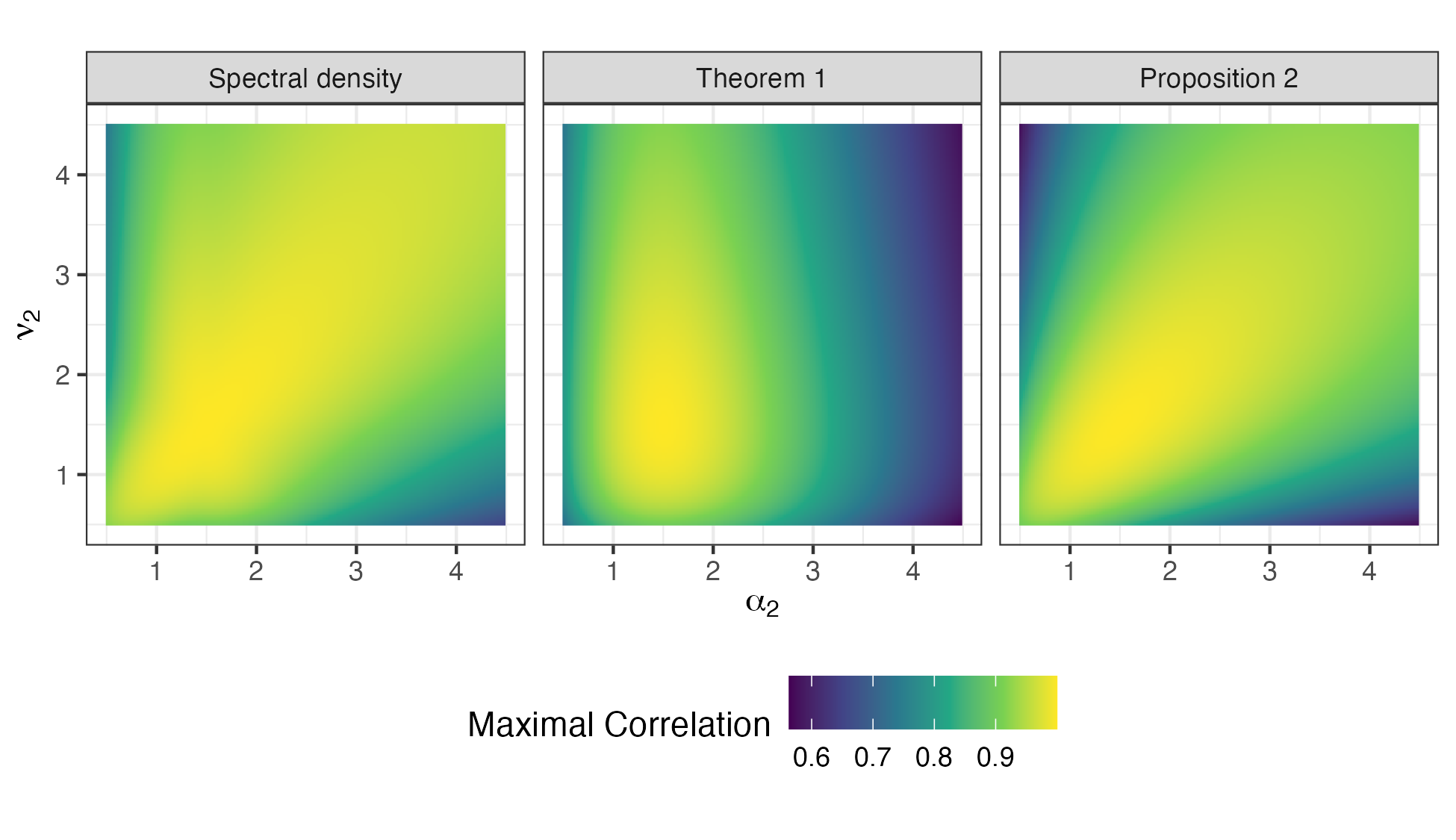}    \caption{Maximum marginal correlation between two processes for $d=1$ with $\nu_1 = \alpha_1 = 3/2$, $\nu_{12} = (\nu_1 + \nu_2)/2$, $\alpha_{12} = (\alpha_1 + \alpha_2)/2$, and $\beta_1 =\beta_2 = \beta_{12}$, and the parameters $\nu_2$ and $\alpha_2$ are varying. ``Spectral density'' uses the value of the required bound, based on a fine grid of $10^{-5} \leq \lVert x  \rVert \leq 10^5$, of $|\sigma_{12}|/(\sigma_{11}\sigma_{22})^{1/2} \leq \min_{10^{-5} < \lVert x  \rVert < 10^5}
    \{f_{\mathcalprime{CH}}(x; \nu_{1}, \alpha_{1}, \beta_{1},1)f_{\mathcalprime{CH}}(x; \nu_{2}, \alpha_{2}, \beta_{2}, 1)\}^{1/2}/f_{\mathcalprime{CH}}(x; \nu_{12}, \alpha_{12}, \beta_{12}, 1)$. Theorem 1 and Proposition 2 give the bounds denoted by (5) and (6), respectively
    }
    \label{fig:max_correlation}
\end{figure}

One may also formulate Theorem \ref{thm:pars_like} in terms of the maximal possible correlation coefficient between two processes. 
Consider the case where $p = 2$. 
Then Theorem \ref{thm:pars_like} implies that the marginal correlation between $Y_1(s)$ and $Y_2(s)$ satisfies \begin{align}
\begin{split}
    \frac{\left\lvert\sigma_{12}\right\rvert}{\left(\sigma_{11}\sigma_{22}\right)^{\frac{1}{2}}} &\leq \frac{\beta_1^{\frac{\alpha_1}{2}}\beta_2^{\frac{\alpha_2}{2}}}{\left(\frac{\beta_{1}^2 + \beta_{2}^2}{2}\right)^{\frac{\alpha_{1} + \alpha_{2}}{2}}}
     \frac{\left\{\Gamma(\nu_1+ \frac{d}{2})\Gamma(\nu_2 + \frac{d}{2})\right\}^{\frac{1}{2}}}{\Gamma(\frac{\nu_{1} + \nu_{2}}{2} +\frac{d}{2})}\\
     &~~~~~~~~~~~~~~~ \times
    \frac{\Gamma(\frac{\nu_{1} + \nu_{2}}{2})}{\left\{\Gamma(\nu_1)\Gamma(\nu_2)\right\}^\frac{1}{2}}\frac{\Gamma(\frac{\alpha_{1} + \alpha_{2}}{2})}{\left\{\Gamma(\alpha_1)\Gamma(\alpha_2)\right\}^\frac{1}{2}}.
    \end{split}\label{eq:cor_thm1}
\end{align}
When $\beta_1 = \beta_2$ and $\alpha_1 = \alpha_2$, one recovers the condition of Theorem 1 of \cite{gneiting2010matern}. 
The maximal correlation for some specific values of parameters is plotted in Fig. \ref{fig:max_correlation}. 

\begin{figure}[!ht]
    \centering
    \begin{subfigure}[t]{0.49\textwidth}
\includegraphics[width = .99\textwidth]{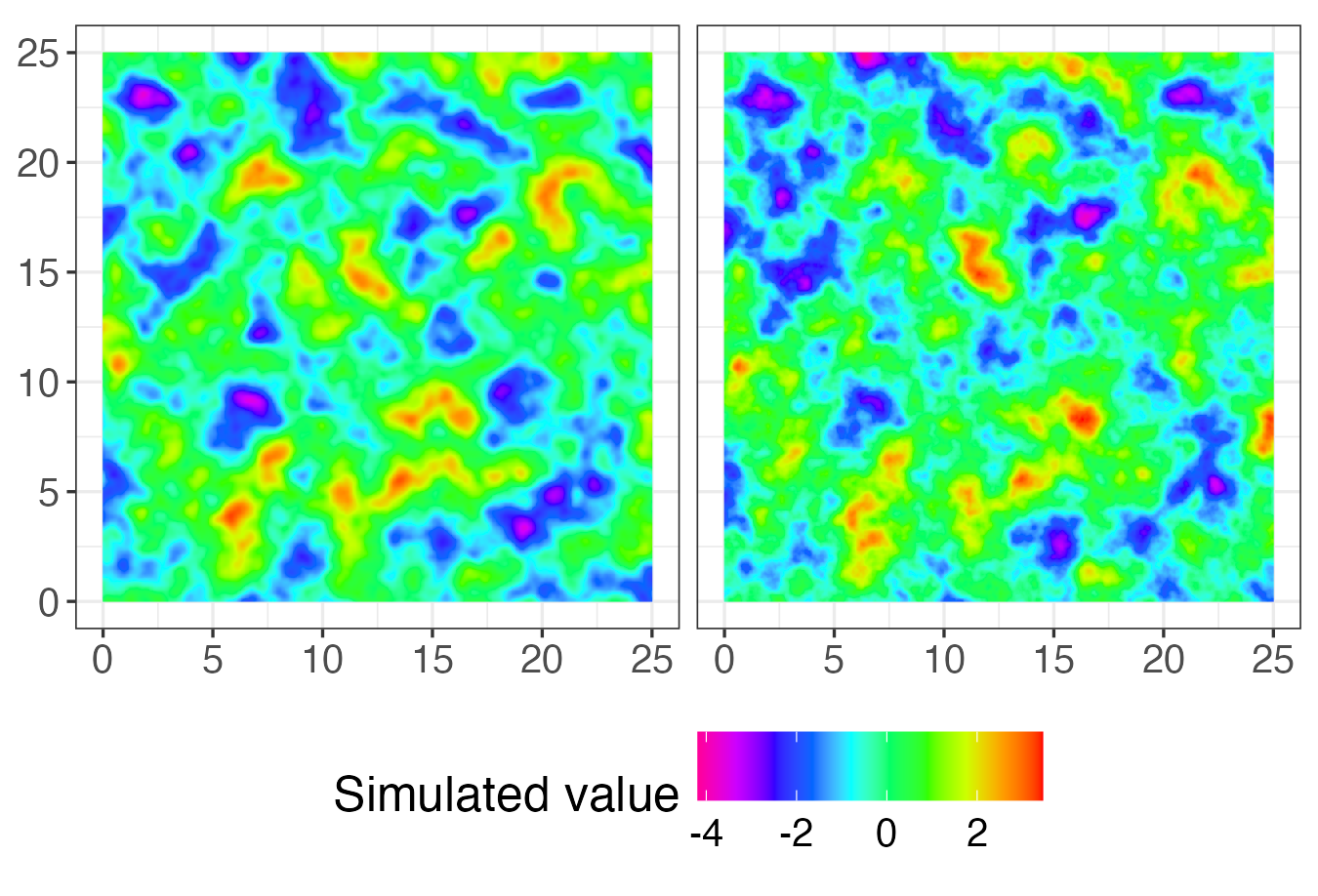}
    \caption{}
    \end{subfigure}
        \begin{subfigure}[t]{0.49\textwidth}
\includegraphics[width = .99\textwidth]{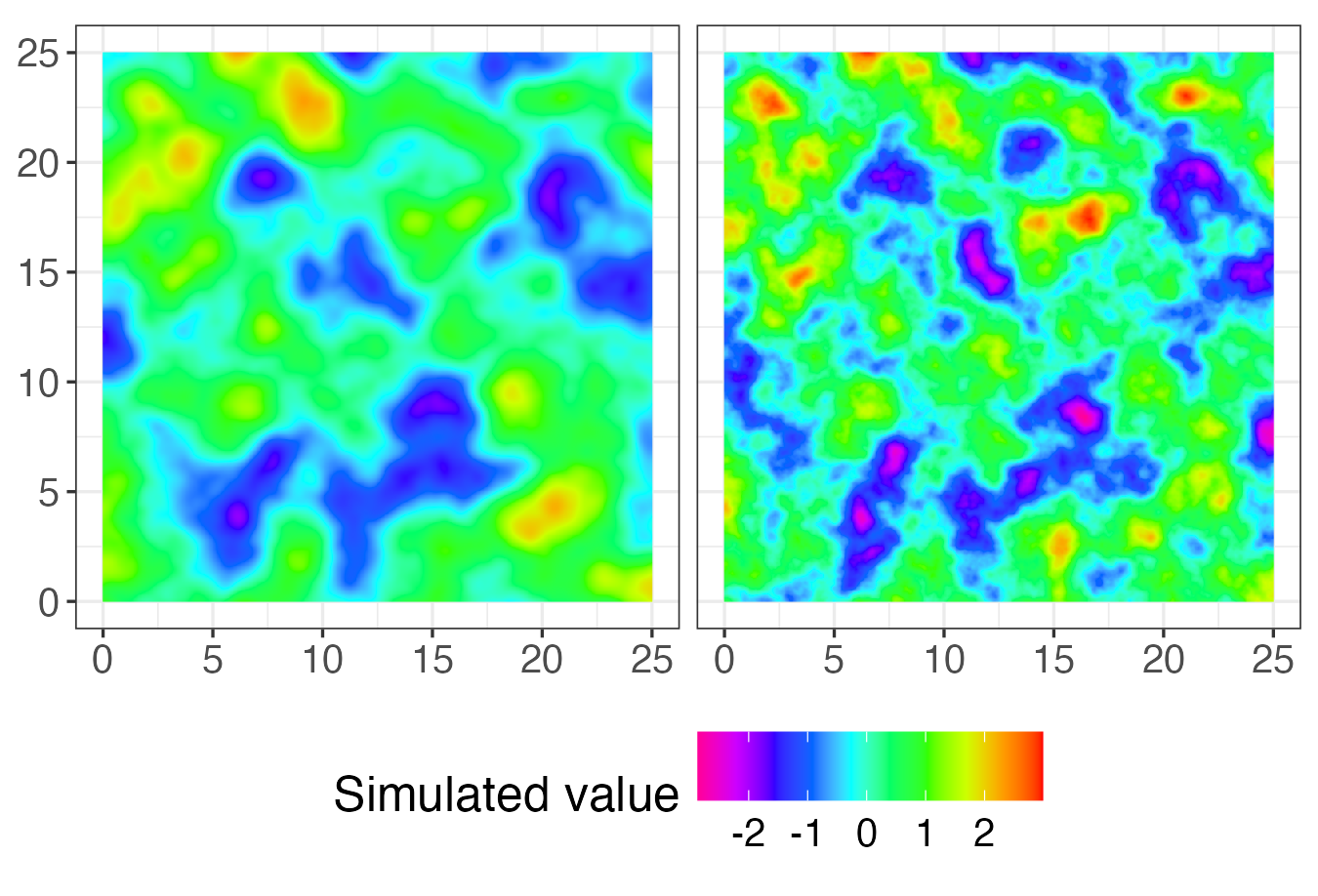}
    \caption{}
    \end{subfigure}
    \caption{(a) Bivariate CH processes with $\nu_1 = 2.5$, $\nu_2 = 1.5$, $\alpha_1 = 3$, $\alpha_2 = 1.5$, $\beta_j = \beta_k = 1$, $\sigma_{jj} = \sigma_{kk} = 1$, and $\sigma_{jk} = 0.8$ (b) Bivariate Mat\'ern processes with $\phi_{jk} =\phi_{jj} = \phi_{kk}=1$, $\nu_1 = 2.5$, $\nu_2 = 1.5$, $\nu_{12} = 2$, $\sigma_{jk} = 0.8$ }
    \label{fig:2d_sim}
\end{figure}

\cite{emery2016improved} provides an approach using the spectral density to simulate processes efficiently. 
Two simulations of bivariate processes are plotted in Fig. \ref{fig:2d_sim} and compared to simulated bivariate Mat\'ern processes based on this approach. 
Similar to results in \cite{ma2022beyond}, some visual properties of the fields are not very intuitive. 
For fixed range parameters $\beta_{jk} = \phi_{jk}= 1$, the CH covariance decays faster near $h = 0$ compared to the Mat\'ern covariance, leading to nearby contrasting values of $Y_j(s)$. 
Also, as in \cite{ma2022beyond}, the influence of $\nu_{j}$ is less visually apparent for CH compared to Mat\'ern processes.

\subsection{Flexible Conditions for the Multivariate Confluent Hypergeometric Model}\label{sec:other_conditions}

The model in Theorem \ref{thm:pars_like} gives full flexibility to the parameters of the marginal processes $\{\alpha_j, \nu_j, \beta_j, \sigma_{jj}\}$ while only slightly increasing the number of parameters needed to estimate the model: the marginal covariances $\sigma_{jk}$ for $j\neq k$, potential values of which are restricted based on $\{\alpha_j, \nu_j, \beta_j, \sigma_{jj}\}$. 
In most cases it would be a suitable model. 
However, in the multivariate Mat\'ern literature, considerable research has been focused on finding more flexible conditions for validity \citep{apanasovich2012valid, emery2022new}. 
For example, one may want to have $\nu_{jk} \neq (\nu_j + \nu_k)/2$. 
Next, the more-recent approach in \cite{emery2022new} is extended to the multivariate CH. 
\begin{thm}\label{th:easyconditions}
Consider a multivariate CH covariance $\mathcalprime{CH}(h; \mb{\nu}, \mb{\alpha}, \mb{\beta}, \mb{\sigma})$, and note the definition of a conditionally negative semidefinite matrix reviewed in the Supplement and \cite{emery2022new}. 
If the following Conditions \ref{cond:one}--\ref{cond:four} hold, then the multivariate CH model is valid.
        \begin{condition}\label{cond:one}
            $\mb{\nu}$ is conditionally negative semidefinite.
        \end{condition}
        \begin{condition}\label{cond:two}
             $\mb{\beta}^2$ is conditionally negative semidefinite.
        \end{condition}
        \begin{condition}\label{cond:three}
            $\alpha_{jk} = (\alpha_{j} +\alpha_{k})/2$ for all $j$ and $k$. 
        \end{condition}
        \begin{condition}\label{cond:four}
            $\mb{\sigma} \mb{\nu}^{\mb{\nu} + d/2}{\normalfont \textrm{exp}}(-\mb{\nu}) \mb{\beta}^{2\mb{\alpha}}/\{\Gamma(\mb{\nu})\Gamma(\mb{\alpha})\}$ is positive semidefinite. 
        \end{condition}
\end{thm}
A proof is provided in the Supplement. 
Consider a comparison of Theorems \ref{th:easyconditions} and \ref{thm:pars_like}.
Set, in the context of Theorem \ref{th:easyconditions}, $\nu_{jk} = (\nu_j + \nu_k)/2$, $\beta^2_{jk} = (\beta_{j}^2 + \beta_{k}^2)/2$, and $\alpha_{jk} = (\alpha_j + \alpha_k)/2$.
Then the matrices $\nu$ and $\beta^2$ are conditionally negative semidefinite, and Conditions \ref{cond:one}, \ref{cond:two}, and \ref{cond:three} are met. 
This results in a similar form to Theorem \ref{thm:pars_like}, with $\nu^{\nu + d/2}\textrm{exp}(-\nu)$ replacing $\Gamma(\nu +d/2)$. 
Theorem \ref{th:easyconditions} provides expanded options for the parameters, in particular $\nu_{jk} > (\nu_j + \nu_k)/2$ and $\beta^2_{jk} > (\beta_j^2 + \beta_k^2)/2$.

As an example, consider the following construction when $p = 2$ and $d=2$. 
Suppose that $\nu_{1} = 1/2$, $\nu_2= 1$, $\nu_{12} = 1$, $\beta^2_{1} = 1$, $\beta^2_2 = 2$, $\beta^2_{12}= 2$, $\alpha_1 = 1/2$, $\alpha_2 = 3/2$, and $\alpha_{12} = 1$. 
Then Conditions \ref{cond:one}, \ref{cond:two}, and \ref{cond:three} are met. 
For these values of the parameters, Condition \ref{cond:four} implies that $\lvert\sigma_{12}\rvert / (\sigma_{11}\sigma_{22})^{1/2} \leq 0.3847$
ensures validity of the multivariate covariance. 
Alternatively, set $\nu_{12} = (\nu_{1} +\nu_2)/2 = 3/4$ and $\beta^2_{12} = (\beta_1^2 + \beta_2^2)/2 = 3/2$ to fall under the purview of Theorem \ref{thm:pars_like}, and this results in a corresponding bound $\lvert\sigma_{12}\rvert / (\sigma_{11}\sigma_{22})^{1/2} \leq 0.8434$, and Theorem \ref{th:easyconditions} gives $\lvert\sigma_{12}\rvert / (\sigma_{11}\sigma_{22})^{1/2} \leq 0.8100$.
This suggests that Theorem \ref{th:easyconditions} allows this flexibility in $\beta^2_{jk}$ and $\nu_{jk}$ for $j\neq k$ at the expense of the possible strength of correlation between the processes. 
This also suggests that Theorem \ref{thm:pars_like} is slightly stronger than Theorem \ref{th:easyconditions} when their conditions intersect.

The next proposition accomplishes two goals when some simplifications of the model are made.
First, the conditions for validity can be made weaker in some cases, allowing for higher correlation between processes compared to the conditions presented in Theorem \ref{thm:pars_like}.
Second, $\alpha_{jk}$ may be chosen more flexibly: $\alpha_{jk} \geq (\alpha_j + \alpha_k)/2 > d/2$. 

\begin{prop}\label{prop:parsimonious}
Consider a bivariate ($p=2$) CH covariance $\mathcalprime{CH}(h; \mb{\nu}, \mb{\alpha}, \mb{\beta}, \mb{\sigma})$ with the restrictions $\nu_{12} = \nu_{21}= (\nu_1 + \nu_2)/2$, $\alpha_j > d/2$ for $j \in \{1,2\}$, $\alpha_{12} = \alpha_{21}\geq (\alpha_1 + \alpha_2)/2$, and $\beta_{jk} = \beta^* $ for all $j$ and $k \in \{1,2\}$.
This covariance is valid if the matrix $\mb{\sigma}/B(\mb{\alpha}, \mb{\nu})$ is positive semidefinite.
\end{prop}
The proof is presented in the Supplement and relies on the spectral density of the CH covariance. 
In terms of the maximum correlation between bivariate processes, one may write \begin{align}
     \frac{\left\lvert\sigma_{12}\right\rvert}{\left(\sigma_{11}\sigma_{22}\right)^{\frac{1}{2}}} \leq 
     \frac{\left\{\Gamma(\nu_1+ \alpha_1)\Gamma(\nu_2 + \alpha_2)\right\}^\frac{1}{2}}{\Gamma(\frac{\nu_{1} + \nu_{2}}{2} +\frac{\alpha_1 + \alpha_2}{2})}
    \frac{\Gamma(\frac{\nu_{1} + \nu_{2}}{2})}{\left\{\Gamma(\nu_1)\Gamma(\nu_2)\right\}^{\frac{1}{2}}}\frac{\Gamma(\frac{\alpha_{1} + \alpha_{2}}{2})}{\left\{\Gamma(\alpha_1)\Gamma(\alpha_2)\right\}^{\frac{1}{2}}}.\label{eq:cor_p2}
\end{align}
Consider a comparison of Theorem \ref{thm:pars_like} and Proposition \ref{prop:parsimonious} at their intersection: $p=2$, $\alpha_j>d/2$, $\alpha_{jk} = (\alpha_j + \alpha_k)/2$, and $\beta_{jk}= \beta$ for all $j$ and $k$.
Then, Theorem \ref{thm:pars_like} establishes that the condition of positive semidefiniteness of $\mb{\sigma}\Gamma(\mb{\nu} + d/2)/\{\Gamma(\mb{\nu})\Gamma(\mb{\alpha})\}$, while Proposition \ref{prop:parsimonious} instead uses $
\mb{\sigma}\Gamma(\mb{\nu} + \mb{\alpha})/\{\Gamma(\mb{\nu})\Gamma(\mb{\alpha})\}$. 
These are numerically compared to evaluation of the spectral density in Fig. \ref{fig:max_correlation}.
While Theorem \ref{thm:pars_like} appears relatively sharp when $\lvert\nu_2 - \nu_1\rvert$ is large and $\lvert\alpha_2 - \alpha_1\rvert$ is small, Proposition \ref{prop:parsimonious} is sharper when $\lvert\nu_2 - \nu_1\rvert$ is small and $\lvert\alpha_2 - \alpha_1\rvert$ is large. 
In fact, when $\nu_j = \nu_k = \nu_{jk}= \nu^*$, $\alpha_j > d/2$, $\alpha_{jk} \geq (\alpha_j + \alpha_k)/2$, and $\beta_{jk}= \beta^*$ for all $j$ and $k$, the tail behavior of the spectral density in Eq. \eqref{eq:tail_spec_density} demonstrates that Proposition \ref{prop:parsimonious} is sufficient and necessary, since the matrix-valued tail of the spectral density decays proportionally to $\mb{\sigma}/B(\mb{\alpha}, \mb{\nu}) $.

One cannot construct a valid multivariate CH model when $\nu_{jk} < (\nu_j + \nu_k)/2$, which was also established for the multivariate Mat\'ern in \cite{gneiting2010matern}. 
To see this, one may look at the the tail behavior of the matrix-valued spectral density in Eq. \eqref{eq:tail_spec_density}, which cannot be positive semidefinite in this case for arbitrarily large $\lVert x \rVert$. 

\subsection{Equivalence of Gaussian Measures Under the Multivariate CH Model}\label{sec:equiv}
Next, the Gaussian equivalence of measures is discussed for the multivariate covariance structure, which implies that the multivariate CH class has asymptotically equivalent predictions under different values of the covariance parameters. 
Let $(\Omega, \mathcalprime{F})$ be a measurable space with sample space $\Omega$ and $\sigma$-algebra $\mathcalprime{F}$. 
Let $\mathcalprime{P}_1(\cdot)$ and $\mathcalprime{P}_2(\cdot)$ be two probability measures on $(\Omega, \mathcalprime{F})$.
One says that $\mathcalprime{P}_1(\cdot)$ and $\mathcalprime{P}_2(\cdot)$ are equivalent on $(\Omega, \mathcalprime{F})$ if $\mathcalprime{P}_1(\cdot)$ is absolutely continuous with respect to $\mathcalprime{P}_2(\cdot)$ on $(\Omega, \mathcalprime{F})$ and vice-versa. 
One says that $\mathcalprime{P}_1(\cdot)$ and $\mathcalprime{P}_2(\cdot)$ are equivalent on the paths of a random process $\mb{Y}(s)$ if they are equivalent on the $\sigma$-algebra generated by $\mb{Y}(s)$. 
\cite{zhang2004inconsistent} has established that the equivalence of Gaussian measures has important implications for parameter estimation.
Here, a multivariate random Gaussian process $\mb{Y}(s)\in \mathbb{R}^p$ with mean $(0, \dots, 0)^\top$ and covariance function $\CovC(\cdot; \mb{\theta})$ on a bounded domain corresponds to a Gaussian measure used to evaluate properties of the covariance. 

The equivalence result for the multivariate CH covariance presented in Sect. \ref{sec:valid_ch} is established here.
Consider the condition in \cite{bachoc_asymptotically_2022}.
\begin{condition}
\label{as:ch_matrix}
    Let $\lambda_1(M)$ for a matrix $M$ be the smallest eigenvalue of $M$. Assume that $\nu_{jk} = \nu^*$ for all $j, k \in \{1, \dots, p\}$ and
    \begin{align*}
        \inf_{u \geq 0}\lambda_1\left[\left\{\sigma_{jk} \frac{\Gamma(\nu^* + \frac{d}{2})\beta_{jk}^d}{B(\alpha_{jk}, \nu^*)}\left(1 + u\right)^{2\nu^* + d}\mathcalprime{U}\left(\nu^*+ \frac{d}{2}, 1 - \alpha_{jk} + \frac{d}{2}, \beta_{jk}^2\frac{ u^2}{2}\right)\right\}_{j,k=1}^p\right]> 0.
    \end{align*}
\end{condition}
This condition is analogous to Eq. (8) of \cite{bachoc_asymptotically_2022}, demonstrating that the diagonal elements of the spectral density decay on the order of $(1 + u)^{-2\nu^* - d}$ for large $u$.
This also implies that after normalization by $(1 + u)^{2\nu^* + d}$ the matrix spectral density is well-conditioned.
This is only slightly more restrictive than multivariate covariance's validity \citep{bachoc_asymptotically_2022}.

\begin{prop}\label{prop:ch_equivalence}
    Let $\mathcalprime{P}_i$ be the Gaussian probability measure corresponding to the multivariate covariance $\mathcalprime{CH}(h; \nu^*, \mb{\alpha}^{(i)}, \mb{\beta}^{(i)}, \mb{\sigma}^{(i)})$ for $i =1, 2$, $d=1, 2,$ or $3$, $\nu^* > 0$, and $\alpha_{jk}^{(i)} > d/2$ for all $j$ and $k$. For both parameter sets $i \in \{1, 2\}$, let Condition \ref{as:ch_matrix} hold. 
    Then $\mathcalprime{P}_1$ and $\mathcalprime{P}_2$ are equivalent on the paths of $\mb{Y}(s)$ if, for all $j$ and $k \in \{1, \dots, p\}$, \begin{align*}
        \sigma_{jk}^{(1)}\frac{\Gamma\left(\nu^* + \alpha_{jk}^{(1)}\right)}{\left(\beta_{jk}^{(1)}\right)^{2\nu^*} \Gamma\left(\alpha_{jk}^{(1)}\right)} =     \sigma_{jk}^{(2)}\frac{\Gamma\left(\nu^*+ \alpha_{jk}^{(2)}\right)}{\left(\beta_{jk}^{(2)}\right)^{2\nu^*} \Gamma\left(\alpha_{jk}^{(2)}\right)}.
    \end{align*}
\end{prop}
The proof is omitted but follows from Theorem 2 of \cite{bachoc_asymptotically_2022} and the tail decay of the spectral density in Eq. \eqref{eq:tail_spec_density}. 
One recovers Theorem 3 of \cite{ma2022beyond} when $p=1$, and for $p > 1$ Proposition \ref{prop:ch_equivalence} requires checking the condition in Theorem 3 of \cite{ma2022beyond} for each of the covariances and cross-covariances. 
Proposition \ref{prop:ch_equivalence} implies that if all processes have the same smoothness parameter $\nu^*$, the parameters $\beta_{jk}$, $\alpha_{jk}$, and $\sigma_{jk}$ are not identifiable under infill asymptotics on a bounded domain \citep{zhang2004inconsistent}. 
If one is interested in a particular parameter of $\beta_{jk}$, $\alpha_{jk}$, and $\sigma_{jk}$, one may fix the other two to sensible values, then estimate the parameter of interest \citep{zhang2004inconsistent}. 
While the estimate of the remaining parameter will remain inconsistent, estimates will be more stable when comparing across optimizations. 
Proposition \ref{prop:ch_equivalence} also suggests that a misspecified multivariate CH model may attain asymptotically efficient prediction.

As the next proposition establishes, the multivariate CH and Mat\'ern covariances are equivalent in terms of Gaussian measures.

\begin{prop}\label{prop:ch_matern_equivalence}
    Let $\mathcalprime{P}_1$ be the Gaussian probability measure corresponding to the multivariate covariance $\mathcalprime{CH}(h; \nu^*, \mb{\alpha}, \mb{\beta}, \mb{\sigma}^{(1)})$, $d=1, 2,$ or $3$, $\nu^* > 0$, $\alpha_{jk} > d/2$ for all $j$ and $k$, for which Condition \ref{as:ch_matrix} holds. 
    Let $\mathcalprime{P}_{\mathcalprime{M}}$ be the Gaussian probability measure corresponding to the multivariate covariance $\mathcalprime{M}(h; \nu^*, \mb{\phi}, \mb{\sigma}^{(2)})$, for which Eq. (8) of \cite{bachoc_asymptotically_2022} holds. 
    Then $\mathcalprime{P}_1$ and $\mathcalprime{P}_{\mathcalprime{M}}$ are equivalent on the paths of $\mb{Y}(s)$ if, for all $j$ and $k \in \{1,  \dots, p\}$,\begin{align*}
        \sigma_{jk}^{(1)}\frac{2^{\nu^*}\Gamma\left(\nu^* + \alpha_{jk}\right)}{\beta_{jk}^{2\nu^*} \Gamma\left(\alpha_{jk}\right)} =     \sigma_{jk}^{(2)}\frac{1}{\phi_{jk}^{2\nu^* }}.
    \end{align*}
\end{prop}
The proof is omitted but follows from Theorem 2 of \cite{bachoc_asymptotically_2022} and covariances' spectral densities.
One recovers Theorem 4 of \cite{ma2022beyond} when $p=1$, and the result implies checking the condition in \cite{ma2022beyond} for all covariances and cross-covariances. 

\subsection{Spectrally-Generated Multivariate CH Models}\label{sec:spectral}

Thus far, cross-covariances proportional to a CH covariance have been discussed. 
However, as discussed by \cite{yarger2023multivariate}, \cite{alegria2021bivariate}, and \cite{li_approach_2011} among others, cross-covariance functions may be much more flexible. 
For example, they may be asymmetric so that $E\{Y_j(s)Y_k(s^\prime)\} \neq E\{Y_j(s^\prime)Y_k(s)\}$ for $j\neq k$. 
\cite{yarger2023multivariate} proposed using the spectral density to create more flexible multivariate Mat\'ern models. 
In this Section, similar multivariate CH models based on the spectral density in Proposition \ref{prop:spectral_density} are discussed.

\begin{figure}
    \centering
        \begin{subfigure}[t]{0.32\textwidth}
\includegraphics[width = .98\textwidth]{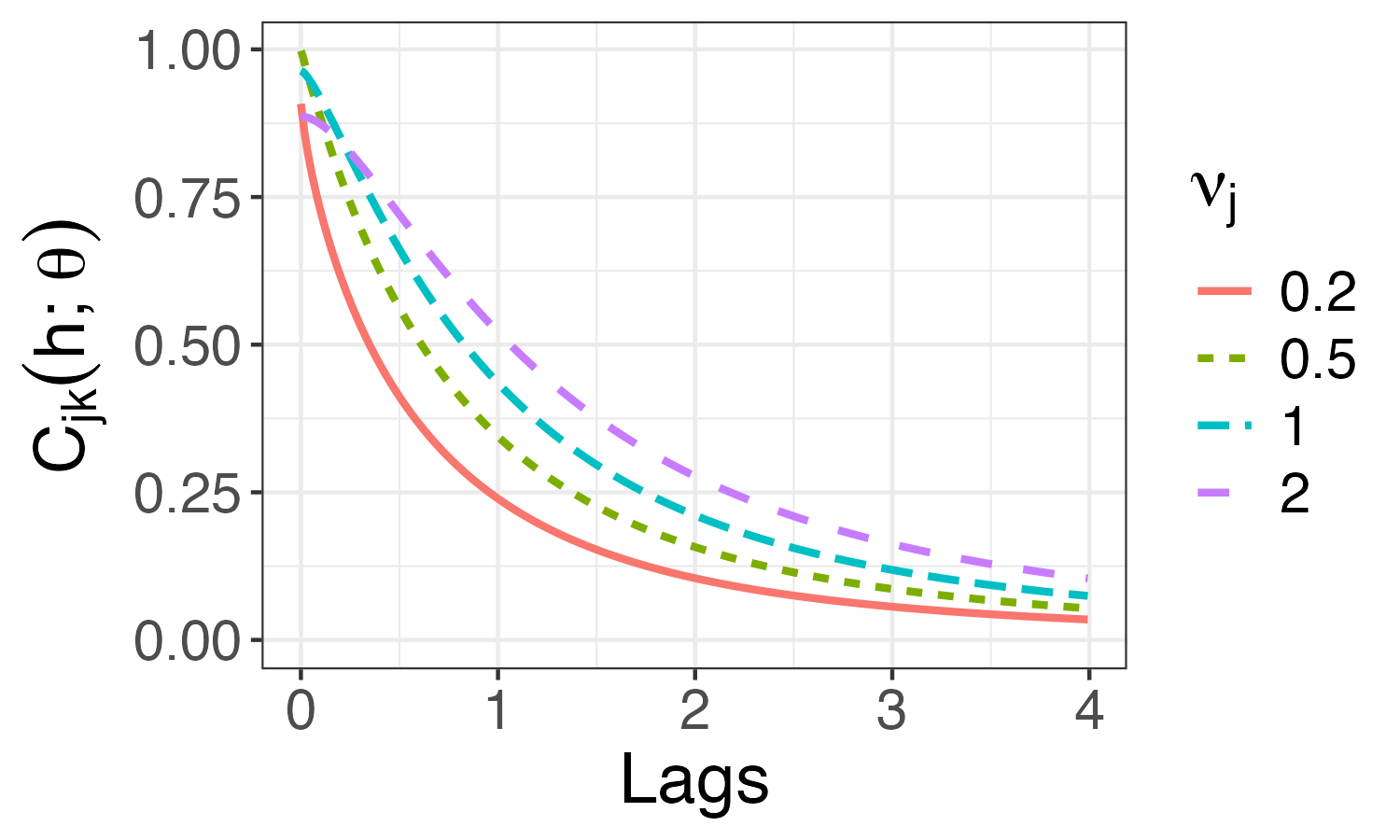}
    \caption{}
    \end{subfigure}
            \begin{subfigure}[t]{0.32\textwidth}
\includegraphics[width = .98\textwidth]{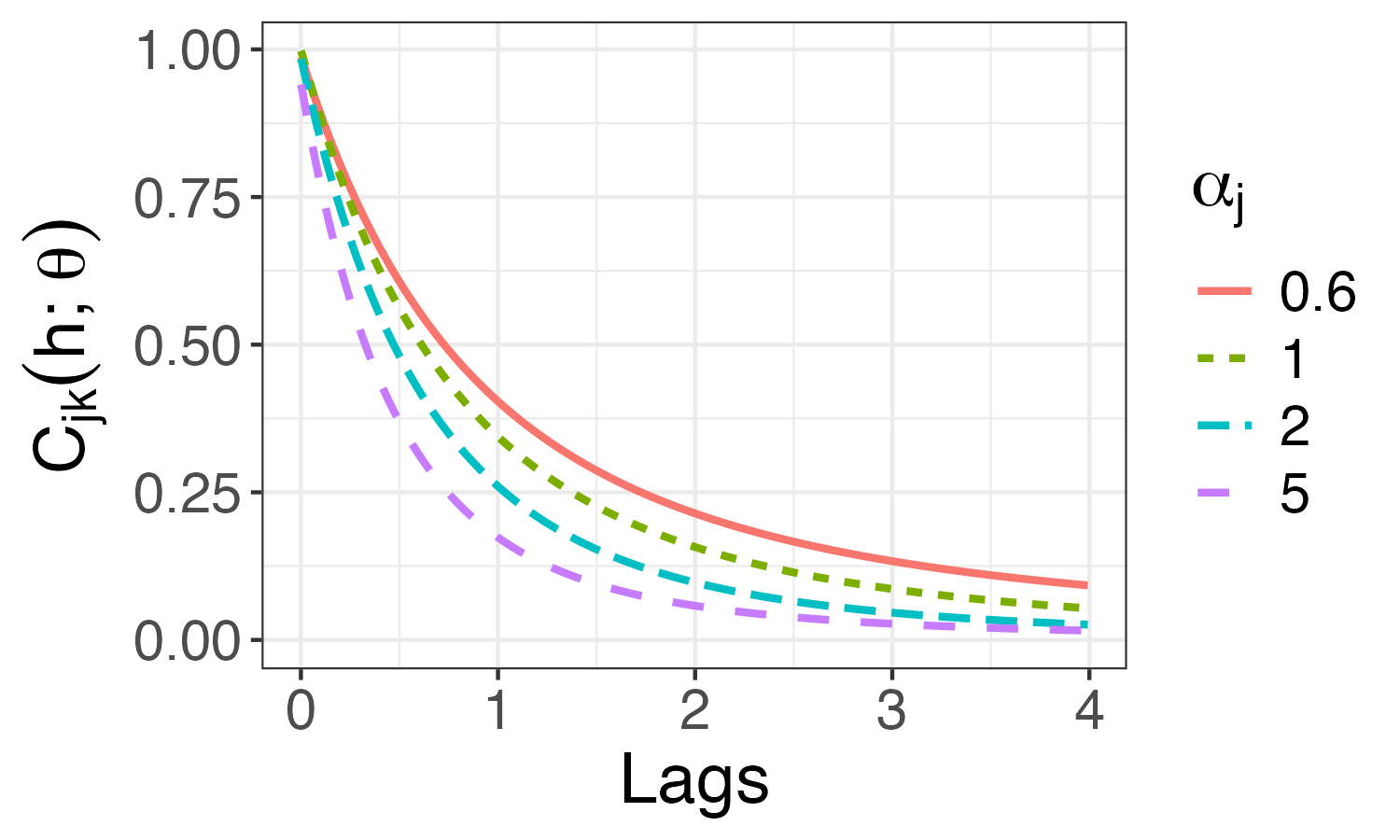}
    \caption{}
    \end{subfigure}
            \begin{subfigure}[t]{0.32\textwidth}
\includegraphics[width = .98\textwidth]{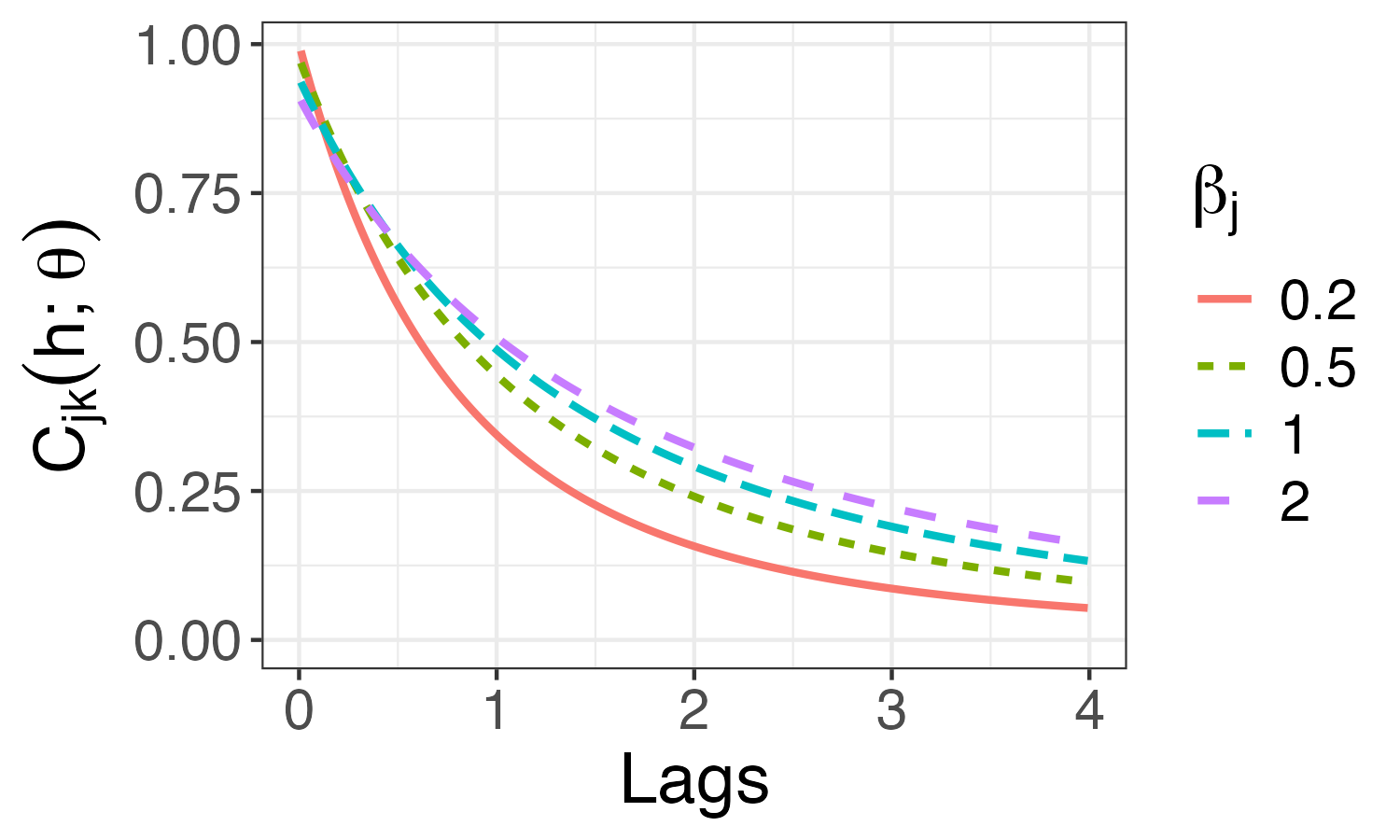}
    \caption{}
    \end{subfigure}

            \begin{subfigure}[t]{0.32\textwidth}
\includegraphics[width = .98\textwidth]{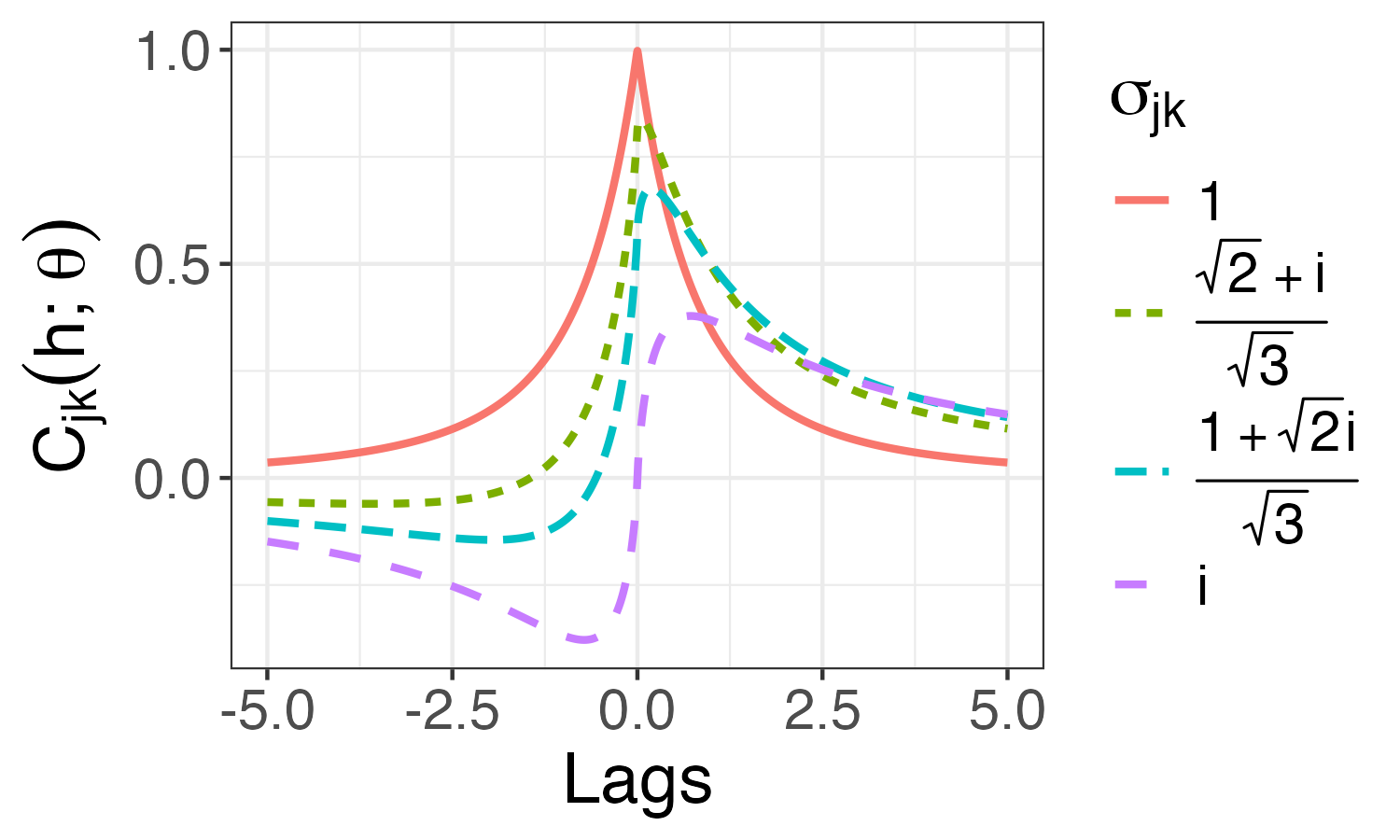}
    \caption{}
    \end{subfigure}
            \begin{subfigure}[t]{0.32\textwidth}
\includegraphics[width = .98\textwidth]{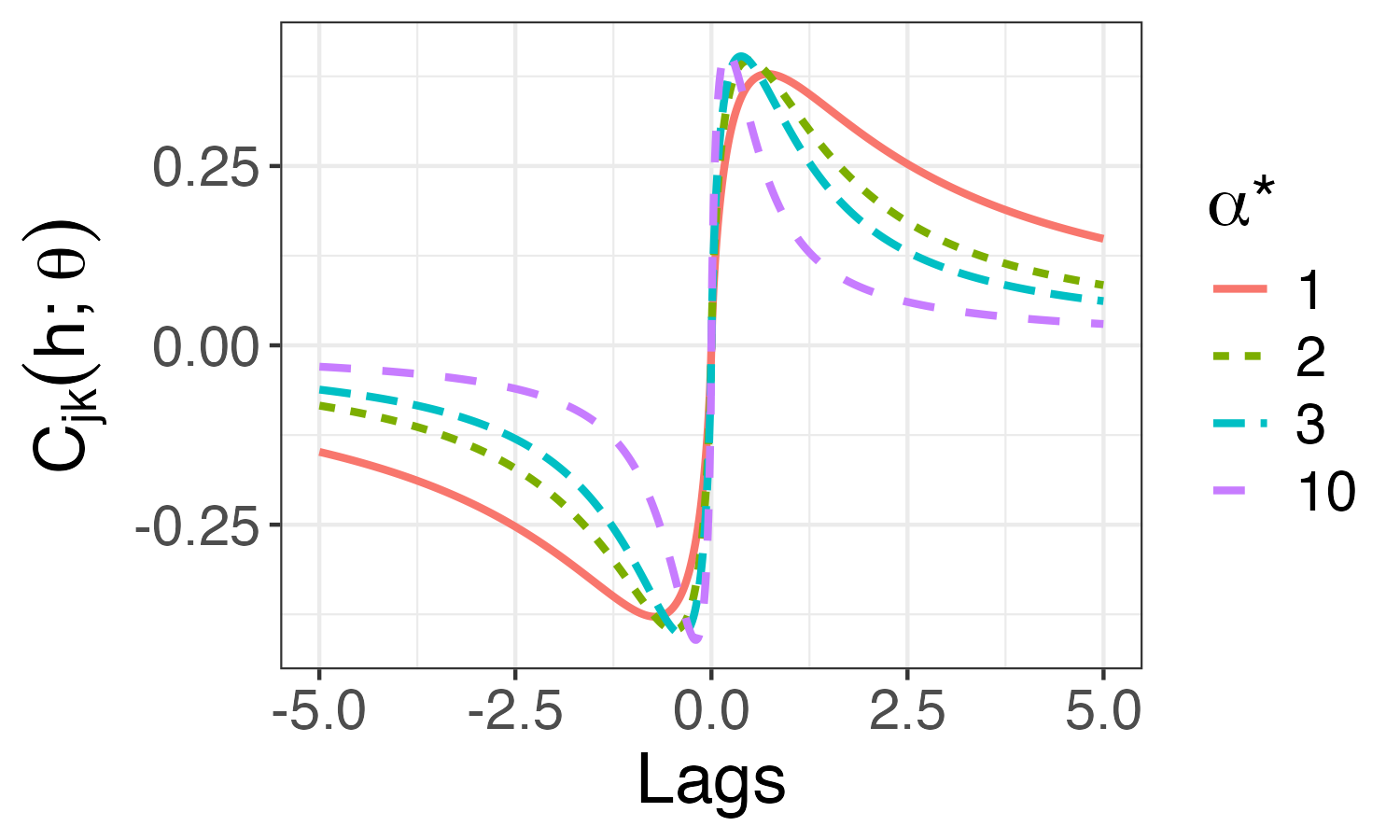}
    \caption{}
    \end{subfigure}
            \begin{subfigure}[t]{0.32\textwidth}
\includegraphics[width = .98\textwidth]{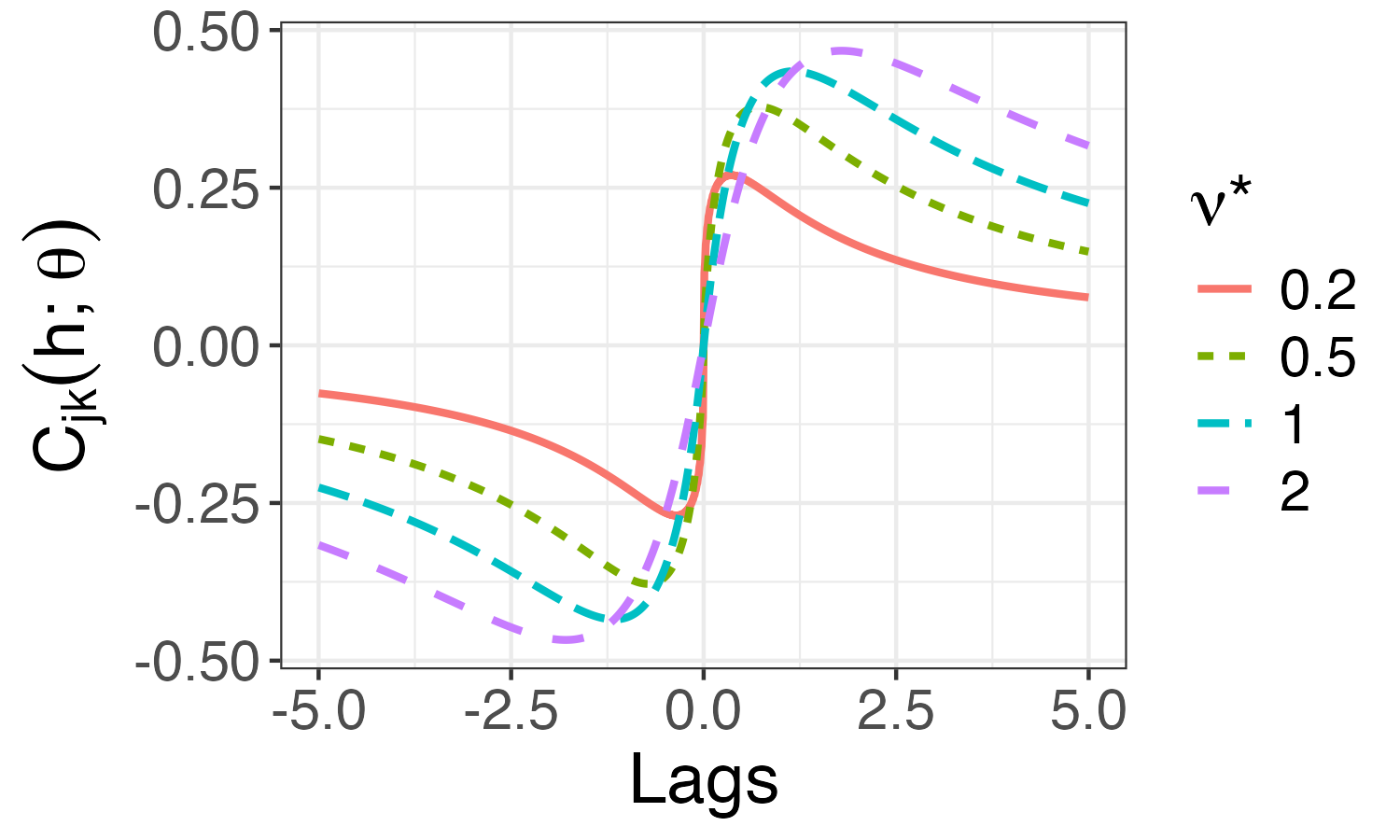}
    \caption{}
    \end{subfigure}
    \caption{Spectrally-generated cross-covariances in $d=1$ and unless otherwise specified, take $\nu_j = \nu_k = 1/2$ and $ \alpha_j = \alpha_k = \beta_j = \beta_k = 1$
    (Top) Isotropic ($\sigma_{jk} = 1$) cross-covariances with (a) $\nu_j$ varying; (b) with $\alpha_j$ varying; (c) with $\beta_j$ varying (Bottom) Asymmetric cross-covariances with (d) $\sigma_{jk}$ varying; 
    (e) $\sigma_{jk} =\I$ and $\alpha^* = \alpha_j = \alpha_k$ varying; (f) $\sigma_{jk} =\I$ and  $\nu^* = \nu_j = \nu_k$ varying }
    \label{fig:spectral}
\end{figure}

Consider an isotropic multivariate covariance with $j$, $k$ entry \begin{align}\begin{split}
    &\int_{\mathbb{R}^d} e^{\I h^\top x } \sigma_{jk} \left\{f_{\mathcalprime{CH}}(x; \nu_j, \alpha_j, \beta_j, 1)\right\}^{\frac{1}{2}}\left\{f_{\mathcalprime{CH}}(x; \nu_k, \alpha_k, \beta_k, 1)\right\}^{\frac{1}{2}}dx .
    \\
    \end{split} \label{eq:real_cc}
\end{align}
For $j=k$, this reduces to the CH covariance with parameters $\nu_j$, $\alpha_j$, $\beta_j$, and $\sigma_{jj}$, assuming that $\alpha_j > d/2$ for all $j$. 
Since one may represent the matrix-valued spectral density as $\mb{P}(x) \mb{\sigma} \mb{P}(x)^\top$ (under matrix multiplication), where $$\mb{P}(x) = \textrm{diag}\left\{f_{\mathcalprime{CH}}(x; \nu_j, \alpha_j, \beta_j, 1)^{\frac{1}{2}}, j=1, \dots, p\right\},$$ the matrix-valued spectral density is positive semidefinite for any $x$ (and thus the model is valid) when $\mb{\sigma}$ is positive semidefinite. 
While the integral in Eq. \eqref{eq:real_cc} likely does not have a closed form, one can compute it efficiently using fast Fourier or $1$-dimensional Hankel transforms \citep{stein1999interpolation}.
Like the class of models presented in Theorem \ref{thm:pars_like}, additional parameters $\nu_{jk}$, $\alpha_{jk}$, and $\beta_{jk}$ for $j\neq k$ do not need to be estimated; based on the form \eqref{eq:real_cc}, these parameters do not exist in this model. 
In Figs. \ref{fig:spectral}(a), \ref{fig:spectral}(b), and \ref{fig:spectral}(c), cross-covariances for different parameter values are plotted. 
Predictably, the parameters $\nu_j$, $\alpha_j$, and $\beta_j$ still have influence over the origin behavior, the tail behavior, and the scale of the cross-covariance, respectively.

To create asymmetric forms in the cross-covariance, $\sigma_{jk}$ is replaced with a complex-valued parameterization. 
In the setting $d=1$, consider the multivariate covariance with $j$, $k$ entry \begin{align*}
    \int_{\mathbb{R}} e^{\I  h x  }& \left\{\Re( \sigma_{jk}) - \textrm{sign}(x)\I \Im(\sigma_{jk})\right\} \left\{f_{\mathcalprime{CH}}(x; \nu_j, \alpha_j, \beta_j, 1)\right\}^{\frac{1}{2}}\\
    &~~~~~~~~\times \left\{f_{\mathcalprime{CH}}(x; \nu_k, \alpha_k, \beta_k, 1)\right\}^{\frac{1}{2}}dx.
\end{align*}The matrix $\sigma$ is now a positive semidefinite and Hermitian matrix with potentially complex entries on the off-diagonal. 
On the diagonals, $\Im(\sigma_{jj})=0$ and one has Eq. \eqref{eq:real_cc} with a CH covariance. 
However, on the off-diagonals, if $\Im(\sigma_{jk}) \neq 0$, asymmetry in the cross-covariance is obtained. 
If $\Re(\sigma_{jk})= 0 $, then the cross-covariance is an odd function, that is, $E\{Y_j(s)Y_k(s^\prime)\} = - E\{Y_j(s^\prime)Y_k(s)\}$, as it is the Fourier transform of an odd function \citep[or, alternatively, the Hilbert transform of a even function, see][]{king2009hilbert}. 
In Figs. \ref{fig:spectral}(d), \ref{fig:spectral}(e), and \ref{fig:spectral}(f), various examples are plotted. 
By varying $\Re(\sigma_{jk})$ and $\Im(\sigma_{jk})$ together, the model's flexibility covers the symmetric case as well as a variety of asymmetric cases. 
For $d>1$, one may extend this model using polar coordinates similar to Sect. 4 of \cite{yarger2023multivariate}. 
While somewhat challenging computationally, this construction can  comprehensively describe processes' origin behavior \citep[see the supplement of][]{yarger2023multivariate}.

The spectral approach also enables construction of cross-covariances in more general settings. 
For example, suppose that $Y_1(s)$ has Mat\'ern covariance function $\mathcalprime{M}(h; \nu_1, \phi^*, \sigma_{11})$ and $Y_2(s)$ has CH covariance function $\mathcalprime{CH}(h; \nu_2, \alpha^*, \beta^*, \sigma_{22})$ with $\alpha^* > d/2$. 
Construct the cross-covariance generated by $$
    \int_{\mathbb{R}^d} e^{\I h^\top x } \sigma_{12} \{f_{\mathcalprime{M}}(x; \nu_1, \phi^*, 1)\}^{\frac{1}{2}}\{f_{\mathcalprime{CH}}(x; \nu_2, \alpha^*, \beta^*, 1)\}^{\frac{1}{2}} dx$$ for $\sigma_{12} \in\mathbb{R}$, which is valid whenever $\lvert\sigma_{12}\rvert\leq (\sigma_{11}\sigma_{22})^{1/2}$. 
Constructing a cross-covariance between two processes that have a different class of marginal covariance functions has previously received attention in \cite{maleki2017joint} and \cite{porcu2018shkarofsky}. 

One limitation of these models is that spectral densities of the covariances are required. Thus, when $\alpha_j <d/2$ for a CH process (the long-range dependence case), such constructions are not available. 
In this case, shift-based asymmetries may be more feasible \citep{li_approach_2011}.

\section{Simulation Experiments}\label{sec:sims}

Consider samples from a bivariate spatial process consisting of $Y_1(s)$ at locations $\mb{s}_1 = (s_{11}, \dots, s_{1n_1})$ and $Y_2(s)$ at locations $\mb{s}_2 = (s_{21}, \dots, s_{2n_2})$. 
That is, take $n_1$ samples of $Y_1(s)$ and $n_2$ samples of $Y_2(s)$, with $s_{ji} \in \mathbb{R}^2$ for $j\in\{1,2\}$. 
If $n_1 = n_2$ and $s_{1i} = s_{2i}$ for all $i \in \{1, \dots, n_1\}$, the processes are observed at the same locations, and the samples of $Y_1(s)$ and $Y_2(s)$ are colocated. 
Consider the prediction of both $Y_1(s)$ and $Y_2(s)$ at a set of additional locations $\mb{s}_{out}$ of size $n_{out}$. 
The locations $\mb{s}_1$ are a uniform random sample of points on $[0,1] \times [0,1]$ with $n_1 = 200$.
The locations $\mb{s}_2$ are similarly generated with $n_2 = 400$, so that $\mb{s}_1$ and $\mb{s}_2$ are entirely distinct. 
Finally, $\mb{s}_{out}$ is sampled similarly with $n_{out} = 200$, so that one predicts at entirely different locations than in $\mb{s}_1$ and $\mb{s}_2$. 

For simpler comparison with the Mat\'ern model, assume that one knows that $\beta_1 = \beta_2$ and takes $\nu_{12} = (\nu_1 + \nu_2)/2$ and $\alpha_{12} = (\alpha_1 + \alpha_2)/2$ as in Theorem \ref{thm:pars_like}.
A bivariate parsimonious Mat\'ern model is also estimated with $\nu_{12} = (\nu_1 + \nu_2)/2$ and $\phi_{12} = \phi_1 = \phi_2$. 
The CH model thus has two additional parameters, $\alpha_1$ and $\alpha_2$, compared to the multivariate Mat\'ern. 
For the CH model, the parameters $\nu_1$, $\nu_2$, $\alpha_1$, $\alpha_2$, $\beta_1$, $\beta_2$, $\sigma_{11}$, $\sigma_{22}$, and $\rho_{12} = \sigma_{12}/\sqrt{\sigma_{11}\sigma_{22}}$ are estimated. 
Finally, as in \cite{ma2022beyond}, generalized Cauchy (GC) covariance is estimated, with multivariate form $
    \sigma_{jk}\{1 + (\lVert h \rVert /\phi_{jk})^{\alpha_{jk}}\}^{-\beta_{jk}/\alpha_{jk}}.$
Since conditions for validity of the multivariate Generalized Cauchy model remain quite technical \citep[see][]{moreva2023bivariate, emery_schoenberg_2023}, one takes that $\alpha_{jk} = \alpha$, $\beta_{jk} = \beta$, and $\phi_{jk} = \phi$ for all $j$ and $k$, which considerably simplifies the estimated covariance. 
All parameters are estimated numerically by maximum likelihood, using the standard L-BFGS-B algorithm over the unknown parameters \citep{byrd1995limited}. 
Before each evaluation of the likelihood, the validity of the model is ensured, using \eqref{eq:cor_thm1} for the CH model, for example. 
If validity was not ensured, an unreasonably small value for the likelihood was returned. 

Let $Y_1(\mb{s}_1) = \{Y_1(s_{1j})\}_{j=1}^{n_1}$ and likewise for similar expressions. 
To predict the response variable $Y_1(\mb{s}_{out})$, conditional expectations and variances given $Y_2(\mb{s}_2)$, and vice-versa, are used in the Gaussian process setting, to emphasize the cross-covariance over the marginal covariances. 
These are compared to the baseline null prediction of $0$ at each location (``Prediction of 0'') that represents no covariance used or estimated. 
Data is generated from these three multivariate covariances classes, and consider two settings, settings A and B, for parameters. 
These two settings differ primarily in the range parameter $\beta_1 = \beta_2$ or $\phi_1 = \phi_{2}$: setting A takes these parameters smaller to focus on the tail behavior of the covariance, while setting B takes these parameters larger to focus on the origin behavior of the covariance.
Approaches are compared over $100$ simulations for each setting and true covariance.

\begin{figure}
    \centering
        \begin{subfigure}[t]{0.48\textwidth}
\includegraphics[width = .98\textwidth]{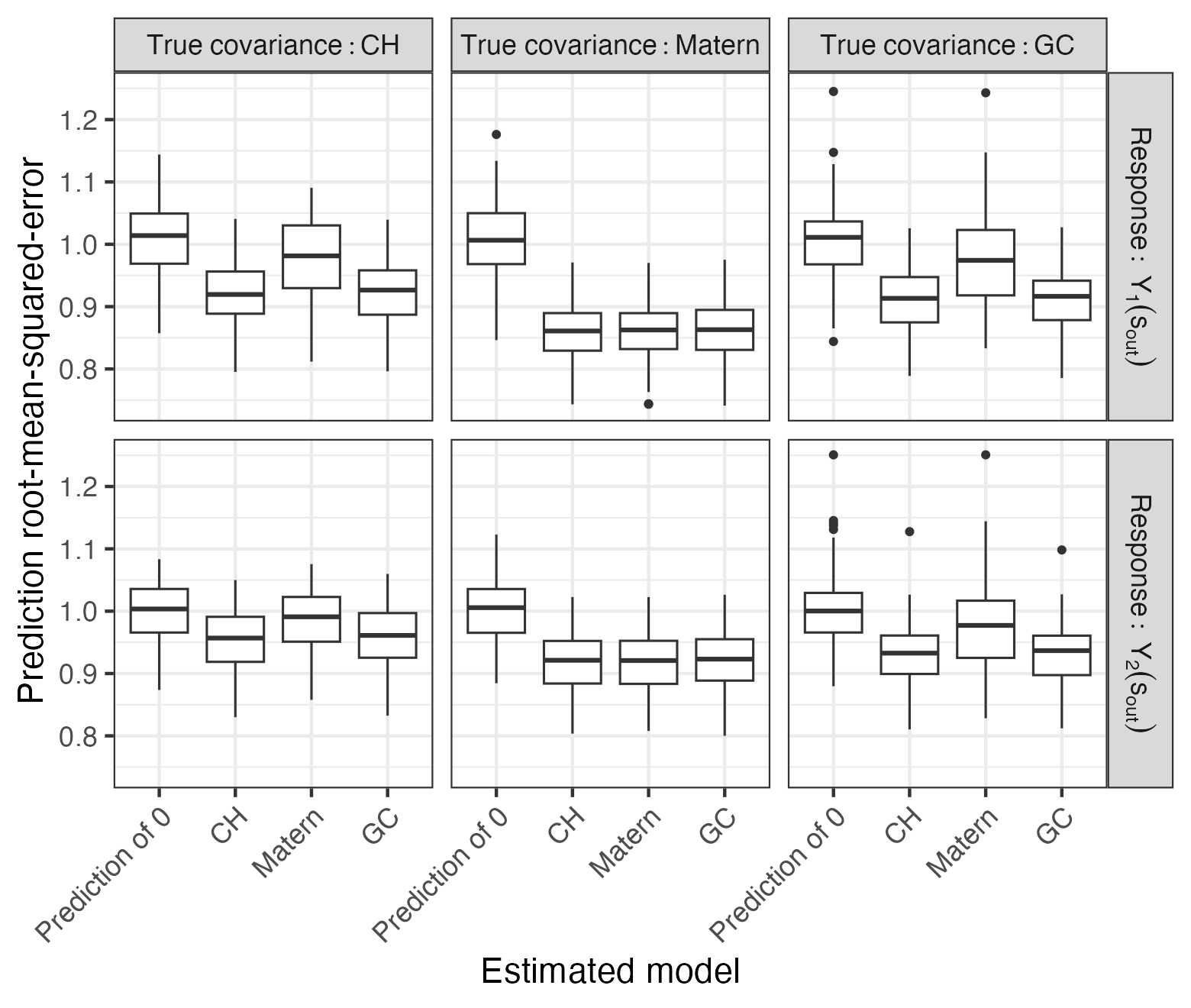}
    \caption{}
    \end{subfigure}
            \begin{subfigure}[t]{0.48\textwidth}
\includegraphics[width = .98\textwidth]{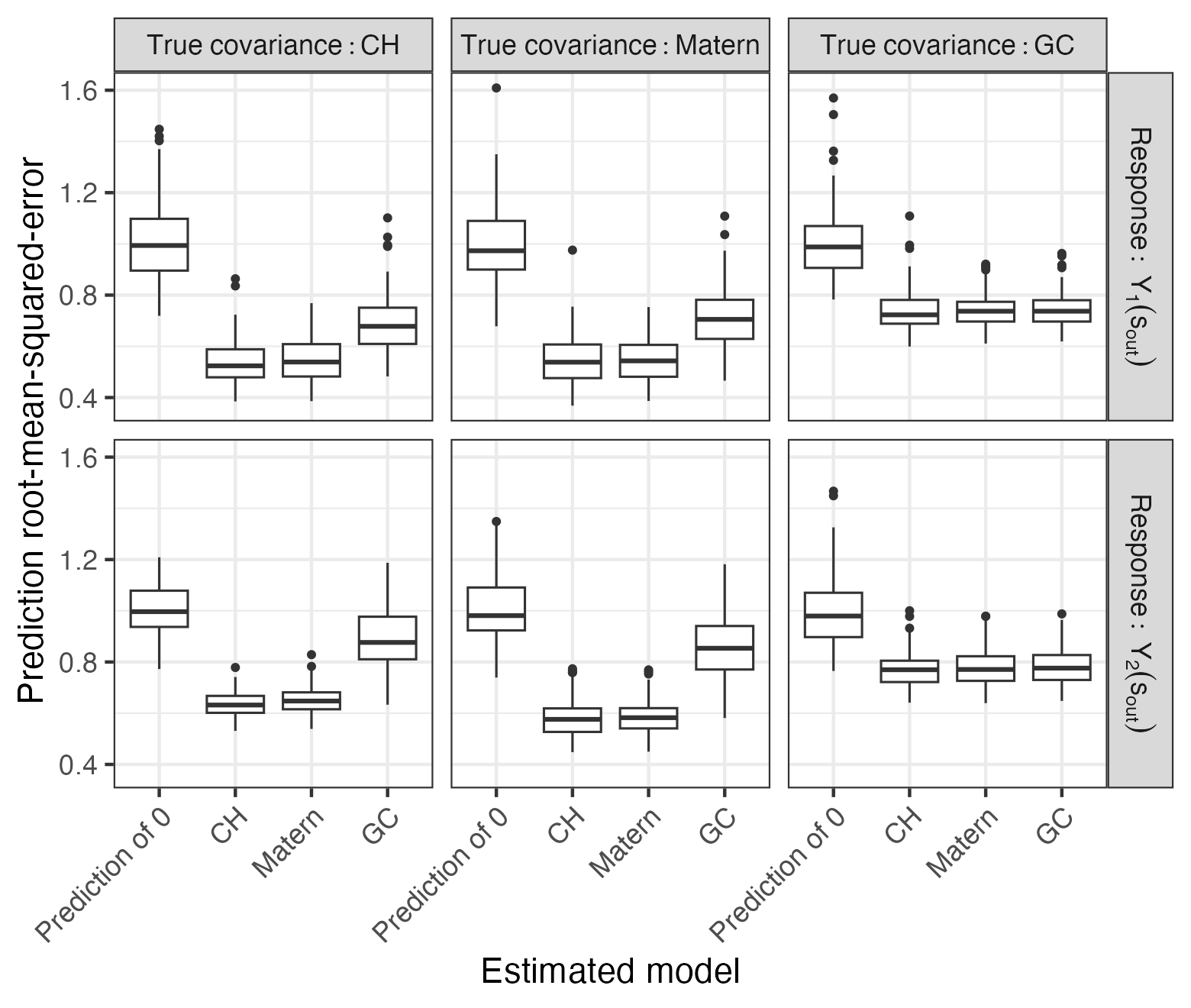}
    \caption{}
    \end{subfigure}
    \caption{Simulation prediction error results for (a) setting A and (b) setting B, using boxplots formed from the prediction RMSE of each of the 100 simulations}
   \label{fig:ch_initial_simulation}
\end{figure}

For each generated covariance, the variables are standardized with $\sigma_{11}=\sigma_{22} = 1$, and the processes are given strong correlation $\sigma_{jk} = 0.80$. 
When the true multivariate covariance is confluent hypergeometric, setting A takes $\nu_1 = 1.75$, $\nu_2 = 1.25$, and  $\nu_{12} = 1.5$ to have different smoothness parameters. 
The tail parameters are $\alpha_1 = 1.1$, $\alpha_2 = 1.9$, and $\alpha_{12} = 1.5$ with relatively heavy tail decay of the covariances and cross-covariances. 
The values for $\beta_1=\beta_2=\beta_{12} = 0.015$ are relatively small to again accentuate the data-sparse setting. 
For setting B, all parameters are the same, except $\nu_1 = 2.25$ and $\nu_{12} = 1.75$, to increase the gap between smoothing parameters, and $\beta_1=\beta_2=\beta_{12} = 0.075$, to effectively making observations closer together. 
Both settings result in the conditions of Theorem \ref{thm:pars_like} being met.
When the true multivariate covariance is Mat\'ern, setting A takes $\phi_{12} = \phi_{11} = \phi_{22} =0.015$, $\nu_1 = 1.75$, $\nu_2 = 1.25$, and $\nu_{12} = 1.50$, with parameters chosen similarly to when the true covariance is CH. 
As with the CH, setting B takes $\nu = {2.25}$, $\nu_{12} = 1.75$, and $\phi_{12} = \phi_{11} = \phi_{22} =0.075$.
For the GC model, setting A takes $\alpha_{jk} = 1$, $\beta_{jk} = 1$, and $\phi_{jk} = 0.015$ for all $j$ and $k$, again aiming to have similar parameters compared to the simulated CH processes. 
For setting B, $\phi_{jk}$ is set to be $0.075$ to be similar to the Mat\'ern and CH settings.

Prediction error results are plotted in Fig. \ref{fig:ch_initial_simulation}.
When the correct cross-covariance class is specified, performance is as good or better than any other class estimated. 
However, the Mat\'ern cross-covariance performs worse in setting A when the true covariance has heavy tails. 
Similarly, the GC cross-covariance performs worse in setting B when the CH or Mat\'ern is the true covariance. 
This follows expectations that the Mat\'ern model is challenged by heavy tail behavior and the GC model lacks flexibility in the origin behavior. 
In contrast, the CH cross-covariance performs approximately as well as any other model in each setting, demonstrating its robustness in flexibly capturing tail and origin behavior of the cross-covariances. 

Results are presented in Table \ref{tab:ch_interval} from the uncertainty estimates of the predictions based on the conditional variances, averaged over both response variables $Y_1(s_{out})$ and $Y_2(s_{out})$. 
In general, the results match the prediction results, now in terms of interval coverage rates, as well as a summary of the interval lengths (the lower interval bound subtracted from the upper interval bound). 
The Mat\'ern model predicts shorter intervals and has lagging coverage of 75\% in setting A when the true covariance has heavier tails. 
In setting B, the GC model cannot appropriately handle the true CH or Mat\'ern covariance, with longer intervals, of average length 2.8 compared to 2.3, and coverage near 90\%. 
The multivariate CH model is adaptable to both settings, performing well on coverage and length comparable to when using the true covariance, for both settings A and B and for each of the three data-generating covariances.

\begin{table}[t]
    \centering
    \begin{tabular}{|c|c|ccc|ccc|}\hline
    \multirow{2}*{Setting}&\multirow{2}*{True Cov}
    & \multicolumn{3}{c}{95\% CI coverage} & \multicolumn{3}{|c|}{Average length} \\ 
        & & CH & M & GC & CH & M& GC \\ \hline 
       A & CH & 94.8 & 75.8 & 94.8 & 3.66 & 2.54 & 3.67\\ 
        A                         & M & 94.8 & 94.8 & 94.8 & 3.46 & 3.46 & 3.48  \\ 
         A                        & GC & 94.9 & 75.1 & 95.0 & 3.61 & 2.46 & 3.61\\ 
        B& CH & 95.3 & 93.4 & 89.8 & 2.33 & 2.20 & 2.77 \\ 
         B                      & M & 96.1 & 95.3 & 89.8 & 2.30 & 2.24 & 2.84\\ 
          B                    & GC & 95.2 & 94.2 & 95.0 & 3.00 & 2.91 & 3.00\\ \hline
    \end{tabular}
    \caption{Prediction interval summaries under different true covariances, comparing estimated CH, Mat\'ern (M), and generalized Cauchy (GC) multivariate covariances}
    \label{tab:ch_interval}
\end{table}

In the Supplement, additional simulations are provided: results for predictions using $Y_2(\mb{s}_{out})$ when predicting $Y_1(\mb{s}_{out})$ and vice versa, as well as a colocated simulation design.

\section{Analysis of Oceanography Data}\label{sec:real}

\subsection{Data and Model Estimation}

The data analysis focuses on an oceanography dataset of temperature, salinity, and oxygen from the Southern Ocean carbon and climate observations and modeling (SOCCOM) project \citep{johnson_soccom_2020}, consisting of measurements of these variables at different depths in the ocean collected by autonomous devices called floats. 
This project is part of the larger Argo project dedicated to float-based monitoring of the oceans \citep{wong2020argo, argo_argo_2023}. 
Most Argo floats collect only temperature and salinity data, referred to as Core Argo floats. 
However, some floats also collect biogeochemical (BGC) variables including oxygen, pH and nitrate; these are referred to as BGC Argo floats.  
The multivariate relationship of temperature, salinity, and oxygen can inform how one uses available Core Argo data to predict oxygen. 
The problem of estimating oxygen based on temperature and salinity data has received interest in \cite{giglio2018} and \cite{korte2022multivariate}. 
The data comes from an area in the Southern Ocean bounded by 100 and 180 degrees of longitude from a depth of 150m in the ocean. 
Data from the months of February, March, and April over the years 2017--2023 are used, resulting in 436 total locations. 
The salinity, temperature, and oxygen data are plotted in Fig. \ref{fig:argo_intro}.

Based on the original data in Fig. \ref{fig:argo_intro}, the mean of the data depends on longitude and latitude, and a spatially-varying mean is estimated using local linear smoothing with a bandwidth of 1{,}000 kilometers.
The resulting residuals, plotted in Fig. \ref{fig:argo_intro} are treated like in \cite{kuusela_locally_2018}: locally stationary, so that a stationary covariance model in this region is used, and the data from different years is assumed to be independent, so that the overall log-likelihood is the sum of the Gaussian log-likelihoods from each of the seven years. 
Isotropic trivariate CH and Mat\'ern covariances of the form from Sect. \ref{sec:valid_ch} are estimated, with a nugget variance parameter $\tau$ for each of the three processes, which is also done in previous analysis in the literature \cite[for example,][]{gneiting2010matern, kuusela_locally_2018}. 
Again, the likelihood is maximized using the L-BFGS-B algorithm \citep{byrd1995limited}, and unreasonably low likelihood values are returned when the covariance is determined invalid based on the minimum eigenvalue of the matrix in Theorem 1. 

\begin{table}[t]
    \centering
    \begin{tabular}{|ccc|ccc|} 
    \hline
        Parameter & CH Estimate & M Estimate & Parameter & CH Estimate & M Estimate \\ \hline 
        $\nu_S$ & 0.388 & 0.376 & $\sigma_S$ & 0.0257 & 0.0251 \\ 
        $\nu_O$ & 0.154 & 0.158 & $\sigma_O$ & 92.327 & 91.708\\ 
        $\nu_T$ & 0.432 & 0.432 & $\sigma_T$ & $~1.584$ & ~1.670\\ 
        $\alpha_S$ & 0.768 & - & $\sigma_{SO}/\left(\sigma_S\sigma_O\right)^{1/2}$ & $-0.705$ & $-0.691$ \\ 
        $\alpha_O$ & 0.475 & - & $\sigma_{ST}/\left(\sigma_S\sigma_T\right)^{1/2}$ & ~0.700 & ~0.741\\ 
        $\alpha_T$ & 2.700 & - & $\sigma_{TO}/\left(\sigma_T\sigma_O\right)^{1/2}$ & $-0.543$ & $-0.559$\\ 
        $\beta_S$ or $\phi_S$ & 472.7 & 506.2 & $\tau_S/\sigma_S$ & 0.0577 & 0.0461\\ 
        $\beta_O$ or $\phi_O$ & 584.5 & 769.0 & $\tau_O/\sigma_O$ & 0.0003 & 0.0001 \\ 
        $\beta_T$ or $\phi_T$ & 721.6 & 442.8 &  $\tau_T/\sigma_T$ & 0.0021 & 0.0005\\ \hline
    \end{tabular}
    \caption{Estimated parameters for the CH and Mat\'ern (M) models for the Argo data, with the subscripts S, O, and T represent salinity, oxygen, and temperature, respectively} 
    \label{tab:est_argo_params}
\end{table}

Since the locations are colocated, exploratory marginal correlation estimates between the processes are straightforwardly computed: $0.694$ (temperature and salinity), $-0.551$ (temperature and oxygen), and $-0.690$ (salinity and oxygen).  
The parameter estimates resulting from maximum likelihood estimation are presented in Table \ref{tab:est_argo_params}. 
The estimates of the smoothness parameters $\nu_S$, $\nu_O$, and $\nu_T$ are similar between the CH and Mat\'ern classes. 
For the other parameters, there may not be similarities due to the result in Proposition \ref{prop:ch_matern_equivalence}. 
For salinity and oxygen, the CH covariance estimates heavy tails for the covariance functions with smaller $\alpha_S$ and $\alpha_O$.

\subsection{Prediction}

\begin{figure}
    \centering
    \begin{subfigure}[t]{0.48\textwidth}
\includegraphics[width = .98\textwidth]{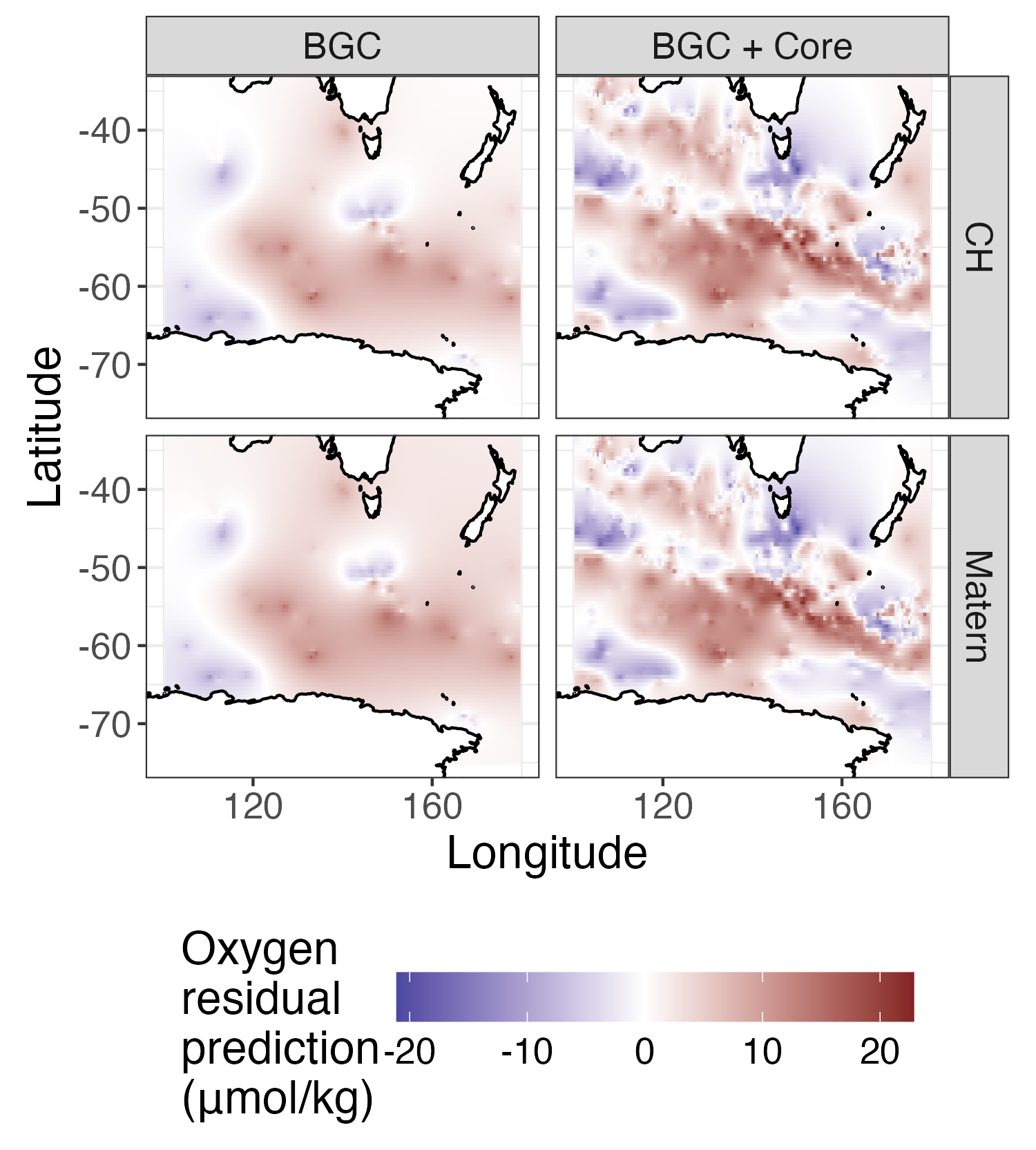}
    \caption{}
    \end{subfigure}
        \begin{subfigure}[t]{0.48\textwidth}
\includegraphics[width = .98\textwidth]{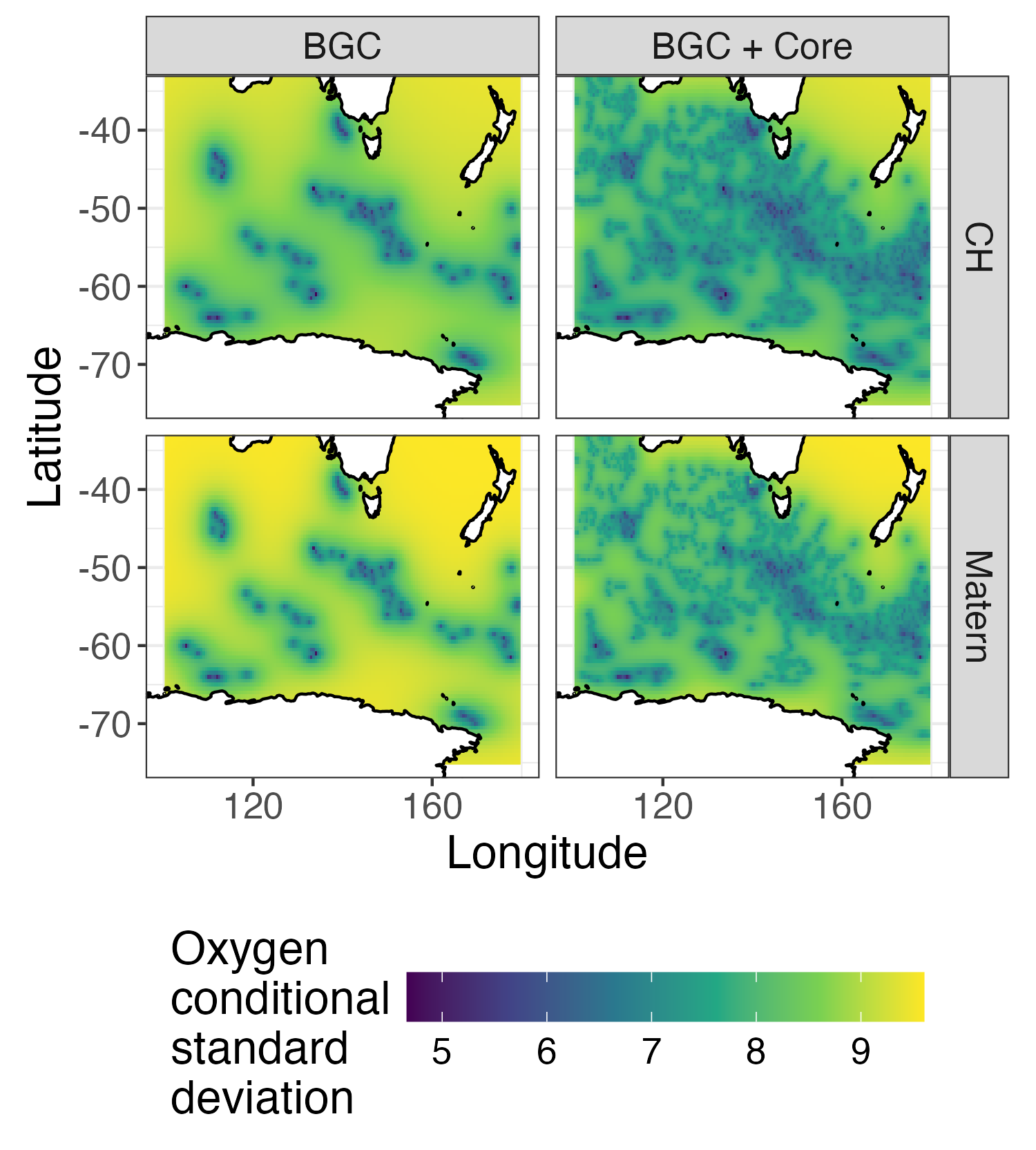}
    \caption{}
    \end{subfigure}

    \begin{subfigure}[t]{0.8\textwidth}
\includegraphics[width = .98\textwidth]{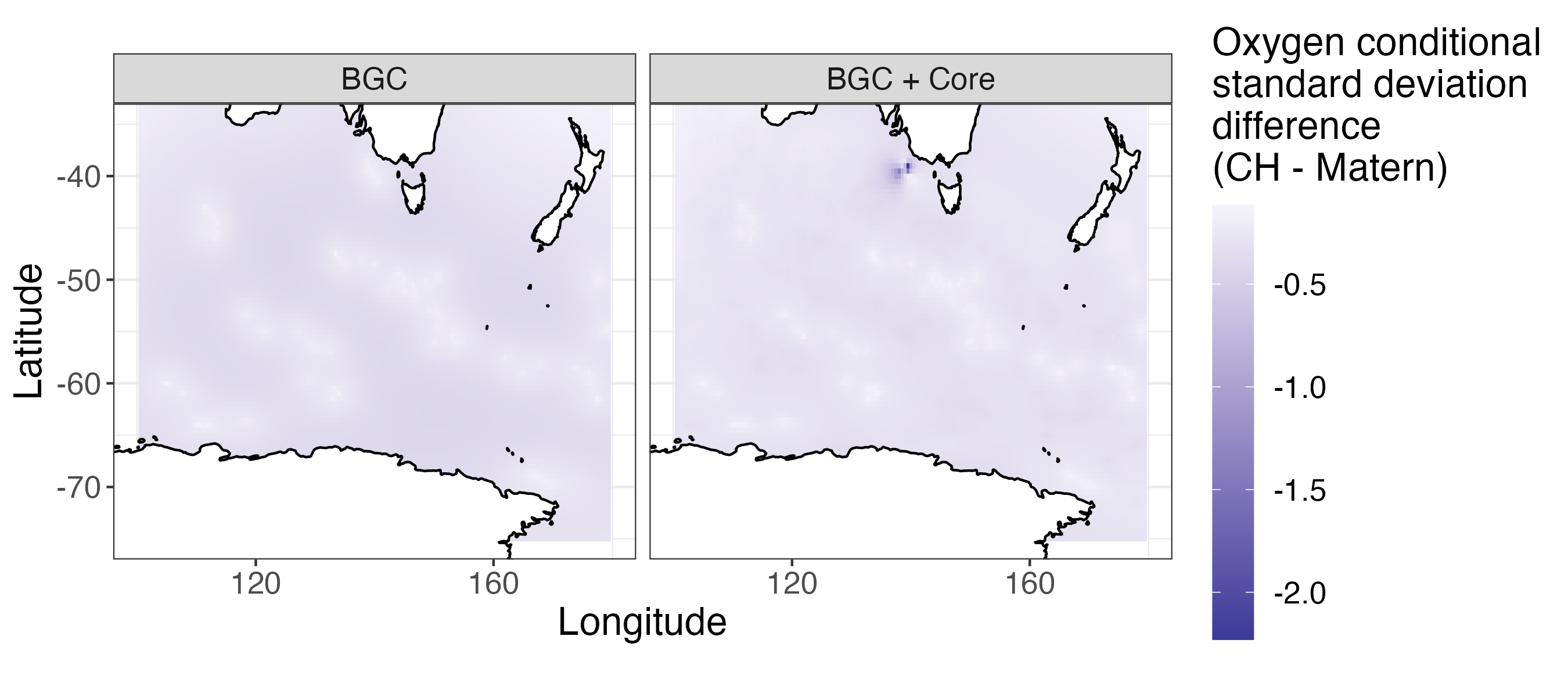}
    \caption{}
    \end{subfigure}
    \caption{(a) Prediction of oxygen in 2019 based on CH and Mat\'ern multivariate covariances (b) Conditional standard deviations of oxygen in 2019 (c) Difference of conditional standard deviations of the CH and Mat\'ern multivariate covariances}
    \label{fig:argo_cond_exp}
\end{figure}

The estimated covariance functions are used to predict oxygen, temperature, and salinity on a regular $1/2$ degree grid in this area, and present results from oxygen here. 
The conditional expectation and variance are used to provide predictions and uncertainties.
In addition to the 436 profiles used in the training of the data, 8,264 additional nearby Argo Core profiles that only have temperature and salinity data are used. 
Ideally, the much more abundant Core data improves oxygen prediction using the estimated cross-covariance between the variables. 
For each year, oxygen is predicted with and without these additional Core data. 
The predictions for 2019 are provided in Fig. \ref{fig:argo_cond_exp}(a). 
The CH and Mat\'ern covariances give similar predictions, and including Core data increases the granularity of the predictions. 

The corresponding conditional standard deviations are given in Fig. \ref{fig:argo_cond_exp}(b). 
In general, the CH standard deviations appear smaller than the Mat\'ern ones, especially further away from any observed location. 
This matches results found in the univariate case in \cite{ma2022beyond}. 
When using Core data, the additional Core profiles considerably decrease the prediction standard deviations. 
On the other hand, the areas with the lowest standard deviations are near locations where BGC data was collected. 
The difference in conditional variances is plotted in Fig. \ref{fig:argo_cond_exp}(c), again confirming the lower conditional variances for the CH covariance.
As the BGC Argo program expands to a global array \citep{matsumoto2022global}, providing predictions and uncertainty estimates for both data-sparse and data-dense settings will be useful. 

To further evaluate performance, the prediction error is estimated using cross-validation in two settings: two-fold cross-validation by float, which represents a data-sparse situation, as well as leave-one-float-out cross-validation. 
Results are summarized in Table \ref{tab:argo_pred}.
As expected, prediction errors for both the CH and Mat\'ern covariance functions were less when using additional Core Argo data. 
In 2-fold cross-validation, the CH multivariate covariance performs similarly to the multivariate Mat\'ern when using Core data, but generally performs better than the Mat\'ern when using BGC data only. 
This reflects that the CH covariance will be most useful in data-sparse settings where the tails of the covariance function will matter more. 
In the leave-float-out setting, the CH and Mat\'ern covariances perform similarly when using BGC data only, while when using the BGC + Core data, the most data-dense setting, the Mat\'ern covariance performs slightly better than the CH covariance. 
Throughout the settings, the CH multivariate covariance provides generally shorter prediction intervals with comparable coverage to the Mat\'ern.

\begin{table}[t]
    \centering
    \begin{tabular}{|c|cccc|cccc|}
    \hline
          & \multicolumn{4}{c}{2-fold} & \multicolumn{4}{|c|}{Leave-float-out} \\ \hline
         Prediction & RMSE & MAE & Cov & I Len & RMSE & MAE & Cov & I Len \\ \hline \hline
         CH, BGC & 9.08 & 6.50 & 94.7 & 9.08 & 8.97 & 6.30 &94.5 & 8.65\\ 
         Mat\'ern, BGC & 9.40&  6.45 & 94.9 & 9.43 & 8.93 & 6.33 & 95.4 & 8.99 \\ 
         CH, BGC + core & 7.03& 4.20 & 92.7 & 6.53 & 7.03 & 4.26 & 92.7 & 6.34 \\ 
        Mat\'ern, BGC + core &6.93&  4.18 & 94.0 & 6.78 & 6.94 &  4.23  & 93.6 & 6.69\\ \hline
    \end{tabular}
    \caption{Oxygen prediction results, for root-mean-squared-error (RMSE), median absolute error (MAE), coverage of 95\% confidence intervals (Cov), and median interval length (I Len)}
    \label{tab:argo_pred}
\end{table}

\section{Discussion}\label{sec:disc}

There are some areas for future work. 
Extensions of the work here could include space-time models \citep[cf.][]{porcu202130, chen2021space} and nonstationary models, similarly to, for example, \citet{kleiber2012nonstationary}. 
If there are sudden nonstationarities in the data, one of the chief benefits of the multivariate CH covariance, its heavy tail decay, may instead degrade performance if a stationary model is assumed. 
In this setting, a mixture model, for example, \citet{bolin2019latent} may be appropriate. 
Alternately, a locally-stationary approach \citep{kuusela_locally_2018} also helps manage the big-$n$ challenge in spatial statistics, where likelihood evaluations and predictions have $O(n^3)$ computational cost and $O(n^2)$ memory storage. 
This work has not aimed to address the big-$n$ challenge directly, but the CH model is adaptable to work that does address it. 
This work has also not explored coherence \citep{kleiber2017coherence}, defined by the form: \begin{align*}
    \frac{f_{jk}(x; \mb{\theta})}{\left\{f_{jj}(x; \mb{\theta})f_{kk}(x; \mb{\theta})\right\}^{\frac{1}{2}}}.
\end{align*}
\cite{kleiber2017coherence} argues that flexible coherence in $x$ is an advantageous property of multivariate covariances, allowing the dependence strength to be different at different spectral frequencies. 
The spectral density's form in terms of 
$\mathcalprime{U}(\nu + d/2, 1 - \alpha + d/2, \cdot)$ makes interpretable and detailed expressions of the coherence challenging. 
However, numerical study suggests that, like for the multivariate Mat\'ern as analyzed in \cite{kleiber2017coherence}, the squared coherence may peak at $x\to 0$, $0 < x < \infty$, or $x \to \infty$. 
As with other multivariate covariances, considering highly-multivariate processes may make application of this model possible in more settings, similar to the problems considered by \citet{dey_graphical_2022} and \citet{mitchell_l_krock_modeling_2023}.

\section*{Acknowledgments}
Data were collected and made freely available by the Southern Ocean carbon and climate observations and modeling (SOCCOM) Project funded by the National Science Foundation, Division of Polar Programs, (NSF PLR-1425989 and OPP-1936222)
supplemented by NASA, and by the International Argo Program and the NOAA programs that contribute to it. (\url{http://www.argo.ucsd.edu}, \url{https://www.ocean-ops.org/board}). The Argo Program is part of the Global Ocean Observing System.

We would like to thank the reviewers of this work who provided helpful suggestions that improved the quality of the paper.

\section*{Supplementary Material}
\label{SM}
The supplementary material includes additional definitions and proofs, simulations, and data analysis results. 
Code is included at \url{https://github.com/dyarger/multivariate_confluent_hypergeometric}. 

\section*{Declarations}

\subsection*{Competing Interests} 

The authors have no competing interests in respect to this work.

\subsection*{Funding}

Bhadra was supported by U.S. National Science Foundation Grant DMS-2014371.

\subsection*{Availability of data and material}

The SOCCOM \citep{johnson_soccom_2020} and Argo \citep{argo_argo_2023} data are publicly available. 

\subsection*{Code availability}

All code used to present results of the paper is included at \url{https://github.com/dyarger/multivariate_confluent_hypergeometric}.

\newpage

\bibliographystyle{MG}       
{\footnotesize
\bibliography{{mch_rev2.bib}}}   

\end{document}


\title{Supplementary Material to ``Multivariate Confluent Hypergeometric Covariance Functions with Simultaneous Flexibility over Smoothness and Tail Decay''\footnote{{\it Mathematical Geosciences}}} 
\author{D. YARGER \and A. BHADRA\footnote{{Department of Statistics, Purdue University, 150 N. University St., West Lafayette, Indiana 47907-2066, U.S.A.
{anyarger@purdue.edu, bhadra@purdue.edu}}}}
\date{}


\maketitle







\setcounter{equation}{0}
\setcounter{page}{1}
\setcounter{table}{0}
\setcounter{section}{0}
\setcounter{figure}{0}

\numberwithin{table}{section}
\renewcommand{\theequation}{S.\arabic{equation}}
\renewcommand{\thesection}{S.\arabic{section}}
\renewcommand{\theproposition}{S.\arabic{proposition}}
\renewcommand{\thedefinition}{S.\arabic{definition}}
\renewcommand{\thesubsection}{S.\arabic{section}.\arabic{subsection}}
\renewcommand{\thepage}{S.\arabic{page}}
\renewcommand{\thetable}{S.\arabic{table}}
\renewcommand{\thefigure}{S.\arabic{figure}}

\renewcommand {\S}{\textsection} 

\section{Definitions and Proofs}\label{sec:additional_conditions}

\subsection{Conditionally negative semidefinite matrices}

\begin{definition}
A $p\times p$ matrix $\mb{A}= [a_{jk}]$ is said to be conditionally negative semidefinite, if, for any $p$-vector $x = (x_1, \dots, x_p)^\top $ such that $\sum_{j=1}^p x_j = 0$, then: \begin{align*}
    x^\top \mb{A} x &= \sum_{j=1}^p \sum_{k=1}^p x_jx_k a_{jk} \leq 0.
\end{align*}
\end{definition}
See \cite{emery2022new} for discussion and references relating to this definition, which they use to show validity conditions for the multivariate Mat\'ern model. 

Throughout the remainder of Section \ref{sec:additional_conditions}, let $\mb{\nu}$, $\mb{\alpha}$, $\mb{\beta}$, and $\mb{\sigma}$ be matrices of parameters, and take all matrix operations elementwise. 
In the following, we will most essentially use the relationship that $
    \textrm{exp}\left(-t\mb{A}\right) \textrm{ is positive semidefinite for all }t > 0$ if and only if $\mb{A} \textrm{ is conditionally negative semidefinite}$ \citep{berg1984harmonic, emery2022new}.

\subsection{Proofs}\label{sec:proof}

We first discuss the general necessary and sufficient condition for validity when all $\alpha_{jk} > d/2$ using the spectral density. 
This will be used in some of the later proofs. 
\begin{proposition}\label{prop:mult_spec_validity}
    Consider a multivariate covariance on $\mathbb{R}^d$ defined by $\mathcal{CH}(h; \mb{\nu}, \mb{\alpha}, \mb{\beta}, \mb{\sigma})$. Suppose that $\alpha_{jk} > d/2$ for all $j$ and $k$. 
    Then the multivariate model is valid if and only if: \begin{align*}
        \mb{\sigma} \frac{\Gamma(\mb{\nu}+ \frac{d}{2}) \mb{\beta}^d}{B(\mb{\alpha}, \mb{\nu})} \mathcal{U}\left(\mb{\nu} + \frac{d}{2}, 1 - \mb{\alpha} + \frac{d}{2}, \mb{\beta}^2\frac{u }{2}\right),
    \end{align*}is positive semidefinite for all $u > 0$.
\end{proposition}
\begin{proof}
    This follows from checking that the matrix-valued spectral density is positive semidefinite for all of its inputs. 
\end{proof}
Since the function $\mathcal{U}(\cdot, \cdot, \cdot)$ is relatively opaque, finding simpler necessary and sufficient conditions is likely not possible. 
However, for specific values of some parameters, this relationship may be used to establish valid models. 

We now discuss the proofs in order of presentation in the main paper. 

\begin{proof}[Proof of Proposition 1]
Throughout, we use the shorthand of $f_{\mathcal{CH}}(x) = f_{\mathcal{CH}}(x; \nu, \alpha, \beta ,\sigma)$.
    Since the covariance only depends on $h$ through its length $\lVert h \rVert$, one may use the inversion-type formulas on page 46 of \cite{stein1999interpolation} when $\alpha > d/2$: 
    \begin{align}
        f_{\mathcal{CH}}(x) &= (2\pi)^{-\frac{d}{2}} \int_0^\infty (w \lVert x\rVert )^{-\frac{d-2}{2}}J_{\frac{d-2}{2}}(w \lVert x\rVert) w^{d-1}  \mathcal{CH}(w; \nu, \alpha, \beta,  \sigma) dw\\
        &=\sigma \frac{\Gamma(\nu +\alpha)}{\Gamma(\nu)(2\pi)^{\frac{d}{2}}}\lVert x\rVert^{-\frac{d-2}{2}} \int_0^\infty J_{\frac{d-2}{2}}(w \lVert x\rVert) w^{\frac{d}{2}} \mathcal{U}\left(\alpha, 1- \nu, \frac{w^2}{2\beta^2}\right) dw.
    \end{align}
    Here, $J_{\kappa}(z)$ is the Bessel function of the first kind.
    Consider the change of variables $u = w^2/(2\beta^2)$ with $dw = \beta/(2u)^{1/2} du$, so that \begin{align*}
        f_{\mathcal{CH}}(x) 
        &=\sigma \frac{\Gamma(\nu +\alpha)}{\Gamma(\nu)(2\pi)^{\frac{d}{2}}}\lVert x\rVert^{-\frac{d-2}{2}} 2^{\frac{d}{4} - \frac{1}{2}}\beta^{\frac{d}{2} +1}\int_0^\infty  u^{\frac{d}{4} - \frac{1}{2}} \mathcal{U}\left(\alpha, 1- \nu, u\right) J_{\frac{d-2}{2}}\left(\beta(2 u)^{\frac{1}{2}}\lVert x\rVert\right) du.
    \end{align*}
    We now apply 13.10.15 of \cite{NIST:DLMF}, which states \begin{align*}
        \int_0^\infty u^{\frac{\mu}{2}} \mathcal{U}(a, b, u) J_\mu\left(2(yu)^{\frac{1}{2}}\right) dt = \frac{\Gamma(\mu - b  +2)}{\Gamma(a)} y^{\frac{1}{2}\mu} \mathcal{U}(\mu - b + 2, \mu - a + 2, y).
    \end{align*}
    If we set $\mu = (d-2)/2$, $a = \alpha$, $b = 1 - \nu$, and $y = \beta^2\lVert x\rVert^2/2$, the restrictions given in  13.10.15 of \cite{NIST:DLMF} are satisfied, and we have \begin{align*}
        f_{\mathcal{CH}}(x) 
        &=\sigma \frac{\Gamma(\nu +\alpha)}{\Gamma(\nu)(2\pi)^{\frac{d}{2}}}\lVert x\rVert^{-\frac{d-2}{2}} 2^{\frac{d}{4} - \frac{1}{2}}\beta^{\frac{d}{2} +1}\frac{\Gamma(\nu + \frac{d}{2} )}{\Gamma(\alpha)}\left(\frac{\beta\lVert x\rVert}{\sqrt{2}}\right)^{\frac{d-2}{2} } \mathcal{U}\left(\nu + \frac{d}{2}, 1 - \alpha  +\frac{d}{2},\frac{\beta^2\lVert x\rVert^2}{2}\right)\\
        &=\sigma \frac{\Gamma(\nu + \frac{d}{2})\beta^d}{B(\nu, \alpha)(2\pi)^{\frac{d}{2}}} \mathcal{U}\left(\nu + \frac{d}{2}, 1 - \alpha  +\frac{d}{2},\frac{\beta^2\lVert x\rVert^2}{2}\right),
    \end{align*}
    which is the desired result.
\end{proof}

\begin{proof}[Proof of Theorem 1]
    In this case, we begin with the valid parsimonious multivariate Mat\'ern, where $\nu_{jk} = (\nu_{j} + \nu_{k})/2$ and $\phi_{jk} = \phi$ for all $j$ and $k$: \begin{align*}
        \mb{\sigma} \frac{2^{1- \mb{\nu}}}{\Gamma(\mb{\nu})}\left(\frac{h}{\phi}\right)^{\mb{\nu}}\mathcal{K}_{\mb{\nu}}\left(\frac{h}{\phi}\right),
    \end{align*}which is known to be valid if the matrix $\mb{\sigma}\Gamma(\mb{\nu} + d/2)/\Gamma(\mb{\nu})$ is positive semidefinite \citep{gneiting2010matern}. 
    We then consider the multivariate covariance of:
    \begin{align*}
        \int_0^\infty \mb{\sigma} \frac{2^{1- \mb{\nu}}}{\Gamma(\mb{\nu})}\left(\frac{h}{\phi}\right)^{\mb{\nu}}\mathcal{K}_{\mb{\nu}}\left(\frac{h}{\phi}\right) \frac{\mb{\beta}^{2\mb{\alpha}}  }{2^{\mb{\alpha}}\Gamma(\mb{\alpha}) } \phi^{-2\mb{\alpha} - 2}  \textrm{exp}\left\{- \mb{\beta}^2/(2\phi^2)\right\} d\phi^2.
    \end{align*}
    This is a scale-mixture of the parsimonious multivariate Mat\'ern model, and it is valid if: \begin{align}\mb{\sigma}\frac{\Gamma(\mb{\nu} + \frac{d}{2})}{\Gamma(\mb{\nu})} \frac{ \mb{\beta}^{2\mb{\alpha}} }{\Gamma(\mb{\alpha}) } \times \left(2^{\frac{1}{2}}\phi\right)^{-2\mb{\alpha}} \phi^{-2} \textrm{exp}\left\{- \mb{\beta}^2/(2\phi^2)\right\}\label{eq:orig_mix_requirement},\end{align} is positive semidefinite for all $\phi^2 > 0$, combining the validity conditions of the parsimonious multivariate Mat\'ern and rearranging the term $\phi^{-2\mb{\alpha} - 2}/2^{\mb{\alpha}}$. 
    See \cite{emery2022new} and \cite{alegria2019covariance} for examples of constructing valid multivariate models for scale-mixtures of valid covariances. 
    Consider left and right multiplication of \eqref{eq:orig_mix_requirement} by the matrix \begin{align*}
        \textrm{diag}\left\{\left(2^{\frac{1}{2}}\phi\right)^{\alpha_j} \times \phi \times \textrm{exp}\left(\frac{\beta_{j}^2}{4\phi^2}\right), j=1 , \dots, p\right\},
    \end{align*}and instead of checking the positive semidefiniteness of \eqref{eq:orig_mix_requirement}, we may check the positive semidefiniteness of the resulting matrix, which has $j$, $k$ entry:\begin{align*}
        \sigma_{jk} \frac{\Gamma(\nu_{jk} + \frac{d}{2})}{\Gamma(\nu_{jk})} \frac{\beta_{jk}^{2\alpha_{jk}}}{\Gamma(\alpha_{jk})} \times \left(2^{\frac{1}{2}}\phi\right)^{-2\alpha_{jk} + \alpha_j + \alpha_k}\textrm{exp}\left\{-\frac{1}{2\phi^2}\left(\beta_{jk}^2  - \frac{(\beta_j^2 + \beta_k^2)}{2}\right)\right\}.
    \end{align*}Then, the choice of $\alpha_{jk} = (\alpha_{j} + \alpha_{k})/2$ and $\beta_{jk}^2 = (\beta_{j}^2 + \beta_{k}^2)/2$ leads to the terms to the right of $\times$ dropping out, resulting in a matrix with $j$, $k$ entry: \begin{align*}
        \sigma_{jk} \frac{\Gamma(\nu_{jk} + \frac{d}{2})}{\Gamma(\nu_{jk})} \frac{\beta_{jk}^{2\alpha_{jk}}}{\Gamma(\alpha_{jk})},
    \end{align*} giving us the required condition. 
\end{proof}

\begin{proof}[Proof of Theorem 2]
The general approach for construction of this model is to combine Example 2 of \cite{emery2022new} with mixing with respect to a matrix composed of inverse-gamma densities. 
In \cite{emery2022new}, they establish that the multivariate Mat\'ern model: \begin{align*}
    \mb{\sigma} \left(\frac{\lvert h\rvert}{\phi}\right)^{\mb{\nu}}\mathcal{K}_{\mb{\nu}}\left(\frac{\lvert h \rvert}{\phi}\right),
\end{align*}is valid if $\mb{\nu}$ is conditionally negative semidefinite and 
    $\mb{\sigma}\Gamma(\mb{\nu})^{-1} \mb{\nu}^{\nu + \frac{d}{2}}\textrm{exp}(-\mb{\nu})$ is positive semidefinite.
We consider the mixture: \begin{align*}
    \int_0^\infty   \mb{\sigma} \left(\frac{\lvert h\rvert}{\phi}\right)^{\mb{\nu}}\mathcal{K}_{\mb{\nu}}\left(\frac{\lvert h\rvert}{\phi}\right) \frac{\mb{\beta}^{2\mb{\alpha}}}{2^{\mb{\alpha}}\Gamma(\mb{\alpha})} \phi^{-2\mb{\alpha}- 2} \textrm{exp}\left(- \frac{\mb{\beta}^2}{2\phi^2}\right)d\phi^2.
\end{align*}
We are interested when the integrand is positive semidefinite for any $\phi^2 > 0$.
Due to the choice of $\mb{\alpha}$, $\phi^{-2\mb{\alpha}-2}2^{-\mb{\alpha}}$ can be removed due to the Schur product theorem or by left and right multiplication by the diagonal matrix of $2^{\alpha_{j}/2} \phi^{\alpha_j - 1}$ ($j=1, \dots, p$). 
We are then left to check that $\mb{\nu}$ is conditionally negative semidefinite, and, \begin{align*}
        \frac{\mb{\sigma}}{\Gamma(\mb{\nu})} \mb{\nu}^{\nu + \frac{d}{2}}\textrm{exp}(-\mb{\nu}) \frac{\mb{\beta}^{2\mb{\alpha}}}{\Gamma(\mb{\alpha})}  \textrm{exp}\left(- \frac{\mb{\beta}^2}{2\phi^2}\right),
\end{align*}is positive semidefinite for any $\phi > 0$. 
Under the assumption that $\mb{\beta}^2$ is real, symmetric, and conditionally negative semidefinite, the matrix $\textrm{exp}(-t\mb{\beta}^2)$ is positive semidefinite for all $t  > 0$, following Lemma 1 of \cite{emery2022new}.
This is equivalent to the matrix $\textrm{exp}(\{-\mb{\beta}^2/(2\phi^2)\}$ is positive semidefinite for all $\phi^2  > 0$ by taking $t = 1/(2\phi^2)$.
By the Schur product theorem, the conditions imply that the multivariate model is valid. 
\end{proof}

\begin{proof}[Proof of Proposition 2]
    We consider the conditions under which the $2$ by $2$ matrix of  \begin{align*}
        f_{\mathcal{CH}}\left(x;\frac{\nu_j + \nu_k}{2}, \alpha_{jk}, \beta, \sigma_{jk}\right),
    \end{align*}is positive semidefinite for all $x$. 
    Here, let $\beta \in \mathbb{R}$ be the range parameter shared by the processes. 
That is, we examine the matrix of: \begin{align*}
    \sigma_{jk}  \frac{\Gamma(\frac{\nu_j+\nu_k}{2} + \frac{d}{2})\beta^{d}}{(2\pi)^{d/2}B(\alpha_{jk}, \frac{\nu_j + \nu_k}{2}) } \mathcal{U}\left(\frac{\nu_j + \nu_k}{2} + \frac{d}{2} , 1 - \alpha_{jk} + \frac{d}{2}, \frac{\beta^2\lVert x\rVert^2}{2}\right) .
\end{align*}
By left and right multiplying by the matrix: \begin{align*}
    \textrm{diag}\left[ \frac{(2\pi)^{\frac{d}{4}}}{\left\{\beta^{d}\Gamma(\nu_j + \frac{d}{2})\mathcal{U}\left(\nu_j + \frac{d}{2} , 1 - \alpha_j + \frac{d}{2}, \frac{\beta^2\lVert x\rVert^2}{2}\right)\right\}^{\frac{1}{2}}}, j=1, \dots, p\right],
\end{align*}we see that this is equivalent to the checking the positive semidefiniteness of the matrix: \begin{align*}
        \sigma_{jk}  \begin{cases}
     \frac{1}{B(\alpha_j, \nu_j)},  & j = k,    \\
     \frac{1}{B(\alpha_{jk}, \frac{\nu_j + \nu_k}{2}) } \frac{\Gamma(\frac{\nu_j+\nu_k}{2} + \frac{d}{2})\mathcal{U}\left(\frac{\nu_j + \nu_k}{2} + \frac{d}{2} , 1 - \alpha_{jk}+ \frac{d}{2}, \frac{\beta^2\lVert x\rVert^2}{2}\right) }{\left\{\Gamma(\nu_j + \frac{d}{2})\mathcal{U}\left(\nu_j + \frac{d}{2} , 1 - \alpha_j + \frac{d}{2}, \frac{\beta^2\lVert x\rVert^2}{2}\right)\Gamma(\nu_k + \frac{d}{2})\mathcal{U}\left(\nu_k + \frac{d}{2} , 1 - \alpha_k + \frac{d}{2}, \frac{\beta^2\lVert x\rVert^2}{2}\right)\right\}^{\frac{1}{2}}},& j \neq k,
     \end{cases}
\end{align*}for each $x$. 
Theorem 2.3 of \cite{bhukya2018some} establishes that:\begin{align*}
\lvert\mathcal{U}(a,c_1,x)\rvert \leq \lvert\mathcal{U}(a, c, x)\rvert
\end{align*} for $c_1 < c$, $a > 0$, $x > 0$.
Applying this to the cross-spectral density, we have:
\begin{align*}
    \left\lvert\mathcal{U}\left(\frac{\nu_j+\nu_k}{2} + \frac{d}{2} , 1 - \alpha_{jk} + \frac{d}{2}, \frac{\beta^2\lVert x\rVert^2}{2}\right)\right\rvert \leq \left\lvert \mathcal{U}\left(\frac{\nu_j+\nu_k}{2} + \frac{d}{2} , 1 - \frac{\alpha_j + \alpha_k}{2}+ \frac{d}{2}, \frac{\beta^2\lVert x\rVert^2}{2}\right) \right\rvert,
\end{align*}where $\alpha_{jk} \geq (\alpha_j + \alpha_k)/2$ by assumption. 
From the representation (13.4.4) of \cite{NIST:DLMF}, one may remove the absolute value signs above, as both expressions are positive. 
Therefore, we may check if the $2\times 2$ matrix \begin{align*}
    \sigma_{jk}  \begin{cases}
     \frac{1}{B(\alpha_j, \nu_j)},  & j = k,    \\
     \frac{1}{B(\alpha_{jk}, \frac{\nu_j + \nu_k}{2}) } \frac{\Gamma(\frac{\nu_j+\nu_k}{2} + \frac{d}{2})\mathcal{U}\left(\frac{\nu_j + \nu_k}{2} + \frac{d}{2} , 1 - \frac{\alpha_j +\alpha_k}{2}+ \frac{d}{2}, \frac{\beta^2\lVert x\rVert^2}{2}\right) }{\left\{\Gamma(\nu_j + \frac{d}{2})\mathcal{U}\left(\nu_j + \frac{d}{2} , 1 - \alpha_j + \frac{d}{2}, \frac{\beta^2\lVert x\rVert^2}{2}\right)\Gamma(\nu_k + \frac{d}{2})\mathcal{U}\left(\nu_k + \frac{d}{2} , 1 - \alpha_k + \frac{d}{2}, \frac{\beta^2\lVert x\rVert^2}{2}\right)\right\}^{\frac{1}{2}}},& j \neq k,
     \end{cases},
\end{align*}is positive definite for each $x$.

\cite{baricz2013turan} establish in Remark 3 that the function $(a,c) \to \Gamma(a)\mathcal{U}(a,c,z)$ is logarithmically convex for $a > 0$, $z> 0$, and $c \in \mathbb{R}$, which gives that  \begin{align*}
\Gamma\left(\frac{a_1 + a_2}{2}\right)\mathcal{U}\left(\frac{a_1 + a_2}{2}, \frac{c_1 + c_2}{2}, z\right) \leq \left\{\Gamma\left(a_1\right)\mathcal{U}\left(a_1, c_1, z\right)\Gamma\left(a_2\right)\mathcal{U}\left(a_2, c_2, z\right)\right\}^{\frac{1}{2}}.
\end{align*}when $z > 0$ and $a_1, a_2 > 0$. Therefore, by the Schur product theorem, we need only check that the matrix with entries
$
            \sigma_{jk} 
     B(\alpha_{jk}, (\nu_j + \nu_k)/2)^{-1}$ is positive semidefinite, as desired.                  
\end{proof}

\section{Additional Data Analysis Plots}

\begin{figure}[ht]
    \centering
    \includegraphics[width = .6\textwidth]{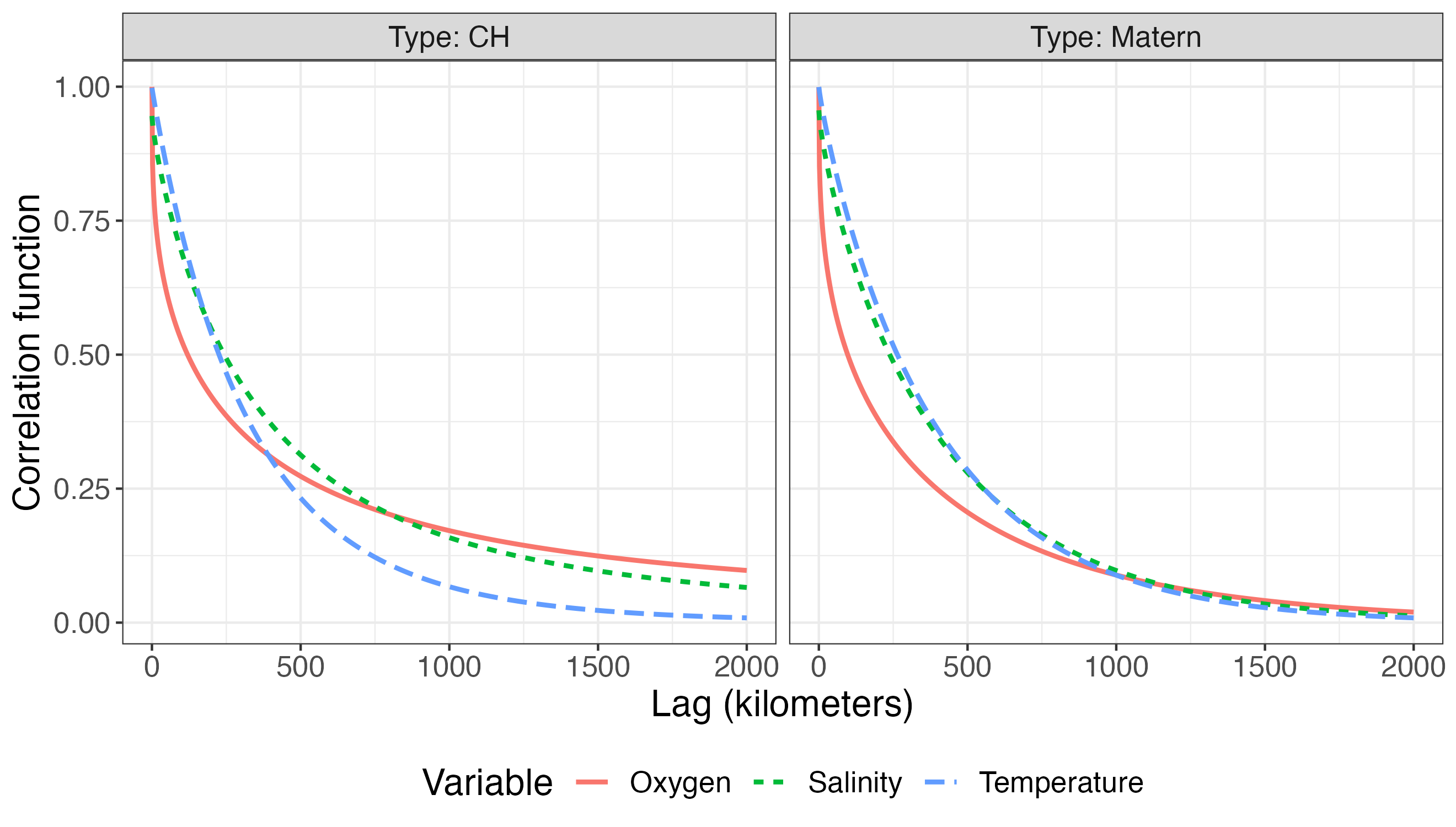}
    
    \includegraphics[width = .6\textwidth]{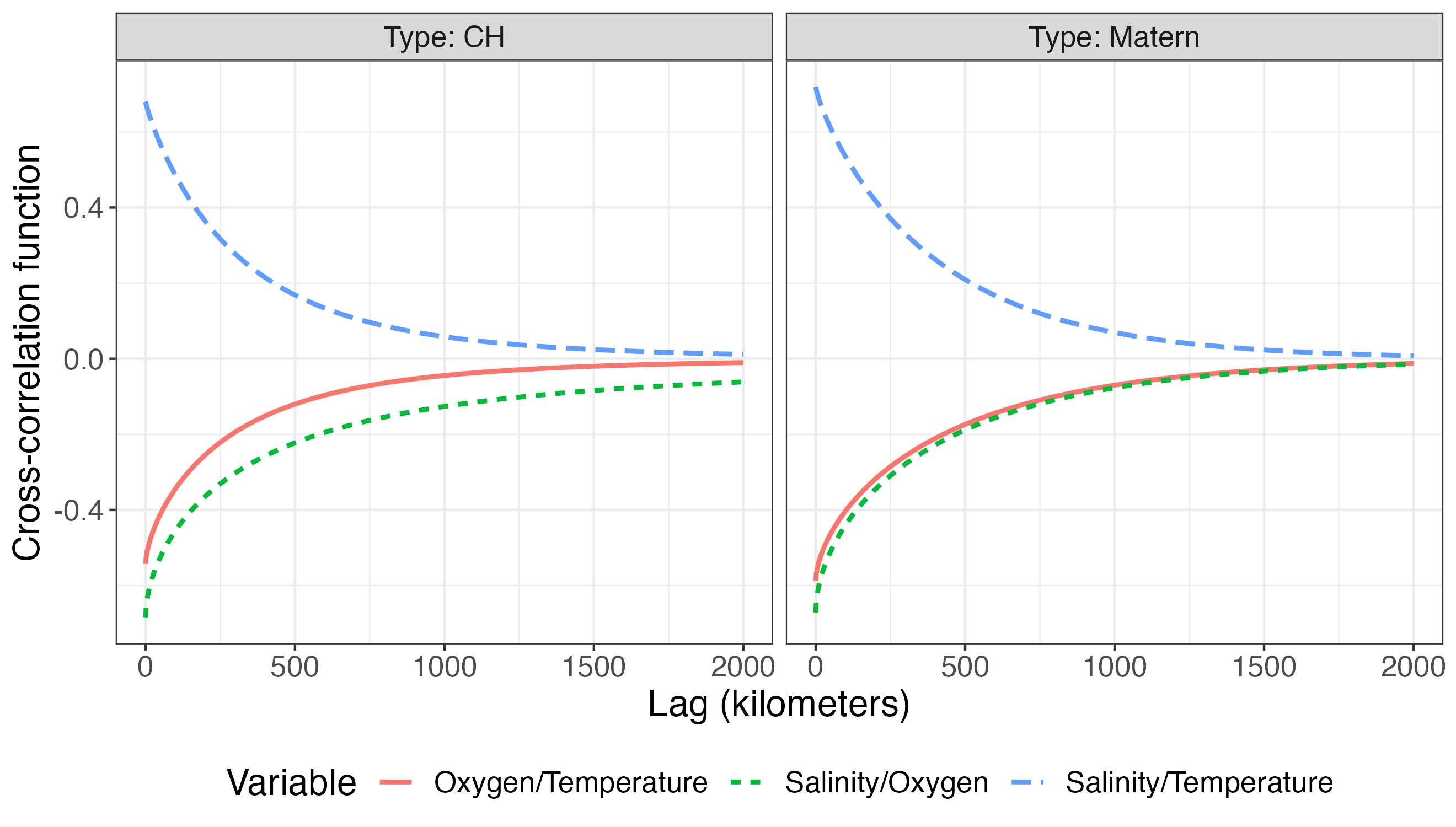}
    \caption{Estimated correlation (Top) and cross-correlation (Bottom) functions for the Argo data based on the CH and Mat\'ern cross-covariances. Note that the colors and type of line represent different things in the two panels. }
    \label{fig:argo_est_cov}
\end{figure}

In Figure \ref{fig:argo_est_cov}, we plot the estimated correlation and cross-correlation functions. 
While the Mat\'ern and CH functions behave similarly near the origin, they behave quite differently in their tails.

\section{Additional simulation results}\label{app:simulation}

\subsection{Using other variable's data at prediction sites}

\begin{figure}[ht]
    \centering
     \includegraphics[width = .85\textwidth]{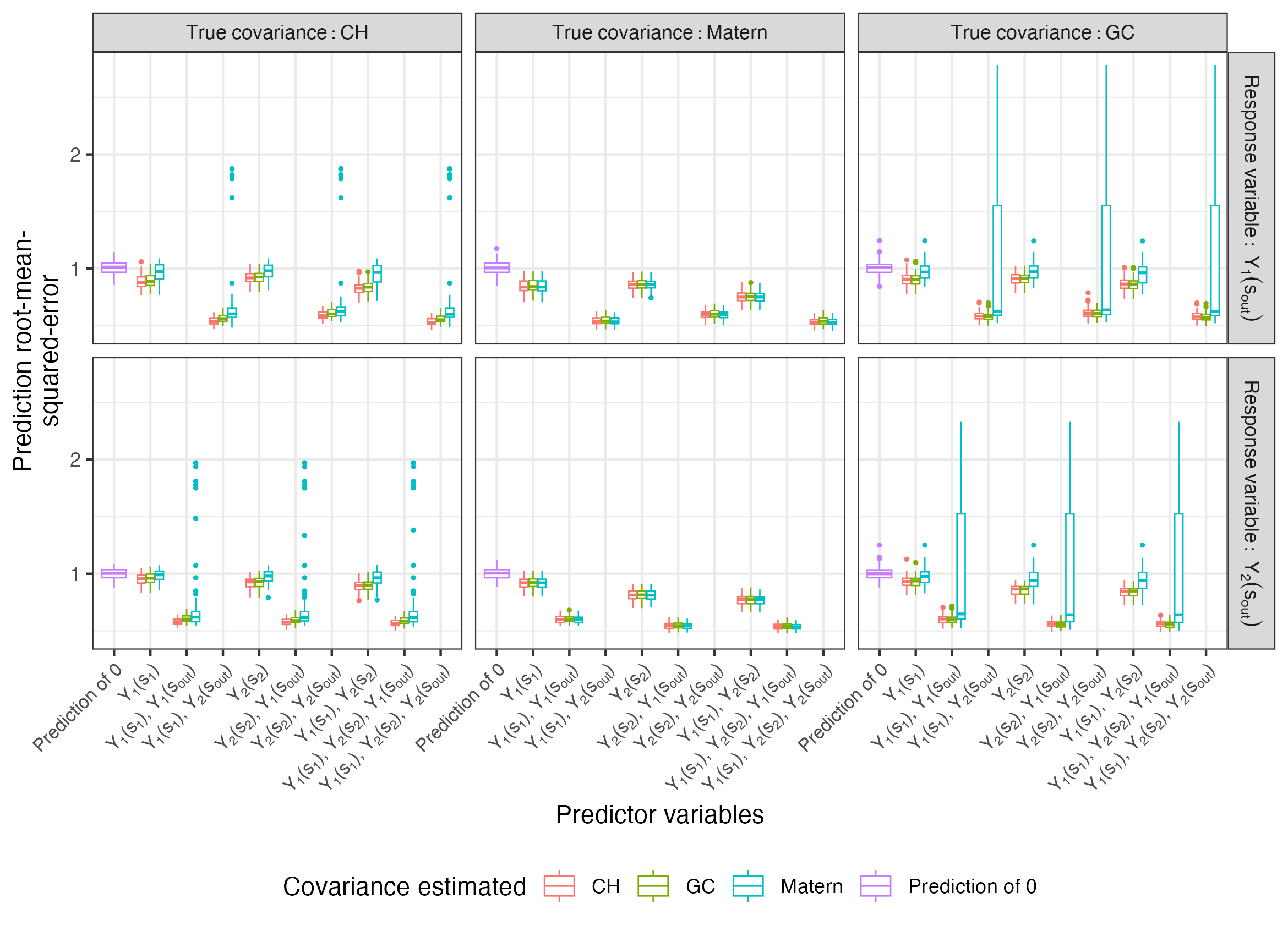}
     \includegraphics[width = .85\textwidth]{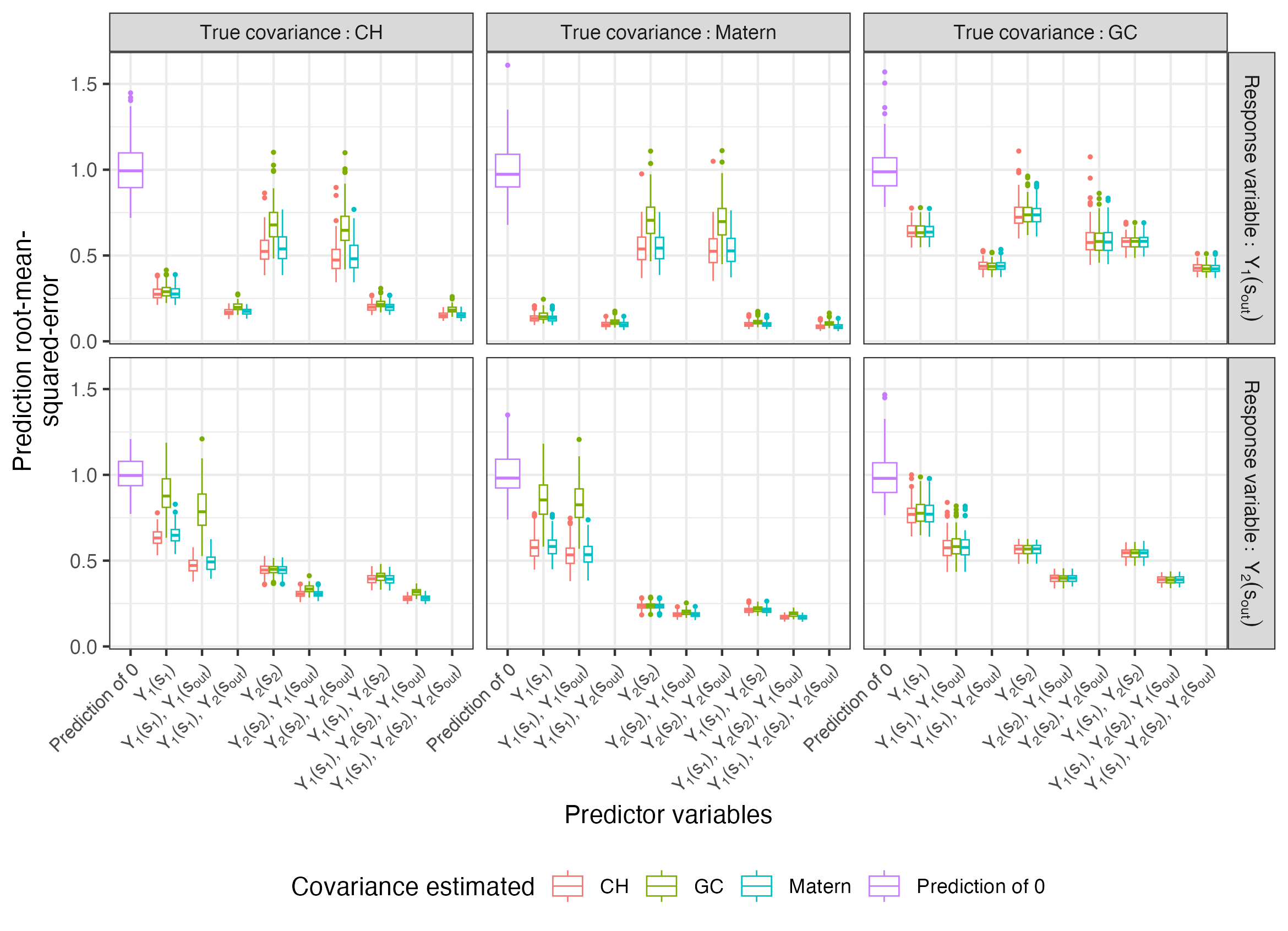}

    \caption{Simulation prediction error results for setting A (Top) and setting B (Bottom), under a variety of conditioning/predictor combinations.}
    \label{fig:sim_w_pred}
\end{figure}

In Figure \ref{fig:sim_w_pred}, we present prediction results both when we use and do not use the other variable for prediction at the prediction sites $\mb{s}_{out}$. 
For setting A, using the other variable at $\mb{s}_{out}$ results in improved predictions when using the CH or generalized Cauchy covariance. 
However, in this setting the multivariate Mat\'ern has much worse performance for some simulations; in these cases, a negative value of $\sigma_{12}$ was estimated for the multivariate Mat\'ern model.
This leads to unreliable predictions when using the other variable at the same site. 
However, when the true covariance in Mat\'ern, using the other variable at $\mb{s}_{out}$ tends to perform better for both the CH, Mat\'ern, and generalized Cauchy covariances estimated. 
Finally, in the setting where the true covariance is generalized Cauchy, the results are similar to the case where the true covariance is CH, with the Mat\'ern model giving unstable predictions when using data at the same site to predict. 
For setting B, the GC once again lags behind when the true covariance is CH or Mat\'ern.

We similarly provide interval summaries in Tables \ref{tab:ch_interval_w_pred} and \ref{tab:matern_interval_w_pred}, for simplicity using the other variable at observed locations and the other variable at observed locations and prediction locations. 
We again see reduced coverage for the Mat\'ern model under polynomial tail decay of the multivariate covariance in setting A in Table \ref{tab:ch_interval_w_pred}. 
We see that when using the other variable at the predicted locations, the intervals are shorter and do not have as high of coverage. 
Without observing any colocated data, each model may over-optimistically drive down conditional variances when data for the other variable is available at that site.
We next consider the colocated data setting. 

\begin{table}[ht]
    \centering
    \begin{tabular}{ccccccccc} 
    \multirow{2}*{\begin{tabular}{c}True\\ covariance\end{tabular}}& \multirow{2}*{\begin{tabular}{c}Predictor\\ variables\end{tabular}}&\multirow{2}*{\begin{tabular}{c}Response\\ variable\end{tabular}}&\multicolumn{3}{c}{95\% CI coverage} & \multicolumn{3}{c}{Average length} \\ 
         &&  & CH & M & GC  & CH & M& GC  \\ 
        CH &$Y_2(\mb{s}_2)$ & $Y_1(\mb{s}_{out})$ & 94.9 & 74.9 & 94.5 & 3.59 & 2.49 & 3.53\\ 
        CH&$Y_2(\mb{s}_2)$, $Y_2(\mb{s}_{out})$ & $Y_1(\mb{s}_{out})$ & 92.1 & 49.9 & 88.0 & 2.23 & 1.13 & 2.02\\ 
        CH&$Y_1(\mb{s}_1)$ & $Y_2(\mb{s}_{out})$ &94.8 & 76.7 & 95.2 & 3.72 & 2.59 & 3.80 \\ 
        CH&$Y_1(\mb{s}_1)$, $Y_1(\mb{s}_{out})$ & $Y_2(\mb{s}_{out})$ & 91.3 & 49.8 & 88.9 & 2.16 & 1.14 & 2.10 \\ 
        Mat\'ern &$Y_2(\mb{s}_2)$ & $Y_1(\mb{s}_{out})$ &94.8 & 94.8 & 94.4 & 3.35 & 3.35 & 3.31 \\ 
        Mat\'ern&$Y_2(\mb{s}_2)$, $Y_2(\mb{s}_{out})$ & $Y_1(\mb{s}_{out})$ &93.3 & 93.6 & 91.2 & 2.27 & 2.28 & 2.11\\ 
        Mat\'ern&$Y_1(\mb{s}_1)$ & $Y_2(\mb{s}_{out})$ &94.8 & 94.9 & 95.2 & 3.57 & 3.57 & 3.65 \\ 
        Mat\'ern&$Y_1(\mb{s}_1)$, $Y_1(\mb{s}_{out})$ & $Y_2(\mb{s}_{out})$ &93.0 & 93.5 & 91.8 & 2.25 & 2.28 & 2.18\\ 
        GC &$Y_2(\mb{s}_2)$ & $Y_1(\mb{s}_{out})$ &  94.9 & 74.1 & 95.0 & 3.58 & 2.42 & 3.57\\ 
        GC &$Y_2(\mb{s}_2)$, $Y_2(\mb{s}_{out})$ & $Y_1(\mb{s}_{out})$ &  90.1 & 45.4 & 90.1 & 2.23 & 1.11 & 2.19\\ 
        GC &$Y_1(\mb{s}_1)$ & $Y_2(\mb{s}_{out})$ &94.9 & 76.1 & 94.9 & 3.64 & 2.51 & 3.65\\ 
        GC &$Y_1(\mb{s}_1)$, $Y_1(\mb{s}_{out})$ & $Y_2(\mb{s}_{out})$ & 90.1 & 45.6 & 90.3 & 2.21 & 1.11 & 2.18 \\ 
    \end{tabular}
    \caption{Prediction interval summaries for setting A. We compare estimated CH, Mat\'ern (M), and generalized Cauchy (GC) multivariate covariances used for prediction. }
    \label{tab:ch_interval_w_pred}
\end{table}

\begin{table}[ht]
    \centering
    \begin{tabular}{ccccccccc} 
    \multirow{2}*{\begin{tabular}{c}True\\ covariance\end{tabular}}& \multirow{2}*{\begin{tabular}{c}Predictor\\ variables\end{tabular}}&\multirow{2}*{\begin{tabular}{c}Response\\ variable\end{tabular}}&\multicolumn{3}{c}{95\% CI coverage} & \multicolumn{3}{c}{Average length} \\ 
         &&  & CH & M & GC  & CH & M& GC  \\ 
        CH &$Y_2(\mb{s}_2)$ & $Y_1(\mb{s}_{out})$ & 95.6 & 92.8 & 84.1 & 2.29 & 2.08 & 2.06\\ 
        CH&$Y_2(\mb{s}_2)$, $Y_2(\mb{s}_{out})$ & $Y_1(\mb{s}_{out})$ & 95.6 & 92.3 & 82.7 & 1.99 & 1.77 & 1.87\\ 
        CH &$Y_1(\mb{s}_1)$ & $Y_2(\mb{s}_{out})$ & 95.2 & 94.6 & 95.4 & 2.48 & 2.48 & 3.64 \\ 
        CH&$Y_1(\mb{s}_1)$, $Y_1(\mb{s}_{out})$ & $Y_2(\mb{s}_{out})$ & 95.0 & 94.6 & 94.5 & 1.86 & 1.91 & 3.20\\ 
        Mat\'ern &$Y_2(\mb{s}_2)$ & $Y_1(\mb{s}_{out})$ & 96.4 & 95.2 & 81.6 & 2.33 & 2.22 & 1.98 \\ 
        Mat\'ern&$Y_2(\mb{s}_2)$, $Y_2(\mb{s}_{out})$ & $Y_1(\mb{s}_{out})$ & 96.4 & 95.2 & 81.2 & 2.25 & 2.14 & 1.94\\ 
        Mat\'ern&$Y_1(\mb{s}_1)$ & $Y_2(\mb{s}_{out})$ & 95.6 & 95.4 & 97.5 & 2.31 & 2.30 & 3.88 \\ 
        Mat\'ern&$Y_1(\mb{s}_1)$, $Y_1(\mb{s}_{out})$ & $Y_2(\mb{s}_{out})$ & 95.4 & 95.2 & 97.3 & 2.14 & 2.14 & 3.77\\ 
        GC &$Y_2(\mb{s}_2)$ & $Y_1(\mb{s}_{out})$ & 95.1 & 94.1 & 94.9 & 2.95 & 2.84 & 2.93\\ 
        GC &$Y_2(\mb{s}_2)$, $Y_2(\mb{s}_{out})$ & $Y_1(\mb{s}_{out})$ &94.2 & 92.9 & 94.2 & 2.32 & 2.18 & 2.30\\ 
        GC &$Y_1(\mb{s}_1)$ & $Y_2(\mb{s}_{out})$ & 95.2 & 94.4 & 95.0 & 3.04 & 2.98 & 3.06 \\ 
        GC&$Y_1(\mb{s}_1)$, $Y_1(\mb{s}_{out})$ & $Y_2(\mb{s}_{out})$ & 94.0 & 92.8 & 94.2 & 2.25 & 2.16 & 2.29 \\ 
    \end{tabular}
    \caption{Prediction interval summaries for setting B. We compare estimated CH, Mat\'ern (M), and generalized Cauchy (GC) multivariate covariances used for prediction. }
    \label{tab:matern_interval_w_pred}
\end{table}

\subsection{Colocated design}

\begin{figure}[ht]
    \centering
    \includegraphics[width = .48\textwidth]{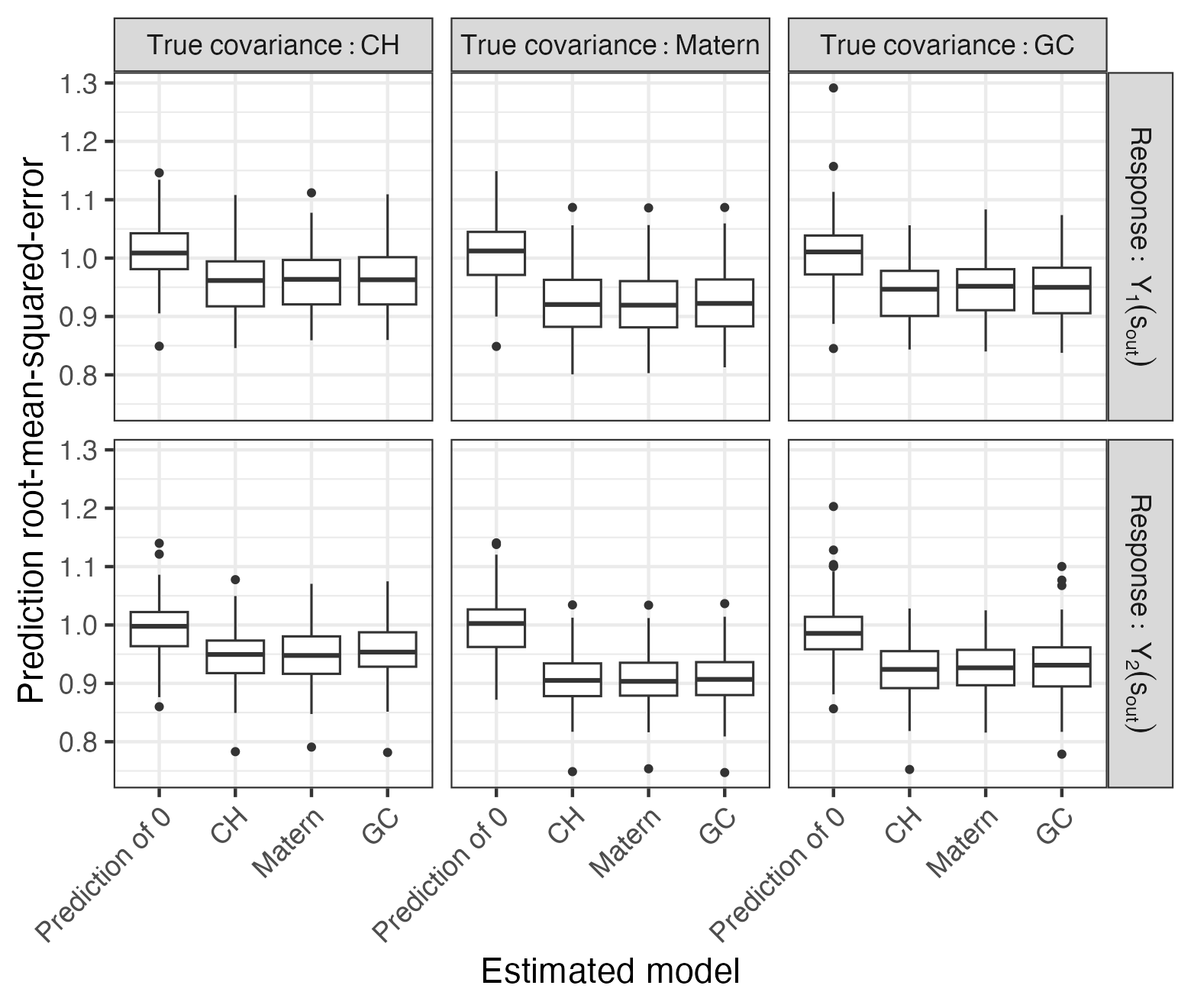}
    \includegraphics[width = .48\textwidth]{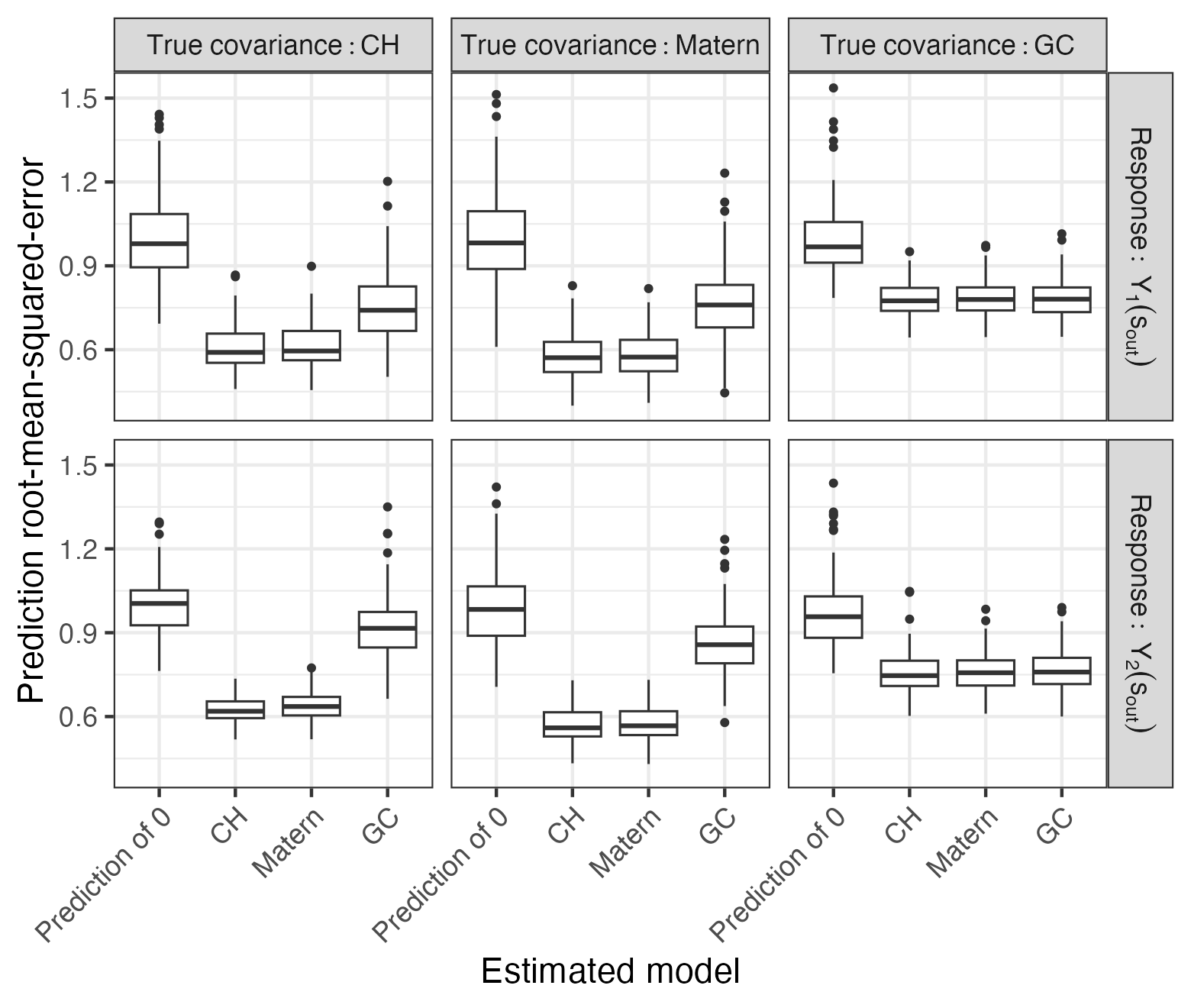}
    \caption{Simulation prediction error results for (Left) setting A and (Right) setting B under the colocated design. 
    The boxplots are formed from the prediction RMSE of 100 simulations.}
   \label{fig:colocated}
\end{figure}

\begin{figure}[ht]
    \centering
    \includegraphics[width = .85\textwidth]{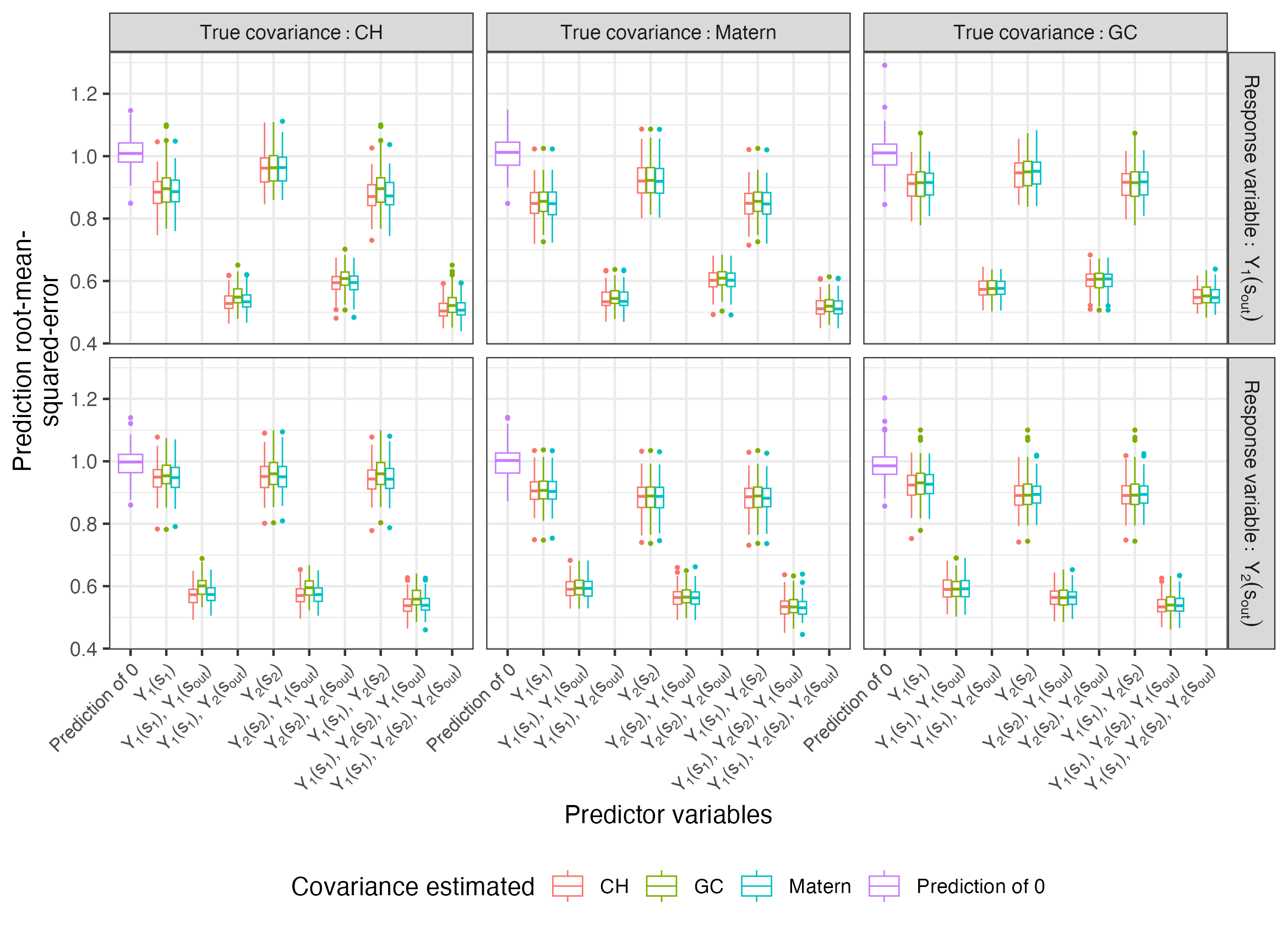}
    
    \includegraphics[width = .85\textwidth]{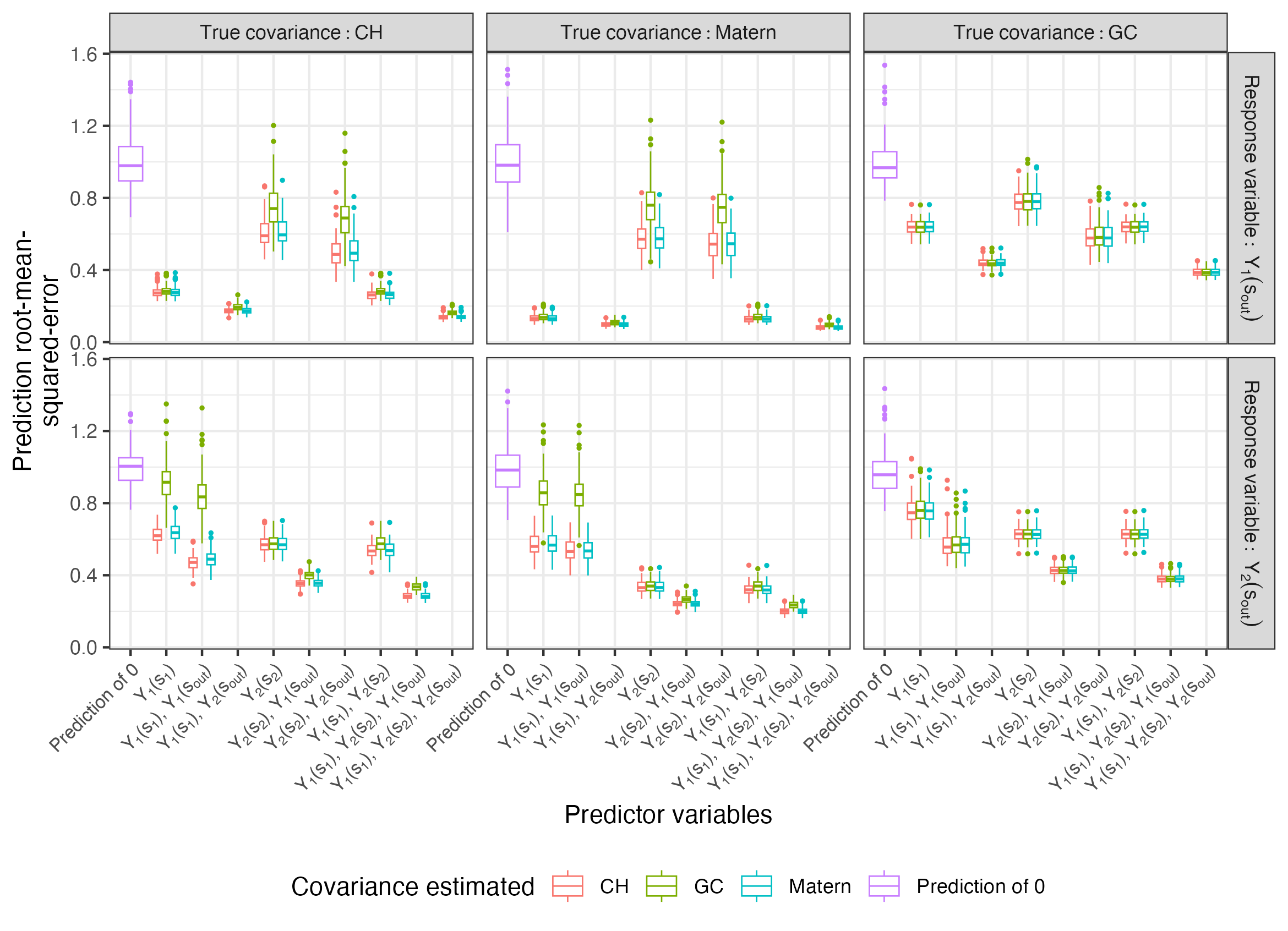}
    \caption{Simulation prediction error results for (Top) setting A and (Bottom) setting B under the colocated design and  a variety of conditioning/predictor combinations. }
   \label{fig:colocated_predvars}
\end{figure}

In this Subsection, we evaluate the setting where $\mb{s}_1 = \mb{s}_2$ instead of having the two variables measured at entirely distinct locations. 
We assume we have $n_1 = n_2 = 200$ locations. 
Point prediction results are summarized in Figures \ref{fig:colocated} and \ref{fig:colocated_predvars}. 
For the most part, the multivariate covariance types perform similarly to each other in setting A, while the generalized Cauchy model still has trouble in setting B. 
Having both variables available at each of the observed data locations means the covariance $\sigma_{12}$ should be easier to estimate. 
We summarize the interval performance in Tables \ref{tab:ch_interval_w_pred_colocated} and \ref{tab:matern_interval_w_pred_colocated}..
Compared to the non-colocated design, interval coverages are better when using the other variable at left out locations.

\begin{table}[ht]
    \centering
    \begin{tabular}{ccccccccc} 
    \multirow{2}*{\begin{tabular}{c}True\\ covariance\end{tabular}}& \multirow{2}*{\begin{tabular}{c}Predictor\\ variables\end{tabular}}&\multirow{2}*{\begin{tabular}{c}Response\\ variable\end{tabular}}&\multicolumn{3}{c}{95\% CI coverage} & \multicolumn{3}{c}{Average length} \\ 
         &&  & CH & M & GC  & CH & M& GC  \\ 
        CH &$Y_2(\mb{s}_2)$ & $Y_1(\mb{s}_{out})$ & 94.9 & 94.6 & 94.1 & 3.74 & 3.72 & 3.65\\ 
        CH&$Y_2(\mb{s}_2)$, $Y_2(\mb{s}_{out})$ & $Y_1(\mb{s}_{out})$ & 94.5 & 94.0 & 93.7 & 2.28 & 2.25 & 2.27\\ 
        CH&$Y_1(\mb{s}_1)$ & $Y_2(\mb{s}_{out})$ &  95.0 & 94.8 & 95.4 & 3.74 & 3.72 & 3.87\\ 
        CH&$Y_1(\mb{s}_1)$, $Y_1(\mb{s}_{out})$ & $Y_2(\mb{s}_{out})$ &94.9 & 94.9 & 95.4 & 2.24 & 2.25 & 2.40 \\ 
        Mat\'ern &$Y_2(\mb{s}_2)$ & $Y_1(\mb{s}_{out})$ &94.9 & 94.9 & 94.6 & 3.60 & 3.60 & 3.55 \\ 
        Mat\'ern&$Y_2(\mb{s}_2)$, $Y_2(\mb{s}_{out})$ & $Y_1(\mb{s}_{out})$ &94.6 & 94.5 & 94.3 & 2.33 & 2.33 & 2.33\\ 
        Mat\'ern&$Y_1(\mb{s}_1)$ & $Y_2(\mb{s}_{out})$ & 95.1 & 95.2 & 95.8 & 3.60 & 3.60 & 3.70 \\ 
        Mat\'ern&$Y_1(\mb{s}_1)$, $Y_1(\mb{s}_{out})$ & $Y_2(\mb{s}_{out})$ &95.0 & 95.0 & 95.6 & 2.32 & 2.33 & 2.42\\ 
        GC &$Y_2(\mb{s}_2)$ & $Y_1(\mb{s}_{out})$ & 94.8 & 94.6 & 94.8 & 3.69 & 3.67 & 3.71\\ 
        GC &$Y_2(\mb{s}_2)$, $Y_2(\mb{s}_{out})$ & $Y_1(\mb{s}_{out})$ & 94.4 & 94.0 & 94.6 & 2.32 & 2.30 & 2.34\\ 
        GC &$Y_1(\mb{s}_1)$ & $Y_2(\mb{s}_{out})$ &95.2 & 95.0 & 95.1 & 3.69 & 3.67 & 3.71\\ 
        GC &$Y_1(\mb{s}_1)$, $Y_1(\mb{s}_{out})$ & $Y_2(\mb{s}_{out})$ &  95.0 & 94.5 & 95.1 & 2.33 & 2.30 & 2.34 \\ 
    \end{tabular}
    \caption{Prediction interval summaries for setting A and the colocated design. We compare estimated CH, Mat\'ern (M), and generalized Cauchy (GC) multivariate covariances used for prediction. }
    \label{tab:ch_interval_w_pred_colocated}
\end{table}

\begin{table}[ht]
    \centering
    \begin{tabular}{ccccccccc} 
    \multirow{2}*{\begin{tabular}{c}True\\ covariance\end{tabular}}& \multirow{2}*{\begin{tabular}{c}Predictor\\ variables\end{tabular}}&\multirow{2}*{\begin{tabular}{c}Response\\ variable\end{tabular}}&\multicolumn{3}{c}{95\% CI coverage} & \multicolumn{3}{c}{Average length} \\ 
         &&  & CH & M & GC  & CH & M& GC  \\ 
        CH &$Y_2(\mb{s}_2)$ & $Y_1(\mb{s}_{out})$ &  95.4 & 93.0 & 83.5 & 2.40 & 2.22 & 2.10\\ 
        CH&$Y_2(\mb{s}_2)$, $Y_2(\mb{s}_{out})$ & $Y_1(\mb{s}_{out})$ & 95.4 & 92.5 & 82.2 & 1.99 & 1.80 & 1.89\\ 
        CH &$Y_1(\mb{s}_1)$ & $Y_2(\mb{s}_{out})$ & 95.4 & 95.0 & 95.9 & 2.49 & 2.49 & 3.76 \\ 
        CH&$Y_1(\mb{s}_1)$, $Y_1(\mb{s}_{out})$ & $Y_2(\mb{s}_{out})$ & 95.2 & 94.8 & 95.4 & 1.87 & 1.92 & 3.37\\ 
        Mat\'ern &$Y_2(\mb{s}_2)$ & $Y_1(\mb{s}_{out})$ & 96.1 & 95.0 & 80.5 & 2.37 & 2.26 & 2.01 \\ 
        Mat\'ern&$Y_2(\mb{s}_2)$, $Y_2(\mb{s}_{out})$ & $Y_1(\mb{s}_{out})$ &  96.1 & 95.1 & 80.2 & 2.26 & 2.14 & 1.96\\ 
        Mat\'ern&$Y_1(\mb{s}_1)$ & $Y_2(\mb{s}_{out})$ &  95.5 & 95.6 & 98.0 & 2.30 & 2.31 & 4.05 \\ 
        Mat\'ern&$Y_1(\mb{s}_1)$, $Y_1(\mb{s}_{out})$ & $Y_2(\mb{s}_{out})$ & 95.3 & 95.1 & 98.0 & 2.13 & 2.15 & 3.95\\ 
        GC &$Y_2(\mb{s}_2)$ & $Y_1(\mb{s}_{out})$ &  95.1 & 94.4 & 94.9 & 3.10 & 3.02 & 3.09\\ 
        GC &$Y_2(\mb{s}_2)$, $Y_2(\mb{s}_{out})$ & $Y_1(\mb{s}_{out})$ &95.2 & 94.3 & 94.9 & 2.34 & 2.25 & 2.34\\ 
        GC &$Y_1(\mb{s}_1)$ & $Y_2(\mb{s}_{out})$ & 95.5 & 95.2 & 95.6 & 3.06 & 3.01 & 3.09\\ 
        GC&$Y_1(\mb{s}_1)$, $Y_1(\mb{s}_{out})$ & $Y_2(\mb{s}_{out})$ &  95.3 & 94.6 & 95.3 & 2.30 & 2.24 & 2.33 \\ 
    \end{tabular}
    \caption{Prediction interval summaries for setting B and the colocated design. We compare estimated CH, Mat\'ern (M), and generalized Cauchy (GC) multivariate covariances used for prediction. }
    \label{tab:matern_interval_w_pred_colocated}
\end{table}

\clearpage
\bibliographystyle{apalike}
\bibliography{mch_rev2.bib}